\documentclass[twocolumn,trackchanges,astrosymb,tighten]{aastex631}

\newcommand{\mrm}[1]{\mathrm{#1}}
\newcommand{\mat}[1]{\textsf{\textbf{#1}}}

\newcommand{\x}{$\times$}

\newcommand{\ie}{i.e.}
\newcommand{\eg}{e.g.}

\definecolor{orange}{rgb}{1.0, 0.49, 0}

\defcitealias{P18A8}{P18}
\defcitealias{hem19}{H19}
\defcitealias{hem21}{H21}
\defcitealias{DES_Y1_3x2pt}{DES Y1}

\usepackage{graphicx}	
\usepackage{amsmath}	
\usepackage{color}
\usepackage{bm,cancel}
\usepackage{multirow}

\shorttitle{Baryonic Physics with DES Y1 + Planck 6$\times$2}
\shortauthors{Xu et al.}

\begin{document}

\title{Constraining Baryonic Physics with DES Y1 and Planck data - Combining Galaxy Clustering, Weak Lensing, and CMB Lensing}

\correspondingauthor{Jiachuan Xu}
\email{jiachuanxu@arizona.edu}

\author[0000-0003-0871-8941]{Jiachuan Xu}
\affiliation{Department of Astronomy/Steward Observatory, University of Arizona, 933 North Cherry Avenue, Tucson, AZ 85721-0065, USA}

\author{Tim Eifler}
\affiliation{Department of Astronomy/Steward Observatory, University of Arizona, 933 North Cherry Avenue, Tucson, AZ 85721-0065, USA}

\author{Vivian Miranda}
\affiliation{C. N. Yang Institute for Theoretical Physics, Stony Brook University, Stony Brook, NY 11794, USA}
\affiliation{Department of Physics \& Astronomy, Stony Brook University, Stony Brook, NY 11794, USA}

\author{Xiao Fang}
\affiliation{Department of Astronomy/Steward Observatory, University of Arizona, 933 North Cherry Avenue, Tucson, AZ 85721-0065, USA}
\affiliation{Berkeley Center for Cosmological Physics, UC Berkeley, CA 94720, USA}

\author{Evan Saraivanov}
\affiliation{Department of Physics \& Astronomy, Stony Brook University, Stony Brook, NY 11794, USA}

\author{Elisabeth Krause}
\affiliation{Department of Astronomy/Steward Observatory, University of Arizona, 933 North Cherry Avenue, Tucson, AZ 85721-0065, USA}
\affiliation{Department of Physics, University of Arizona, 1118 E Fourth Str, Tucson, AZ 85721, USA}

\author{Hung-Jin Huang}
\affiliation{Department of Astronomy/Steward Observatory, University of Arizona, 933 North Cherry Avenue, Tucson, AZ 85721-0065, USA}

\author{Karim Benabed}
\affiliation{Sorbonne Université, CNRS, UMR7095, Institut d’Astrophysique de Paris, 98 bis Boulevard Arago, 75014, Paris, France}

\author{Kunhao Zhong}
\affiliation{Department of Physics \& Astronomy, Stony Brook University, Stony Brook, NY 11794, USA}



\begin{abstract}

We constrain cosmology and baryonic feedback scenarios with a joint analysis of weak lensing, galaxy clustering, cosmic microwave background (CMB) lensing, and their cross-correlations (so-called 6$\times$2) using data from the Dark Energy Survey (DES) Y1 and the Planck satellite mission. 
Noteworthy features of our 6$\times$2 pipeline are: We extend CMB lensing cross-correlation measurements to a band surrounding the DES Y1 footprint (a $\sim 25\%$ gain in pairs), and we develop analytic covariance capabilities that account for different footprints and all cross-terms in the 6$\times$2 analysis. We also measure the DES Y1 cosmic shear two-point correlation function (2PCF) down to $0\farcm25$, but find that going below $2\farcm5$ does not increase cosmological information due to shape noise. 
We model baryonic physics uncertainties via the amplitude of Principal Components (PCs) derived from a set of hydro-simulations. Given our statistical uncertainties, varying the first PC amplitude $Q_1$ is sufficient to model small-scale cosmic shear 2PCF.
For DES Y1+Planck 6$\times$2 we find $S_8=0.799\pm0.016$, comparable to the 5$\times$2 result of DES Y3+SPT/Planck $S_8=0.773\pm0.016$. 
Combined with our most informative cosmology priors --- baryon acoustic oscillation (BAO), big bang nucleosynthesis (BBN), type Ia supernovae (SNe Ia), and Planck 2018 EE+lowE, we measure $S_8=0.817\pm 0.011$.
Regarding baryonic physics constraints, our 6$\times$2 analysis finds $Q_1=2.8\pm1.8$. Combined with the aforementioned priors, it improves the constraint to $Q_1=3.5\pm1.3$. 
For comparison, the strongest feedback scenario considered in this paper, the cosmo-OWLS AGN ($\Delta T_\mathrm{heat}=10^{8.7}$ K), corresponds to $Q_1=5.84$.
\end{abstract}

\section{Introduction}
\label{sec:intro}
Baryonic physics, including radiative cooling, star formation, active galactic nucleus (AGN) feedback, and supernovae (SNe) feedback, is one of the main modeling uncertainties in cosmological analyses that attempt to utilize the small-scale information of the (non-linear) matter density field. 
Efficient gas cooling can foster star formation and make the matter distribution more concentrated in the halo center~\citep{M04}, while effective baryonic feedback process from AGN and SNe can heat the intergalactic medium (IGM) or eject gas to the outskirt of halos, thus suppress the power at quasi-linear scales~\citep[e.g.,][]{JZL+06,RZK08}.
In particular, if cosmic shear is included in the data vector, several groups have shown a non-negligible impact of baryonic physics as an astrophysical systematic for stage-III (Dark Energy Survey\footnote{DES: \href{https://www.darkenergysurvey.org}
{\nolinkurl{https://www.darkenergysurvey.org}}}, Hyper Suprime Camera Subaru Strategic Program\footnote{HSC: \href{http://www.naoj.org/Projects/HSC/HSCProject.html}
{\nolinkurl{http://www.naoj.org/Projects/HSC/HSCProject.html}}}, Kilo-Degree Survey\footnote{KiDS: \href{http://www.astro-wise.org/projects/KIDS/}
{\nolinkurl{http://www.astro-wise.org/projects/KIDS/}}}) and stage-IV surveys (Rubin Observatory's Legacy Survey of Space and Time\footnote{LSST: \href{https://www.lsst.org}
{\nolinkurl{https://www.lsst.org}}}, Euclid\footnote{Euclid: \href{https://sci.esa.int/web/euclid}
{\nolinkurl{https://sci.esa.int/web/euclid}}}, Roman Space Telescope\footnote{Roman: \href{https://roman.gsfc.nasa.gov}
{\nolinkurl{https://roman.gsfc.nasa.gov}}})~\citep[e.g.,][]{SHS+11,SHS13,ekd15,MPH+15,ST15,CRD+18,CMJ+19,hem19,STS+19,hem21,MBT+21,AAC+21,FSD+23}
In addition, the current and next generation CMB experiments (South Pole Telescope\footnote{SPT: \href{https://pole.uchicago.edu/public/Home.html}{\nolinkurl{https://pole.uchicago.edu/public/Home.html}}}, 
Atacama Cosmology Telescope\footnote{ACT: \href{https://act.princeton.edu}{\nolinkurl{https://act.princeton.edu}}},
Simons Observatory\footnote{SO: \href{https://simonsobservatory.org}{\nolinkurl{https://simonsobservatory.org}}} and CMB-S4\footnote{S4: \href{https://cmb-s4.org}{\nolinkurl{https://cmb-s4.org}}}) will probe the CMB at higher resolution and sensitivity; baryonic physics models are essential ingredients to interpret signals from CMB lensing, thermal/kinetic Sunyaev-Zel'dovich (tSZ and kSZ) effect, and cosmic infrared background (CIB)~\citep{ogp19,obc19,DESY1_5x2pt_method,DESY1xPlanckxSPT_6x2pt,RAH+21,YWT+21,TMH+22,SGA+22,PLB+23,O22,HFP+23,FKE+23,FKM+23,RNR+23}.

The default choice to mitigate parameter biases due to baryonic effects is discarding or suppressing the small-scale data at the cost of lower constraining power, which is widely used in ongoing stage-III surveys~\citep{DES_Y1_3x2pt,DES_Y3_3x2pt,ATH+20,KiDS_shear20,HSC_Y1_2PCF}. 
However, several ideas have been proposed by the community to improve the modeling/marginalization techniques for baryons and utilize the small-scale information~\citep[see][for a review]{CMJ+19}. 
The first approach is based on halo models, where additional degrees of freedom (d.o.f.) are introduced to depict the impact of baryonic effects. 
\cite{SHS+11,SHS13,F14,MMT+14} break the halo density profile into dark matter, gas, and star components and model them separately. 
\cite{ZRH08,ZSD+13,MPH+15,MBT+21} modify the mass-concentration relation found in dark matter-only (DMO) simulations to include baryonic effects, with higher concentration when baryonic feedback is stronger. 
The \texttt{HMCode}/\texttt{HMx} developed by \cite{MPH+15,MTH+20,MBT+21} has been used in analyzing data from DES~\citep{DES_Y3_WL2,DESY3_KiDS1000_shear}, KiDS~\citep{KiDS_shear20,KiDS1000_3x2pt,JLA+21}, and HSC~\citep{HSC_Y3_LZS+23,HSC_Y3_DLN+23,HSC_Y3_SMM+23,HSC_Y3_MST+23} to mitigate baryonic effects and improve constraints on cosmology.

\cite{DMS20} find that at given scale, the power spectrum suppression due to baryonic effects highly correlates with the baryon fraction inside halos. Thus, they propose a fitting function for the power modification with only one free parameter of the halo baryon fraction.
The third approach, called baryonification or baryonic correction model~\citep{ST15,DFS18,STS+19,AAH+20}, post-processes $N$-body simulations directly and perturbs particles' positions to mimic baryonic effects. Then, matter power spectra are measured from perturbed $N$-body simulations, and fitting functions are derived from the power suppression as a function of baryon-related parameters, depending on the detailed baryonification methods.

Ideally, trustworthy hydrodynamic simulations would provide an emulator that returns the matter power spectrum as a function of cosmological and baryonic physics parameters. Initial/important work in this direction can be found in ~\cite{AZC+21,AAH+20,AAC+21,VAG+21,GS21,SMK+23}. 
\cite{AAZ+23,CAA+23} use an emulator trained with \texttt{BACCO} simulations to constrain cosmology and baryonic feedback with DES Y3 data. 
In the absence of such an emulator, approximate methods can capture the impact of baryonic physics, with the inherent approximation that the dependence of feedback processes on cosmology is negligible. 
The caveat is that cosmological parameters like $\Omega_{m}$ and $\Omega_{b}$ can alter the mean halo baryon fraction and hence affect the baryonic feedback strength~\citep{DAT+23}.

Extracting PCs of suites of hydro-simulations of different baryonic feedback models and cosmologies is another avenue to capture the impact of baryonic physics over a range of scenarios~\citep[][hereafter H19 and H21]{ekd15,hem19,hem21}. The method is inherently limited to the physics range spanned by the simulations, and model complexity is determined by the number of PCs included.  
As such, a PC-based baryonic analysis will not be able to constrain ``first-principle'' baryonic physics parameters, e.g., the density profile of gas or stellar component, ejection mass fraction and radius, and typical halo mass to retain half of its gas~\citep{ST15,STS+19,AAH+20,AAH+21}. Using PCs, one can only make aggregate statements that a particular simulation is in tension or agreement with the data. However, PCs have the advantage that they represent linear combinations of these physical baryonic parameters, and these PCs are ranked as a function of importance for the summary statistics that are used in the analysis. In the presence of limited constraining power, which is true for all stage-III surveys, such a minimal model may be less affected by overfitting problems and unphysical parameter degeneracy/projection effects.

In this paper, we constrain cosmology and baryonic effects in our Universe with a 6$\times$2 joint analysis of DES Y1 galaxies~\citep[][hereafter DES Y1]{DES_Y1_3x2pt} and Planck 2018 CMB lensing data~\citep[][hereafter P18]{P18A8}. We closely follow the DES Y1 3$\times$2 analysis as presented in \citetalias{hem21} and extend their method to 6$\times$2. We find that, given the constraining power of our data, 1 PC is sufficient to capture the baryonic physics uncertainties that otherwise can bias our results.

We start the paper by measuring the cross-correlations between DES Y1 galaxies and Planck CMB lensing convergence. Our measurement extends beyond the DES footprint, taking advantage of the full-sky coverage of the Planck CMB lensing map. We gain $\sim 25\%$ pair counts at $250\arcmin$ compared to using the DES footprint only. We also extend the cosmic shear 2PCF measurement to $0\farcm 25$ to explore whether additional information can be gained from these small scales. 

We build a forward modeling framework for the full 6$\times$2 data vector and develop code to compute the non-Gaussian covariance matrix of the 6$\times$2 data vector, including survey geometry effects and covariance between CMB lensing band-power and other configuration space 2PCFs. We correct for the survey geometry effect to the leading order in the covariance matrix. 

Our paper is structured as follows: we present our measurements and modeling methodology including the 6$\times$2 covariance in Section~\ref{sec:method}. Analysis choices, including systematics, scale cuts, and blinding strategy, as well as model validation, are discussed in Section~\ref{sec:choices}. Results on cosmology parameters are shown in Section~\ref{sec:result_LCDM}, and our constraints on baryonic feedback scenarios are presented in Section~\ref{sec:result_Q1}. We discuss the robustness of our results in Section~\ref{sec:robustness} and conclude in Section~\ref{sec:conclusions}.

\begin{figure*}
    \centering
    \includegraphics[width=\textwidth]{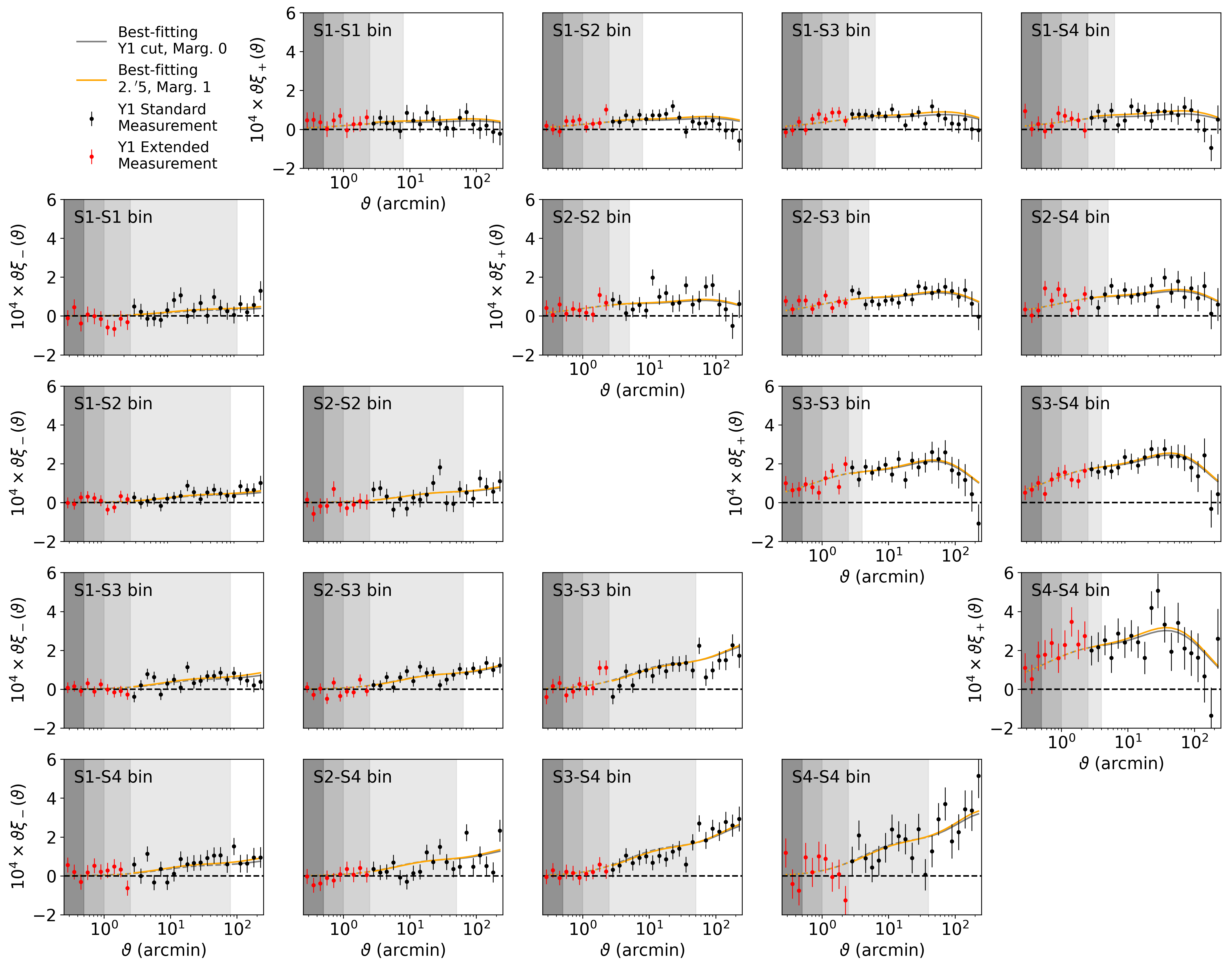}
    \caption{The $\hat{\xi}_\pm^{ab}(\vartheta)$ measurement from the DES Y1 \textsc{metacalibration} shape catalog extended to $\vartheta \geq 0\farcm 25$ (new data points in red). The shaded regions in each panel, from right to left show the data points included depending on scale-cuts: \citetalias{DES_Y1_3x2pt} (white only), \citetalias{hem21} scale cut of $\vartheta \geq 2\farcm5$ (light gray), and three additional shaded gray regions corresponding to $\vartheta \geq 1\arcmin$, $\vartheta \geq 0\farcm5$, and $\vartheta \geq 0\farcm25$. We show the best-fitting model vectors in solid lines within their valid scales and non-solid lines when extrapolated to small scales. The gray lines are the best-fitting model vector assuming the standard Y1 scale cut without baryon PCs marginalization, and the orange lines are from our 6$\times$2 (including baryons) baseline analysis.
    }
    \label{fig:xipm_025}
\end{figure*}

\begin{figure*}
    \centering
    \includegraphics[width=\textwidth]{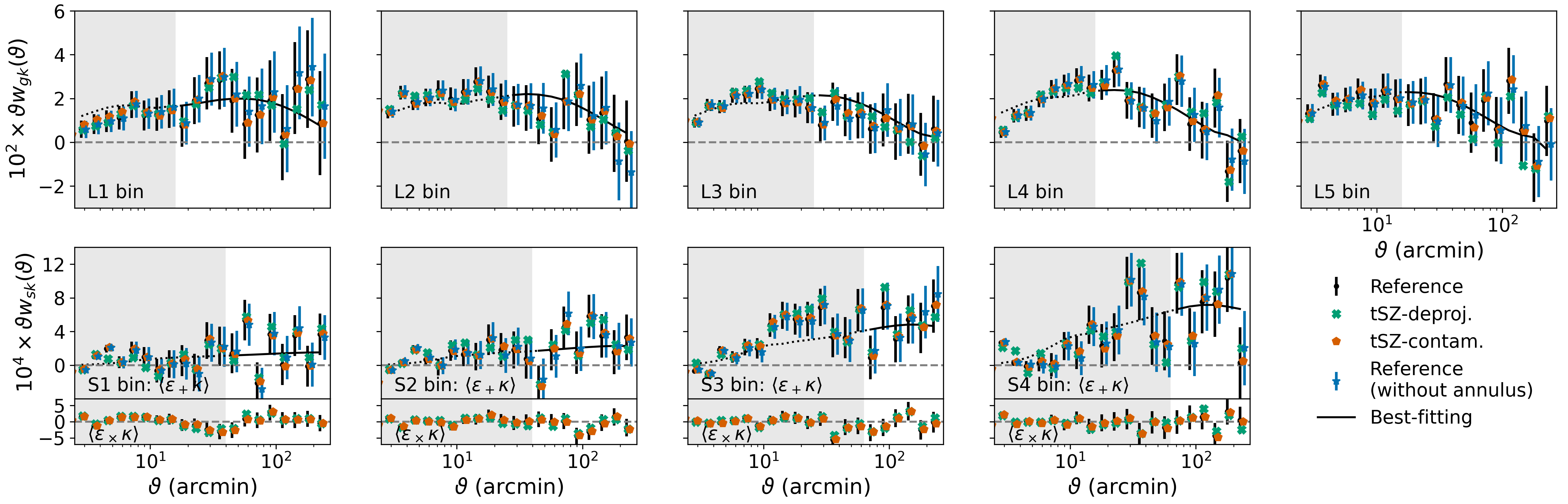}
    \caption{The two-point correlation functions $\hat{w}_{g\kappa}^a(\vartheta)$ (top row) and $\hat{w}_{s\kappa}^a(\vartheta)$ (bottom row, with $\langle\epsilon_+\kappa\rangle$ and $\langle\epsilon_\times\kappa\rangle$ separated) measured between DES Y1 galaxies and reconstructed Planck CMB lensing convergence. 
    The black data points are based on the reference CMB lensing reconstruction; green and orange points are correspondingly measured with tSZ-deprojected and tSZ-contaminated maps. 
    Blue data points are measured using the reference reconstruction but only over the DES Y1 footprint. 
    Error bars of the reference map measurements with and without extended annulus are the square roots of the diagonal elements in the analytic covariance matrix. We omit the error bars of tSZ-variant measurements because they are the same as the reference ones.
    We show the DES Y1 scale-cuts in~\citet{ogp19,obc19} as shaded regions. 
    The best-fitting prediction from our 6$\times$2 baseline analysis is shown as the black lines (solid for scales above and dotted for scales below the scale cut).}
    \label{fig:gksk_agr2}
\end{figure*}
\section{Methodology}
\label{sec:method}
We develop our 6$\times$2 modeling and inference pipeline within \texttt{CoCoA}, the \texttt{Cobaya}-\texttt{CosmoLike} Architecture. \texttt{CoCoA} integrates \texttt{CosmoLike}~\citep{EKS+14,KE17,FES+22} and \texttt{Cobaya}~\citep{TL21} into a precise and efficient pipeline that relies on a highly-optimized \texttt{OpenMP} shared-memory allocation and cache system \citep[also see][for details]{ZSM+23}.

We assume a Gaussian likelihood~\citep[see][for discussions of the validity]{likelihood, SHH18} of the data vector $\bm{D}$ given the parameters $\bm{p}$
\begin{equation}
\label{eq:like}
    \mathcal L(\bm{D}|\bm{p}) \propto  \mathrm{exp}[-\frac{1}{2}(\bm{D}-\bm{M})^\mathrm{T}\cdot \mat{C}^{-1}\cdot(\bm{D}-\bm{M})]\, .
\end{equation}
Below, we detail derivations of data vector $\bm{D}$, model vector $\bm{M}$, and covariance $\mat{C}$.

\subsection{Data Vector}
\subsubsection{Catalogs and Maps}
\label{sec:data_source}
We use the DES Y1 \textsc{redMaGiC} sample~\citep{EP+_redmagic,R+_redmagic} as the lens galaxies and the \textsc{metacalibration} sample~\citep{HM17,SH17,ZSS+18,HGB+18,GVD+18} as the source galaxies\footnote{\href{https://des.ncsa.illinois.edu/releases/y1a1/key-catalogs}{https://des.ncsa.illinois.edu/releases/y1a1/key-catalogs}}. 
Both samples cover an area of 1,321 deg$^2$. 
The lens galaxies are divided into five tomography bins, and the galaxy density of each bin is $(0.0134,\,0.0343,\,0.0505,\,0.0301,\,0.0089)$ arcmin$^{-2}$. The source galaxies are divided into four tomography bins, and the galaxy density is $(1.496,\,1.5189,\,1.5949,\,0.7949)$ arcmin$^{-2}$. The shape noise of the source sample is $\sigma_{\bm{\epsilon}}=0.394$ (including both the shear components). 

For the CMB lensing convergence, we use the Planck PR3\footnote{\href{http://pla.esac.esa.int/pla/\#cosmology}{\nolinkurl{http://pla.esac.esa.int/pla/\#cosmology}}} in the baseline 
analysis~\citepalias{P18A8}. 
PR3 contains multiple variations of CMB lensing convergence reconstruction. The baseline reconstruction is estimated from \texttt{SMICA} DX12 CMB maps~\citep{P18A4} and includes reconstructions from temperature-only (TT), polarization-only (PP), and minimum-variance (MV) estimates. We use the MV map in this work. 

\subsubsection{Two-point Function Estimators}
\label{sec:estimators}
We consider six two-point functions derived from the three observables of lens galaxy overdensity $\delta_g$, source galaxy shape $\epsilon_{1/2}$, and CMB lensing convergence $\kappa$\footnote{Unless mentioned specifically, we use $\kappa$ for CMB lensing convergence and $\kappa_g$ for galaxy lensing convergence in this work.}: 
cosmic shear $\xi_\pm^{ab}(\vartheta)$, 
galaxy-galaxy lensing $\gamma_{t}^{ab}(\vartheta)$, 
galaxy clustering $w_g^a(\vartheta)$, 
galaxy-CMB lensing convergence $w_{g\kappa}^a(\vartheta)$, 
(tangential) galaxy lensing-CMB lensing convergence $w_{s\kappa}^a(\vartheta)$, 
and CMB lensing convergence band-power $C_{L_{b}}^{\kappa\kappa}$. 
The superscript ${a,\,b}$ refers to tomography bin indices, $\vartheta$ is the average angular separation in the real space angular bin, and $L_{b}$ is the average angular multipole in the Fourier space angular bin.

The $C_{L_{b}}^{\kappa\kappa}$ measurement is directly taken from Planck data release~\citepalias{P18A8} while the other five auto- and cross-2PCFs are measured using \texttt{TreeCorr}~\citep{JBJ04}. For conciseness, we first define the catalog average 
\begin{equation}
    \langle X Y \rangle_\mathrm{cat}\equiv \sum_{ij}W_iW_j\Delta_\vartheta(ij)(X_iY_j)/N_{p}(\vartheta) \,,
\end{equation}
where $i$ and $j$ are iterated over the galaxy catalog or \texttt{HEALPix} pixels~\citep{Zonca2019,2005ApJ...622..759G}, the weight $W_i$ is a binary survey footprint mask, $\Delta_\vartheta(ij)$ is a bin-averaged delta function, i.e. $\Delta_\vartheta(ij)=1$ if $\left|\bm{\theta}_i-\bm{\theta}_j\right|\in[\vartheta-\Delta\vartheta/2, \vartheta+\Delta\vartheta/2)$ and $\Delta_\vartheta(ij)=0$ otherwise. 
$N_{p}(\vartheta)$ is the pair counts in the angular bin. $X$ and $Y$, depending on the tracer samples, can be one of position indicator function $\bm{1}^a_\mathrm{D}$ (lens sample data catalog, in tomographic bin $a$), $\bm{1}^a_\mathrm{R}$ (lens sample randomized catalog, in tomography bin $a$), $\hat{\epsilon}^a_{1/2}$ (source sample shape), and $\hat{\kappa}$ (reconstructed CMB lensing convergence field). 

Using these conventions, the 5$\times$2 estimators read as
\begin{subequations}
    \begin{equation}
        \hat{\xi}_\pm^{ab}(\vartheta) \equiv \left(\langle\hat{\epsilon}^a_1\hat{\epsilon}^b_1\rangle_\mathrm{cat}\pm\langle\hat{\epsilon}^a_2\hat{\epsilon}^b_2\rangle_\mathrm{cat}\right)/\left(R^aR^b\right)\,,
    \end{equation}
    \begin{equation}
        \hat{\gamma}_{t}^{ab}(\vartheta) \equiv \left(\langle\bm{1}_\mathrm{D}^a\hat{\epsilon}_+^b\rangle_\mathrm{cat} - \langle\bm{1}_\mathrm{R}^a\hat{\epsilon}_+^b\rangle_\mathrm{cat}\right)/R^b\,,
    \end{equation}
    \begin{equation}
        \hat{w}_{g}^a(\vartheta) \equiv \frac{\langle\bm{1}_\mathrm{D}^a\bm{1}_\mathrm{D}^a\rangle_\mathrm{cat}-2\langle\bm{1}_\mathrm{D}^a\bm{1}_\mathrm{R}^a\rangle_\mathrm{cat}+\langle\bm{1}_\mathrm{R}^a\bm{1}_\mathrm{R}^a\rangle_\mathrm{cat}}{\langle\bm{1}_\mathrm{R}^a\bm{1}_\mathrm{R}^a\rangle_\mathrm{cat}}\,,
    \end{equation}
    \begin{equation}
        \hat{w}_{g\kappa}^a(\vartheta) \equiv \langle\bm{1}_\mathrm{D}^a\hat{\kappa}\rangle_\mathrm{cat} - \langle\bm{1}_\mathrm{R}^a\hat{\kappa}\rangle_\mathrm{cat}\,,
    \end{equation}
    \begin{equation}
        \hat{w}_{s\kappa}^a(\vartheta) \equiv \langle\hat{\epsilon}_+^a\hat{\kappa}\rangle_\mathrm{cat}/R^a\,,
    \end{equation}
\end{subequations}
where $R^a$ is the mean shear response of the source sample in the $a$-th bin~\citep{ZSS+18,HM17,SH17}, and $\hat{\epsilon}_+^a$ is the tangential shear when counting pairs. 

\subsubsection{Measurement of 3$\times$2}
\label{sec:xipm_025}
The measured $\hat{\xi}_\pm^{ab}(\vartheta)$ with 30 logarithmic bins between $\vartheta\in [$0\farcm 25$,\,250^\prime]$ are shown in Fig.~\ref{fig:xipm_025}. The red data points show the new small-scale measurements going beyond the scales $\vartheta\in [$2\farcm 5$,\,250^\prime]$ which were used in previous analyses~\citepalias{DES_Y1_3x2pt,hem21}.
The \citetalias{DES_Y1_3x2pt} analysis imposed a redshift-dependent scale-cut to avoid contamination of baryonic effects, while the re-analysis of \citetalias{hem21} included all scales down to $2\farcm 5$ (c.f. white region in Fig.~\ref{fig:xipm_025} vs light gray shaded region). 

In this paper, we consider three additional scale cuts, indicated by the darker gray-shaded regions in Fig.~\ref{fig:xipm_025}. From right to left, these correspond to $\vartheta \geq 1\arcmin$, $\vartheta \geq 0\farcm5$, and $\vartheta \geq 0\farcm25$. The error bars of all data points are calculated as the square roots of the diagonal elements in the analytic covariance matrix. 

We also show the best-fitting models assuming the standard Y1 scale cut and our 6$\times$2 baseline analysis (including baryon mitigation) as gray and orange lines, respectively. The orange lines are calculated using the \citetalias{hem21} scale cut $\vartheta\in [2\farcm 5,\,250^\prime]$. Both are extrapolated to smaller scales as gray-dashed and orange-dashed
lines, respectively. 

We explore the potential benefits of using more aggressive scale cuts than \citetalias{hem21} in Section \ref{sec:syn_ss_smallscale}.

\subsubsection{CMB Cross-correlation Measurements}
\label{sec:optimal_measurements}

Before measuring $\hat{w}_{g\kappa}^a$ and $\hat{w}_{s\kappa}^a$, we smooth the reconstructed Planck CMB lensing $\hat{\kappa}$ map with a Gaussian beam of an FWHM = $7\arcmin$ to suppress the small-scale lensing reconstruction noise. 
This beam size matches the Planck 143 GHz channel \citep{P15PA7}. 
We further impose a scale cut of $8\leq L\leq 2048$ in Fourier space, which matches the aggressive $L$-cut (\texttt{agr2}) in Planck CMB lensing reconstruction. 
Modes at $L<8$ are sensitive to the mean-field subtraction and could be biased by the fidelity of FFP10 simulations~\citep{P18A3}. 
We then transform the map from Fourier space to \texttt{HEALPix} map with $N_\mathrm{side}=1024$, and mask out foreground contamination. $\hat{w}_{g\kappa}^a$ and $\hat{w}_{s\kappa}^a$ are measured using \texttt{TreeCorr} with 20 logarithmic bins in $\vartheta \in [2\farcm 5,\, 250^\prime]$. 

The $\hat{w}_{g\kappa}^a$ and $\hat{w}_{s\kappa}^a$ measurement results are shown as the black dots in Fig.~\ref{fig:gksk_agr2}. We adopt the scale-cuts of \cite{ogp19,obc19} and similarly define the signal-to-noise (S/N) as 
\begin{equation}
    S/N=\sqrt{\chi^2}=\sqrt{\bm{D}\cdot\mat{C}^{-1}\cdot\bm{D}}\,,
\end{equation}
where $\mat{C}$ is the covariance matrix and data vector $\bm{D}$ is $\hat{w}^a_{g\kappa}$ and $\hat{w}^a_{s\kappa}$.
We find S/N values of $7.6\sigma$ for $\hat{w}_{g\kappa}$ and $6.5\sigma$ for $\hat{w}_{s\kappa}$, respectively. The combined detection S/N for $\hat{w}_{g\kappa}+\hat{w}_{s\kappa}=9.6\sigma$. As a comparison, \cite{ogp19} find $9.9\sigma$ detection of lensing in $\hat{w}_{g\kappa}$ and \cite{obc19} find 5.8$\sigma$ detection of lensing in $\hat{w}_{s\kappa}$. 

\paragraph{tSZ contamination} 
Imperfect component separation when building the CMB temperature map results in residual contamination from tSZ and CIB signals. Those residuals propagate into the lensing potential estimate through the quadratic estimator reconstruction and contribute to the cross-correlation with LSS traced by DES Y1 galaxies, $w_{g\kappa}$ and $w_{s\kappa}$, at small-scales. 

\citetalias{P18A8} find that the CIB residual contributes sub-percent bias in the reconstructed CMB lensing band power. The tSZ bias is larger but still a negligible fraction of the error budget.

We adopt the scale cut used in~\cite{obc19,ogp19}, i.e. $\vartheta_\mathrm{min}^{g\kappa}=(15\farcm8,\, 25\arcmin,\,25\arcmin,\,15\farcm8,\,15\farcm8)$ and $\vartheta_\mathrm{min}^{s\kappa}=(39\farcm6,\,39\farcm6,\,62\farcm8,\,62\farcm8)$. 
Since their combined Planck+SPT map includes smaller scales than our $L_\mathrm{max}=2048$ map, we note that this is a conservative choice.

To estimate the tSZ bias for the 6$\times$2 analyses, we derive three sets of $\hat{w}_{g\kappa}^{a}(\vartheta)$ and $\hat{w}_{s\kappa}^{a}(\vartheta)$ measurements using CMB lensing convergence maps with different tSZ treatments, and see how it affects the posteriors. The three $\hat{\kappa}$ map variants are:
\begin{enumerate}
    \item the baseline Planck lensing reconstruction map\footnote{\texttt{COM\_Lensing\_4096\_R3.00}}, which is built from \texttt{SMICA} DX12 CMB maps. 
    When building the component-separated CMB maps, the input frequency maps have to be pre-processed by masking compact sources and resolved SZ clusters~\citep[see Appendix D in][for more details]{P18A4}. 
    SZ clusters in the 2015 Planck SZ catalog with $S/N>5$ are also masked, as well as a galactic mask blocking the galactic plane. 
    Then the baseline \texttt{SMICA} algorithm is applied to the input frequency maps.
    \item the Planck lensing reconstruction map with tSZ contamination\footnote{\texttt{COM\_Lensing\_Sz\_4096\_R3.00}}. The map is built with the same method as the baseline one, except that the $S/N>5$ SZ clusters are not masked.
    \item the Planck tSZ-deprojected lensing reconstruction map\footnote{\texttt{COM\_Lensing\_Szdeproj\_4096\_R3.00}}, which is reconstructed using tSZ-deprojected \texttt{SMICA} CMB map (TT only). 
    The tSZ-deprojection is implemented by a constrained internal linear combination, requiring that the tSZ signal is canceled out when solving for the multipole weights, which are used to combine the input frequency map multipoles~\citep{P18A4}.
\end{enumerate} 

We calculate $\Delta\chi^2=0.49$ between the reference and the tSZ-contaminated measurements and $\Delta\chi^2=14.06$ between the reference and the tSZ-deprojected data vectors and conclude that this effect is negligible (also see Section \ref{sec:robustness}). The data points and the relevant scale cuts are shown in Fig.~\ref{fig:gksk_agr2}. 

We perform null-test using the cross-component of the ellipticity $\epsilon_\times$ instead of the tangential component $\epsilon_+$ in the data vector, i.e. $\langle\hat{\epsilon}_\times^a\hat{\kappa}\rangle_\mathrm{cat}/R^a$, to check for residual systematics in our measurements. The $\chi^2$-values of data vectors measured with different maps and tomography bins are shown in Table~\ref{tab:sk_null_test}. We compute the probability to exceed (PTE, values in brackets in Table~\ref{tab:sk_null_test}) which quantifies the probability to observe equal or more extreme $\chi^2$ values given our null hypothesis $\langle\hat{\epsilon}_\times^a\hat{\kappa}\rangle_\mathrm{cat}/R^a=0$. The PTE values of the combined samples are close to 1 for all the maps considered; we conclude that our $\chi^2$ values are highly consistent with our null hypothesis.

\begin{figure}
    \centering
    \includegraphics[width=\linewidth]{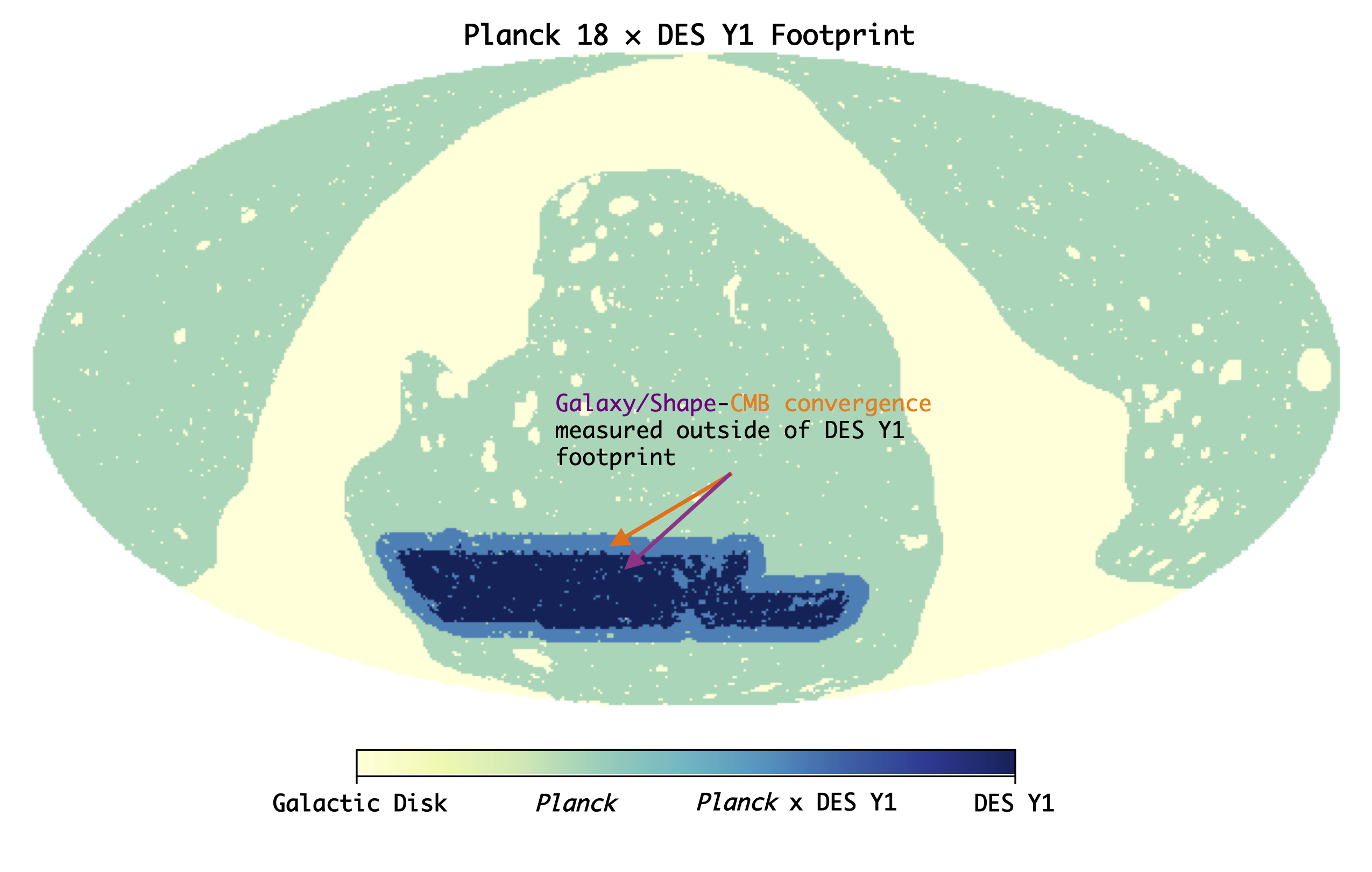}
    \caption{Illustration of the $\hat{w}^a_{g\kappa/s\kappa}$ measurement method and footprint treatment. The yellow region is masked due to e.g. galactic foreground and tSZ contamination from large galaxy clusters. The green region is the footprint of Planck $\hat{\kappa}$ field. The deep blue region is the DES Y1 footprint. When measuring $\hat{w}^a_{g\kappa/s\kappa}$, we include $\hat{\kappa}$ outside the DES Y1 footprint but within the angular binning, which forms an extended annulus (the light blue region) around the DES Y1 footprint.}
    \label{fig:footprint}
\end{figure}

\paragraph{Footprint extension} Our cross-correlation measurements $\hat{w}_{g\kappa}^a$ and $\hat{w}_{s\kappa}^a$ include $\hat{\kappa}$ pixels outside the DES Y1 footprint. This is illustrated in Fig.~\ref{fig:footprint}, where the light blue region around the DES Y1 footprint indicates the additional $\hat{\kappa}$ measurements included in the cross-correlations.
In Fig.~\ref{fig:gksk_agr2} we also show the $\hat{w}_{g\kappa}^a$ and $\hat{w}_{s\kappa}^a$ with and without $\hat{\kappa}$ pixels outside the DES Y1 footprint. While this type of measurement involving two different footprints complicates covariance calculations (see Section \ref{sec:covmat}), it increases $\sim25\%$ pair counts at $250\arcmin$.

\begin{table}
    \centering
    \begin{tabular}{lcccc}
    \toprule
    \multirow{2}{*}{Sample} & \multirow{2}{*}{d.o.f.} & \multicolumn{3}{c}{$\chi^2_\mathrm{null}\equiv \bm{D}\cdot\mat{C}^{-1}\cdot\bm{D}$ (PTE)}\\
    \cline{3-5}
      &  & Reference & tSZ-deproj. & tSZ-contam.\\
    \tableline
    S1  & 8  &  2.98(0.94) &  3.50(0.90) &  3.14(0.93)\\
    S2  & 8  &  5.78(0.67) &  4.62(0.80) &  5.92(0.66)\\
    S3  & 6  &  4.79(0.57) &  4.22(0.65) &  4.28(0.64)\\
    S4  & 6  &  2.38(0.88) &  2.58(0.86) &  2.61(0.86)\\
    \tableline
    All & 28 & 19.28(0.89) & 17.54(0.94) & 19.30(0.89)\\
    \tableline
    \end{tabular}
    \caption{Null test of the cross-correlation of cross-component shape $\epsilon_\times$ and CMB convergence $\kappa$. We show the degree-of-freedom in the second column for samples in each tomography bin as well as all bins combined. The $\chi^2_\mathrm{null}$ and associated probability-to-exceed (PTE) for measurements with different tomography bins and $\hat{\kappa}$ map variants are shown in the right three columns.}
    \label{tab:sk_null_test}
\end{table}

\subsection{Model Vector}
\label{sec:mv}
We adopt the Limber approximation and curved-sky geometry when calculating the model vector. The 6$\times$2 model vector  depends on the angular power spectra of the tracer field $X$ and $Y$, in tomographic bin $a$ and $b$,
\begin{equation}
    \begin{aligned}
        C_{XY}^{ab}(\ell) = \int \mathrm{d}\chi \frac{q_{X}^a(\chi)q_{Y}^b(\chi)}{\chi^2}P_{XY}^{ab}(l/\chi, z(\chi))\,.
    \end{aligned}
\end{equation}
Depending on the tracer field (galaxy overdensity $\delta_g$, galaxy lensing convergence $\kappa_g$, CMB lensing convergence $\kappa$), $q_{X}^a$ reads
\begin{subequations}
    \begin{equation}
        q_{\kappa_g}^a = \frac{3H_0^2\Omega_{m}}{2c^2}\frac{\chi}{a(\chi)}\int_\chi^{\chi_{h}}\mathrm{d}\chi^\prime\frac{n_\mathrm{src}^a(z(\chi^\prime))\mathrm{d}z/\mathrm{d}\chi^\prime}{\bar{n}_\mathrm{src}^a}\frac{\chi^\prime-\chi}{\chi^\prime}\,,
    \end{equation}
    \begin{equation}
        q_{\delta_g}^a = \frac{n_\mathrm{lens}^a(z(\chi))}{\bar{n}_\mathrm{lens}^a}\frac{\mathrm{d}z}{\mathrm{d}\chi}\,,
    \end{equation}
    \begin{equation}
        q_{\kappa} = \frac{3H_0^2\Omega_{m}}{2c^2}\frac{\chi}{a(\chi)}\frac{\chi_\star-\chi}{\chi_\star}\,,
    \end{equation}
\end{subequations}
where $\chi_{h}$ is the comoving distance to the horizon and $\chi_\star$ is the comoving distance to the last scattering surface, $n_\mathrm{src/lens}^a(z)$ is the redshift distribution of the source/lens sample in tomography bin $a$ and $\bar{n}_\mathrm{src/lens}^a$ is its surface density. $P_{XY}^{ab}(k,z)$ is the 3D power spectrum, depending on the tracer ${X}$ and ${Y}$ in bins $a$ and $b$, the expression reads
\begin{subequations} \label{eqn:power_spectrum}
    \begin{equation}\label{eqn:power_spectrum_g}
        P_{\delta_g {Y}}^{ab}(k,z) = b_g^aP_{{m} Y}^b(k,z)\,,
    \end{equation}
    \begin{equation}\label{eqn:power_spectrum_s}
        P_{\kappa_g {Y}}^{ab}(k,z) = P_{{m} Y}^b(k,z)\,,
    \end{equation}
    \begin{equation}\label{eqn:power_spectrum_k}
        P_{\kappa {Y}}^{b}(k,z) = P_{{m} Y}^b(k,z)\,,
    \end{equation}
\end{subequations}
where the subscript ${m}$ stands for matter and $P_{mm}(k,z)$ is the non-linear matter power spectrum from \texttt{Halofit}~\citep[][which is our baseline choice]{TSN+12} or \texttt{EuclidEmulator2}~\citep{EuclidEmu2}. Here, $b_g^a$ is the linear galaxy bias of the lens sample in bin $a$.

\subsubsection{Real Space: 5$\times$2}
\label{sec:mv_real}
We calculate the angular bin-averaged results for the five position-space correlation functions as: 
\begin{subequations}\label{eqn:dv_theory}
    \begin{equation}\label{eqn:dv_xipm}
        \xi_\pm^{ab}(\vartheta) = \sum_\ell \frac{2\ell+1}{2\pi\ell^2(\ell+1)^2}\left[\overline{G_{\ell,2}^+\pm G_{\ell,2}^-}\right]\,C_{\kappa_g\kappa_g}^{ab}(\ell)\,,
    \end{equation}
    \begin{equation}\label{eqn:dv_gammat}
        \gamma_{t}^{ab}(\vartheta) = \sum_\ell \frac{2\ell+1}{4\pi\ell(\ell+1)}\overline{P_\ell^2}\,C_{\delta_g\kappa_g}^{ab}(\ell)\,,
    \end{equation}
    \begin{equation}\label{eqn:dv_wtheta}
        w_{g}^{a}(\vartheta) = \sum_\ell \frac{2\ell+1}{4\pi}\overline{P_\ell}\,C_{\delta_g\delta_g}^{aa}(\ell)\,,
    \end{equation}
    \begin{equation}\label{eqn:dv_w_gk}
        w_{g\kappa}^a(\vartheta) = \sum_\ell \frac{2\ell+1}{4\pi}F(\ell)\,\overline{P_\ell}\,C_{\delta_g\kappa}^{a}(\ell)\,,
    \end{equation}
    \begin{equation}\label{eqn:dv_w_sk}
        w_{s\kappa}^a(\vartheta) = \sum_\ell \frac{2\ell+1}{4\pi\ell(\ell+1)}F(\ell)\,\overline{P_\ell^2}\,C_{\kappa_g\kappa}^{a}(\ell)\,,
    \end{equation}
\end{subequations}
where $\overline{G_{\ell,2}^+\pm G_{\ell,2}^-}$, $\overline{P_\ell}$, and $\overline{P_\ell^2}$ are $\theta$-bin-averaged Legendre polynomials defined in~\cite{FKEM20}.
$F(\ell)$ is a window function incorporating the Gaussian smoothing, pixelization, and scale-cut,
\begin{equation}
\begin{aligned}
    \label{eqn:beam_kernel}
    F(\ell) =&\,\widetilde{W}_{N_\mathrm{side}}(\ell)\Theta(\ell-L_\mathrm{min})\Theta(L_\mathrm{max}-\ell)\times\\
    &\mathrm{exp}(-\ell(\ell+1)/\ell_\mathrm{beam}^2)\,,
\end{aligned}
\end{equation}
where $\Theta(\ell)$ is the Heaviside function and $\ell_\mathrm{beam}\equiv\sqrt{16\mathrm{ln}2}/\theta_\mathrm{FWHM}$. We assume $\theta_\mathrm{FWHM}=7\arcmin$ in this work. 
$(L_\mathrm{min},L_\mathrm{max})=(8,2048)$ are the scale-cut in the CMB lensing convergence map. $\widetilde{W}_{N_\mathrm{side}}(\ell)$ is the pixel window function of \texttt{HEALPix}. 

\subsubsection{Fourier Space: CMB Lensing Band-power}
\label{sec:mv_fourier}
We adopt the band-power estimates in~\citetalias{P18A8} as the data vector for CMB lensing convergence auto-correlation,
\begin{equation}
    \label{eqn:dv_bp}
    C_{L_{b}}^{\kappa\kappa} = \sum_L \mathcal{B}_i^LC_L^{\kappa\kappa,\mathrm{th}} = \sum_L \frac{L^2(L+1)^2}{4} \mathcal{B}_i^LC_L^{\phi\phi,\mathrm{th}}\,,
\end{equation}
where $\mathcal{B}_i^L$ is the binning function for optimal band power estimates, and $L_{b}$ is the weighted mean $L$-mode in angular bin $i$ (equation 12 in \citetalias{P18A8}).
$C_L^{\kappa\kappa, \mathrm{th}}$ is the theoretical expectation of the the quadratic estimator $\hat{C}^{\kappa\kappa}_L$ while $C^{\phi\phi,\mathrm{th}}_L$ is the lensing potential equivalent.
Note that $\hat{C}^{\kappa\kappa}_L$ also responds to the primary CMB information. 
To remove the impact of primary CMB, we follow~\citetalias{P18A8}, linearize the dependency of $C_{L_{b}}^{\kappa\kappa}$ on the primary CMB power spectra and marginalize over them:
\begin{equation}
    \mathcal{B}_i^LC_L^{\phi\phi,\mathrm{th}} \approx \mathcal{B}_i^L\left.C_L^{\phi\phi}\right|_{\bm{\theta}} + M_i^{a,\ell^\prime}\left(\left.C_{\ell^\prime}^a\right|_{\bm{\theta}}-\left.C_{\ell^\prime}^a\right|_\mathrm{fid}\right)\,,
\end{equation}
where $\left.\right|_{\bm{\theta}}$ means quantities evaluated at cosmological parameter $\bm{\theta}$ and $\left.\right|_\mathrm{fid}$ are evaluated at the FPP10 fiducial cosmology, $a$ sums over $C_L^{\phi\phi}$ and other primary CMB power spectra. $M_i^{a,\ell^\prime}$ is pre-computed in the fiducial model and is provided in the public lensing likelihood data~\citepalias[see equations 31\&32 in][]{P18A8}. Effectively the model vector $C_{L_{b}}^{\kappa\kappa}$ is calculated as
\begin{equation}
    \label{eqn:dv_bp_marg}
    \begin{aligned}
    C_{L_{b}}^{\kappa\kappa} =& \sum_L\left(\mathcal{B}_i^L+M_i^{\phi,L}\right)\left.C_L^{\kappa\kappa}\right|_{\bm{\theta}} - \Delta C_{L_{b}}^{\kappa\kappa,\mathrm{offset}}\,,
    \end{aligned}
\end{equation}
where $\Delta C_{L_{b}}^{\kappa\kappa,\mathrm{offset}}$ is a constant offset pre-computed at the FFP10 fiducial cosmology
\begin{equation}
\begin{aligned}
    \Delta C_{L_{b}}^{\kappa\kappa,\mathrm{offset}} \equiv &\sum_L  \frac{L^2(L+1)^2}{4}\left[M_i^{\phi,L}\left.C_L^{\phi\phi}\right|_{\mathrm{fid}}+\right.\\
    &\left.M_i^{X,\ell}\left(\left.C_\ell^X\right|_\mathrm{fid}-\hat{C}_\ell^{X}\right)\right]\,,
\end{aligned}
\end{equation}
here $X$ sums over the eight MV lensing estimators (TT, EE, TE, TB, EB, ET, BT, and BE). 

\subsection{Covariance Matrix}
\label{sec:covmat}

\subsubsection{Covariance Modeling of 6$\times$2}
\label{sec:6x2pt_covmat_modeling}

We extend the DES Y1/Y3 3$\times$2 covariance matrix routines as described in~\cite{DESY1_method,FKEM20} to compute an analytic 6$\times$2 covariance (for a curved-sky geometry). This covariance includes all cross-covariance terms (e.g., covariance between the real-space 2PCFs and $C^{\kappa\kappa}_{L_{b}}$). 
For convenience of description we use $\Theta^{ab}$ to denote any probes in $\{\xi^{ab}_\pm,\,\gamma^{ab}_t,\,w_g^{a}\,(a=b)\}$ and $\Xi^{a}$ to denote any probes in $\{w^a_{g\kappa},\,w^a_{s\kappa}\}$.
\begin{subequations}
    \begin{equation}
        \label{eqn:covmat_1}
        \begin{aligned}
        \mathrm{Cov}\left( \Theta^{ab}(\vartheta),\, \Xi^{c}(\vartheta^\prime) \right)=& \sum_{\ell,\ell^\prime=\ell_\mathrm{min}}^{\ell_\mathrm{max}} \overline{P^\Theta_\ell} \,\overline{P^\Xi_{\ell^\prime}} F(\ell^\prime)\times\\
        &\mathrm{Cov}\left(C_\Theta^{ab}(\ell),\,C_\Xi^{c}(\ell^\prime)\right)\,,
        \end{aligned}
    \end{equation}
    \begin{equation}
        \label{eqn:covmat_2}
        \begin{aligned}
        \mathrm{Cov}\left( \Xi^{a}(\vartheta),\, \Xi^{b}(\vartheta^\prime) \right)=& \sum_{\ell,\ell^\prime=\ell_\mathrm{min}}^{\ell_\mathrm{max}} \overline{P^\Xi_\ell}F(\ell) \, \overline{P^\Xi_{\ell^\prime}}F(\ell^\prime)\times\\
        &\mathrm{Cov}\left(C_\Xi^{a}(\ell),\,C_\Xi^{b}(\ell^\prime)\right)\,,
        \end{aligned}
    \end{equation}
\end{subequations}
here functions $\overline{P_\ell^{\Theta/\Xi}}$ are the combination of $\ell$-factors and Legendre polynomials in equations~(\ref{eqn:dv_xipm}--\ref{eqn:dv_w_sk}), which are also defined in~\cite{FKEM20}. 
The corresponding angular power spectra are also described in equations~(\ref{eqn:dv_xipm}--\ref{eqn:dv_w_sk}). 

The covariance matrices related to $C_{L_{b}}^{\kappa\kappa}$ are
\setcounter{equation}{13}
\begin{subequations}
    \setcounter{equation}{2}
    \begin{equation}
        \label{eqn:covmat_3}
        \begin{aligned}
        \mathrm{Cov}\left(\Theta^{ab}(\vartheta),\,C_{L_{b}}^{\kappa\kappa}\right)=&\sum_{\ell=\ell_\mathrm{min}}^{\ell_\mathrm{max}} 
        \overline{P_{\ell}^{\Theta}}
        \sum_{L=L_\mathrm{min}}^{L_\mathrm{max}} \mathcal{B}_i^L\times\\
        &\mathrm{Cov}\left(C_\Theta^{ab}(\ell),\,C^{\kappa\kappa}_L\right)\,,
        \end{aligned}
    \end{equation}
    \begin{equation}
        \label{eqn:covmat_4}
        \begin{aligned}
        \mathrm{Cov}\left(\Xi^{a}(\vartheta),\,C_{L_{b}}^{\kappa\kappa}\right)=&\sum_{\ell=\ell_\mathrm{min}}^{\ell_\mathrm{max}} 
        \overline{P_{\ell}^{\Xi}}\,F(\ell)
        \sum_{L=L_\mathrm{min}}^{L_\mathrm{max}} \mathcal{B}_i^L\times\\
        &\mathrm{Cov}\left(C_\Xi^{a}(\ell),\,C^{\kappa\kappa}_L\right)\,.
        \end{aligned}
    \end{equation}
\end{subequations}
In equations~(\ref{eqn:covmat_1}--\ref{eqn:covmat_4}), we choose $\ell_\mathrm{min}=2$, $\ell_\mathrm{max}=50,000$, $L_\mathrm{min}=8$ and $L_\mathrm{max}=2048$. The cosmological and nuisance parameters where the covariance is evaluated are shown in Table~\ref{tab:params_table}.

An analytic CMB lensing band-power covariance is non-trivial to model because of the primary CMB marginalization. 
Therefore, we adopt the ``lensing-only'' covariance matrix provided by Planck PR3, which is derived from FFP10 simulations, and the impact of $C_\ell^\mathrm{CMB}$ is marginalized over assuming a Gaussian covariance~\citepalias[see equation 34 in][]{P18A8}.

Equations (\ref{eqn:covmat_1}--\ref{eqn:covmat_4}) depend on the Fourier space covariance matrix, which breaks into three parts: the Gaussian covariance $\mathrm{Cov}^\mathrm{G}$, the non-Gaussian covariance ignoring the survey geometry $\mathrm{Cov}^\mathrm{NG,0}$, and the super-sample covariance $\mathrm{Cov}^\mathrm{SSC}$. 
We refer readers to the appendix of ~\citet{KE17} for our implementation of the 3$\times$2 Fourier space covariance. The other 6$\times$2 covariance blocks follow a similar methodology; details can be found in Appendix~\ref{sec:appd_a}.

\subsubsection{Survey Geometry Effects}

When considering the DES Y1 footprint only, the leading order correction for survey boundary effects is the well-known pair count reduction that affects the shot/shape noise calculation~\citep{TKC+18}:
\begin{equation}
\label{eq:paircounts}
    N_{p}(\vartheta) \approx \bar{n}_X\,\bar{n}_Y\,2\pi\,\mathrm{sin}(\vartheta)\,\Delta\vartheta\,\Omega_s^{X\cap Y}\,\bar{\xi}^{W_XW_Y}(\vartheta)\,,
\end{equation}
where $N_{p}(\vartheta)$ denotes the $X$-$Y$ pair counts in angular bin $\theta\in[\vartheta-\Delta\vartheta/2, \vartheta+\Delta\vartheta/2)$. 
$\Omega_s^{X\cap Y}$ is the area of the tracers' footprint intersection, $\bar{n}_{X/Y}$ is the mean surface density. As can be seen from equation (\ref{eq:paircounts}) the normalized 2PCF of the tracers footprints $\bar{\xi}^{W_XW_Y}(\vartheta)$ modulates the number of pairs; ignoring survey boundary effects corresponds to  $\bar{\xi}^{W_XW_Y}(\vartheta)\equiv 1$.

As mentioned in Section~\ref{sec:optimal_measurements}, our $\hat{w}_{g\kappa}^{a}$ and $\hat{w}_{s\kappa}^{a}$ measurements include pairs where the CMB lensing convergence is located outside of the DES Y1 footprint. Hence, we measure $\bar{\xi}^{W_XW_Y}(\vartheta)$ for ${W_X,\,W_Y}$=\{DES Y1, DES Y1\}, \{DES Y1, Planck\}, and \{Planck, Planck\}, and incorporate them in the shot noise covariance matrix calculation. 

For 3$\times$2 where the shot noise has a flat spectrum, we include the survey geometry correction as
\begin{equation}
\begin{aligned}
    \frac{2\sigma_\epsilon^2}{\bar{n}^a_\mathrm{src}} &\rightarrow \frac{2\sigma_\epsilon^2}{\bar{n}^a_\mathrm{src}\sqrt{\bar{\xi}^\mathrm{DES\,Y1}(\vartheta)}} \,, \\
    \frac{1}{\bar{n}^a_\mathrm{lens}} &\rightarrow \frac{1}{\bar{n}^a_\mathrm{lens}\sqrt{\bar{\xi}^\mathrm{DES\,Y1}(\vartheta)}} \,.
\end{aligned}
\end{equation}

For $\hat{w}^a_{g\kappa}$ and $\hat{w}^a_{s\kappa}$, we note that the DES Y1 footprint is fully embedded in the Planck mask. This means including extra pairs by allowing $\hat{\kappa}$ outside Y1 footprint \textit{is equivalent to not correcting for the survey boundary effect}, i.e. $\bar{\xi}^{\mathrm{DES\,Y1\times Planck}}(\vartheta)\approx 1$ to leading order (see Appendix~\ref{sec:appd_a} for derivation). 

In addition to the corrections for the shot noise terms, we consider survey boundary effects in the non-Gaussian covariance terms. These can be calculated as changes to the survey window function, which we further detail in Appendix~\ref{sec:appd_a}. 

The survey geometry effect on $C^{\kappa\kappa}_{L_{b}}$ is already taken into account in the PR3 lensing products.

\section{Analysis Choices And Model Validation}
\label{sec:choices}

\begin{table}
	\centering
	\begin{tabular}{lcc}
	\toprule
    \rule{0pt}{5ex}
	Parameters & \shortstack{Fid. Value of \\Covariance} & Prior\\
	\tableline
    \multicolumn{3}{c}{Cosmological Parameters}\\
	$\Omega_{m}$& 0.31735 & flat[0.1, 0.9]\\
	$ A_s\times 10^9 $& 2.119 & flat[0.5, 5.0]\\
	$n_s$& 0.964 & flat[0.87, 1.07]\\
	$\Omega_{b}$& 0.04937 & flat[0.003, 0.07]\\
	$H_0$& 67.0 & flat[55, 91]\\
    \tableline
	\multirow{3}{*}{$\sum m_\nu$/eV} & \multirow{3}{*}{0.06} & baseline: fixed\\
     & & wide: flat[0.046, 0.931]\\
     & & info: flat[0.046, 0.121]\\
	\tableline
	\multicolumn{3}{c}{Systematics Parameters}\\
	$\Delta_{z,\,\mrm{src}}^1\times 10^2$ & 0.0 & $\mathcal{N}(-0.1, 1.6)$ \\
	$\Delta_{z,\,\mrm{src}}^2\times 10^2$ & 0.0 & $\mathcal{N}(-1.9, 1.3)$ \\
	$\Delta_{z,\,\mrm{src}}^3\times 10^2$ & 0.0 & $\mathcal{N}(0.9, 1.1)$ \\
	$\Delta_{z,\,\mrm{src}}^4\times 10^2$ & 0.0 & $\mathcal{N}(-1.8, 2.2)$ \\
	\tableline
	$\Delta_{z,\,\mrm{lens}}^1\times 10^2$ & 0.0 & $\mathcal{N}(0.8, 0.7)$ \\
	$\Delta_{z,\,\mrm{lens}}^2\times 10^2$ & 0.0 & $\mathcal{N}(-0.5, 0.7)$ \\
	$\Delta_{z,\,\mrm{lens}}^3\times 10^2$ & 0.0 & $\mathcal{N}(0.6, 0.6)$ \\
	$\Delta_{z,\,\mrm{lens}}^4\times 10^2$ & 0.0 & $\mathcal{N}(0.0, 1.0)$ \\
	$\Delta_{z,\,\mrm{lens}}^5\times 10^2$ & 0.0 & $\mathcal{N}(0.0, 1.0)$ \\
	\tableline
	$m^a\times 10^2$ & 0.0 &  $\mathcal{N}(1.2, 2.3)$ \\
	\tableline
    \multirow{2}{*}{$Q_1$} & \multirow{2}{*}{0.0} & wide: flat[-3, 12]\\ 
    & & info: flat[0 ,4]\\
	$Q_2$ & 0.0 & flat[-2.5, 2.5]\\
	\tableline 
	$A_\mathrm{IA}$ & 0.0 & flat[-5, 5]\\
	$\beta_\mathrm{IA}$ & 0.0 & flat[-5, 5]\\
	\tableline 
	$b_g^1$ & 1.44 & \multirow{5}{6em}{flat[0.8, 3.0]}\\
    $b_g^2$ & 1.70 & \\
    $b_g^3$ & 1.698 & \\
    $b_g^4$ & 1.997 & \\
    $b_g^5$ & 2.058 & \\
	\tableline
	\end{tabular}
	\caption{This table summarizes the prior ranges used in both the synthetic and real analyses and the fiducial parameter values used when computing the analytic covariance matrix. We express a flat prior as flat[min, max] and a Gaussian prior as $\mathcal{N}(\mu,\sigma)$ where $\mathcal{N}$ means normal distribution, $\mu$ is the mean and $\sigma$ is the standard deviation. We note that $Q_1$ has a wide or informative prior depending on whether we constrain baryonic physics or cosmology with real data. $Q_2$ is not used for real data but in several simulated analyses (see Fig. \ref{fig:ss_scalecut}).
	}
	\label{tab:params_table}
\end{table}

\subsection{Systematics}
\label{sec:systematics}
We closely follow~\citetalias{hem21} and~\citetalias{DES_Y1_3x2pt} choices, which are briefly summarized below and in Table~\ref{tab:params_table}.
\paragraph{Galaxy bias} We use linear galaxy bias in this work and assign a galaxy bias $b_g^a$ with flat priors on $b_g^a\in[0.8,\,3.0]$ for each tomography bin in the lens galaxy sample (c.f. equation~(\ref{eqn:power_spectrum_g})).

\paragraph{Photo-$z$ uncertainty} We model photo-$z$ uncertainty by allowing 
 a shift in the mean of each tomographic redshift distribution in the lens and source galaxy sample, i.e. the true redshift distribution is offset from the measured photo-$z$ distribution by $\Delta_z^a$, $n^a(z)=n_{\mathrm{photo}\text{-}z}^a(z-\Delta_z^a)$. We sample $\Delta_{z,\mathrm{lens/src}}^a$ assuming Gaussian priors listed in Table~\ref{tab:params_table}.
 
\paragraph{Shear calibration bias} Shear calibration bias is parameterized by an additive bias $c^a$ and a multiplicative bias $m^a$ per tomography bin. We fix $c^a\equiv 0$ and sample $m^a$ with Gaussian priors $\mathcal{N}(\mu=1.2\times 10^{-2}, \sigma=2.3\times 10^{-2})$. 

\paragraph{Intrinsic alignment} We use the non-linear alignment (NLA) model to mitigate the intrinsic alignment of source galaxies with its surrounding LSS environment~\citep{hs04,brk07,Krause16_IA}. In NLA, the intrinsic shapes of galaxies are proportional to the tidal field with a redshift-dependent amplitude $A(z)$
\begin{equation}
    \label{eqn:NLA_amp}
    A(z)=-A_\mathrm{IA}C_1\frac{3H_0^2\Omega_{m}}{8\pi G D(z)}\left[(1+z)/1.62\right]^{\eta_\mathrm{IA}}\,,
\end{equation}
where $A_\mathrm{IA}$ is the amplitude at pivot redshift $0.62$ and $\eta_\mathrm{IA}$ is the redshift evolution slope. $C_1=5\times 10^{-14}\,\mathrm{M}_\odot^{-1}\,h^{-2}\,\mathrm{Mpc}^3$ is a normalization constant, $D(z)$ is the linear growth factor. Both $A_\mathrm{IA}$ and $\eta_\mathrm{IA}$ are sampled with flat priors $[-5,\,5]$.

\paragraph{Baryonic physics} 
We model baryonic physics using the PCA method described in~\cite{ekd15} and~\citetalias{hem21}. The PCs are obtained from a series of hydrodynamical simulations: Illustris~\citep{GVS+14,VGS+14}, IllustrisTNG~\citep[TNG100,][]{MVP+18,NPSR+18,NPSW+18,PNH+18,SPP+18}, Horizon-AGN~\citep{DPW+14}, MassiveBlack-II~\citep[MB2,][]{KMC+15,TMM+15}, Eagle~\citep{SCB+15}, three cosmo-OWLS simulations~\citep[cOWLS,][]{BMSP14} with the minimum AGN heating temperature $\Delta T_\mathrm{heat}/\mathrm{K}=10^{8.0},\,10^{8.5},\,10^{8.7}$, and three BAHAMAS simulations~\citep{MSB+17} with $\Delta T_\mathrm{heat}/\mathrm{K}=10^{7.6},\,10^{7.8},\,10^{8.0}$. We refer to~\citetalias{hem19,hem21} for details on PCA implementation. 

The PCs are uncorrelated and ranked with respect to capturing the largest variance of the model vector from baryonic feedback. The full model vector is then computed as  
\begin{equation}
    \bm{M}_\mathrm{bary}(\bm{p},\bm{Q})=\bm{M}(\bm{p})+\sum_{i=1}^{n}Q_i\,\mathbf{PC}_i\,,
\end{equation}
where the amplitudes $Q_i$ are varied during analyses.

\subsection{Simulated Likelihood Analyses}

We conduct a large number of simulated likelihood analyses (also see Appendix \ref{sec:syn_like}) using analytically generated data vectors. These simulations validate our pipeline and determine our final analysis setup (scale cuts, priors, parameterizations).

\subsubsection{Scale Cuts}
\label{sec:syn_ss_smallscale}
We impose the same scale-cuts as~\citetalias{DES_Y1_3x2pt} in $\hat{w}_{g}^{a}(\vartheta)+\hat{\gamma}^{ab}_{t}(\vartheta)$ and as~\cite{ogp19,obc19} in $\hat{w}^{a}_{g\kappa}(\vartheta)$+$\hat{w}^{a}_{s\kappa}(\vartheta)$. Since the $C_{L_{b}}^{\kappa\kappa}$ is directly coming from \citetalias{P18A8}, the only remaining scale cut to determine is that of cosmic shear. 

The cosmic shear scale cut is linked to the number of PCs used in baryonic physics modeling.  \citetalias{hem21} demonstrated that one PC is sufficient to model baryonic physics in the cosmic shear part of a 3$\times$2 analysis for DES Y1 down to 2\farcm 5. As detailed in Section \ref{sec:xipm_025}, we measure $\hat{\xi}_{\pm}^{ab}(\vartheta)$ in 30 logarithmic bins spanning $[0\farcm 25,\,250^{\prime}]$.

In the following, we derive our scale cuts for $\hat{\xi}_{\pm}^{ab}(\vartheta)$ and decide the number of PCs to marginalize over such that the posterior on $\Omega_{m}$-$S_8$ from 6$\times$2 analysis is not biased and the error budget remains as low as possible.

We compute three synthetic data vectors contaminated with baryonic feedback of varying strengths (from weak to strong feedback):  dark-matter only (DMO, $Q_1=Q_2=0$), Eagle ($Q_1=0.42$, $Q_2=0.39$), and Illustris ($Q_1=4.25$, $Q_2=0.19$). We run 39 simulated likelihood analyses using the three data vectors considering four scale cuts ($\vartheta_\mathrm{min}^{\xi_\pm}\in \{2\farcm5,\, 1\farcm0,\, 0\farcm5,\, 0\farcm25\}$) and three types of baryon mitigation strategies (0 PC, 1 PC, 2PCs). We also run simulated analysis for the DES Y1 reference scale cuts~\citep{DESY1_method, DESY1_5x2pt_method, ogp19, obc19, DESY1_5x2pt_results}.

\begin{figure}
    \centering
    \includegraphics[width=\linewidth]{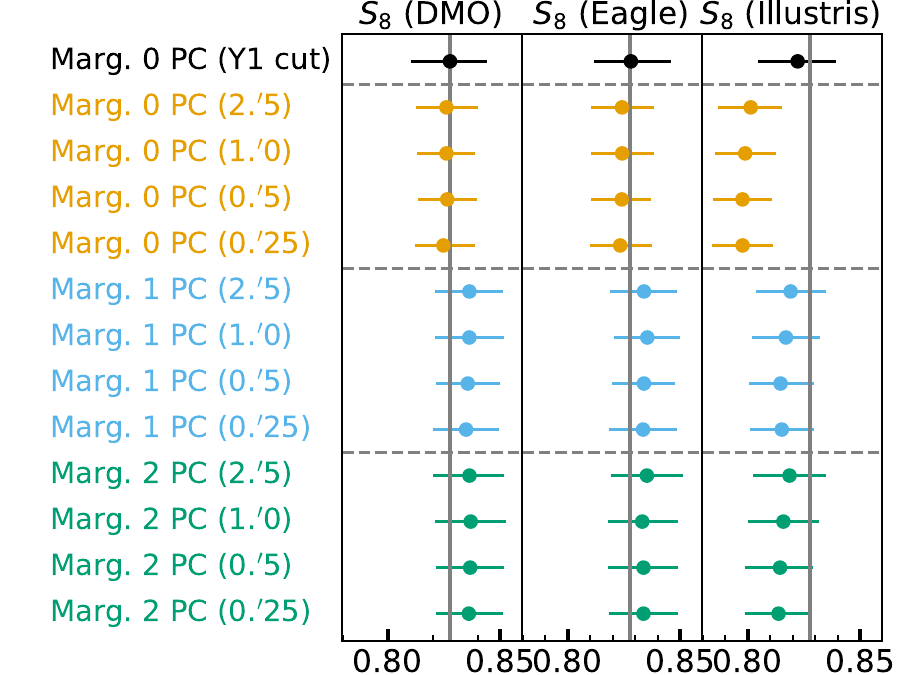}
    \caption{\textbf{Simulated Analysis:} The 1D $S_8$ constraints from 39 simulated likelihood analyses using different baryonic scenarios as input and then analyzing them with different scale cuts and different numbers of PCs in the mitigation scheme. We find that a scale cut of $2\farcm 5$ and marginalizing over one PC offers the best combination in terms of bias mitigation and information gain.
    }
    \label{fig:ss_scalecut}
\end{figure}
 
We summarize results in Fig.~\ref{fig:ss_scalecut} showing the $S_8$ biases and uncertainties for the 39 scenarios. We find
\begin{enumerate}
    \item Compared with the Y1 standard scale cut, the $S_8$ error is reduced by $\sim 10\%$ when small-scale $\xi_\pm^{ab}(\vartheta)$ are included.
    \item Including small-scale $\xi_\pm^{ab}(\vartheta)$ without marginalizing over the baryon PCs leads to high $S_8$ bias ($\sim 2\sigma$) in cases of strong baryonic feedback scenarios like Illustris. Marginalization over the first PC already reduces this bias sufficiently.
    \item Marginalizing over the second PC does not further decrease the bias but increases the $S_8$ uncertainty.
    \item In terms of $S_8$ uncertainty, the most aggressive $0\farcm 25$ cut does not outperform the $2\farcm 5$ cut; instead the $2\farcm 5$ and $1\farcm 0$ cuts are slightly more robust in terms of $S_8$ bias. This can be explained by the fact that scales below $2\farcm 5$ are shot-noise dominated for DES Y1 measurement. We expect these scales in DES Y3/Y6 data to contribute more information.
\end{enumerate}

We require a $S_8$ bias $\leq 0.5\sigma$ for all three synthetic data vectors. We decide to adopt $\vartheta_\mathrm{min}^{\xi_\pm}=2\farcm 5$ and marginalize over the first PC as our baseline analysis setting. As an extended analysis, we also explore the $1\farcm 0$ cut (see Fig.~\ref{fi:robust_tension}).

\subsubsection{Parameterizations and Priors}
The prior ranges for cosmological and systematics parameters are summarized in Table~\ref{tab:params_table}; the systematics parameterizations are described in Section \ref{sec:systematics}. 

Following \citetalias{hem21} we consider two different priors on the amplitude of the baryonic physics PC $Q_1$: a wide prior $Q_1\in[-3,\,12]$ and an informative prior $Q_1\in[0,\,4]$. The former is used when constraining baryonic physics directly (see Section \ref{sec:result_Q1}), and the latter is used when constraining cosmology. We note that $Q_1\in[0,\,4]$  is well-motivated by galaxy formation studies~\citep{HSV+16,BMSP14}, which we consider as an independent source of information. 

Regarding massive neutrinos, we fix $\sum m_\nu = 0.06$ eV for our baseline analysis. In simulated likelihood analyses, we find that varying neutrino mass does not meaningfully impact our constraints on other parameters aside from small shifts due to projection effects in parameter space. We aim to avoid these projection effects and hence opt for a fixed sum of the neutrino mass, a similar setup as the Planck analysis of primary CMB data \citep{P18A6}.

\subsection{Blinding Strategy}
\label{sec:blinding}
Our blinding strategy has five unblinding criteria and can be summarized as follows:
\begin{enumerate}
    \item Our \texttt{CoCoA} 6$\times$2 pipeline is validated against the older DES Y1 \texttt{CosmoLike} pipeline, which has undergone extensive validation during DES Y1 (and Y3).
    Real data vectors are only analysed after extensive code comparison, simulated likelihood analyses, and after all the analyses choices are fixed. 
    \item We conduct consistency tests among different data vector partitions and analysis choices by comparing the 1D marginalized $S_8$. We require that $S_8$ are consistent with each other within $1\sigma$. We design our internal consistency evaluation routine such that it does not show any $S_8$ values or other cosmological parameters during evaluation. 
    \item For the goodness-of-fit, we compute the reduced $\chi^2/\nu$ of the best-fitting point ($\nu$ is the degrees-of-freedom). We require that the probability-to-exceed (PTE) $p\geq 0.05$ given $\chi^2/\nu$ for combining 3$\times$2 and the complementary 3$\times$2 (c3$\times$2, including $w_{g\kappa}^a(\vartheta)$, $w_{s\kappa}^a(\vartheta)$, and $C_{L_b}^{\kappa\kappa}$) into 6$\times$2.
    \item We also visually check that the maxima of the nuisance parameter posteriors are not located near the boundaries of the parameter space. This test is blind to the actual parameter values. 
    \item All chains are converged by passing the $R-1\leq 0.02$ criterion, where $R$ is the Gelman-Rubin diagnostic.
\end{enumerate}

\begin{figure}
    \centering
    \includegraphics[width=0.49\textwidth]{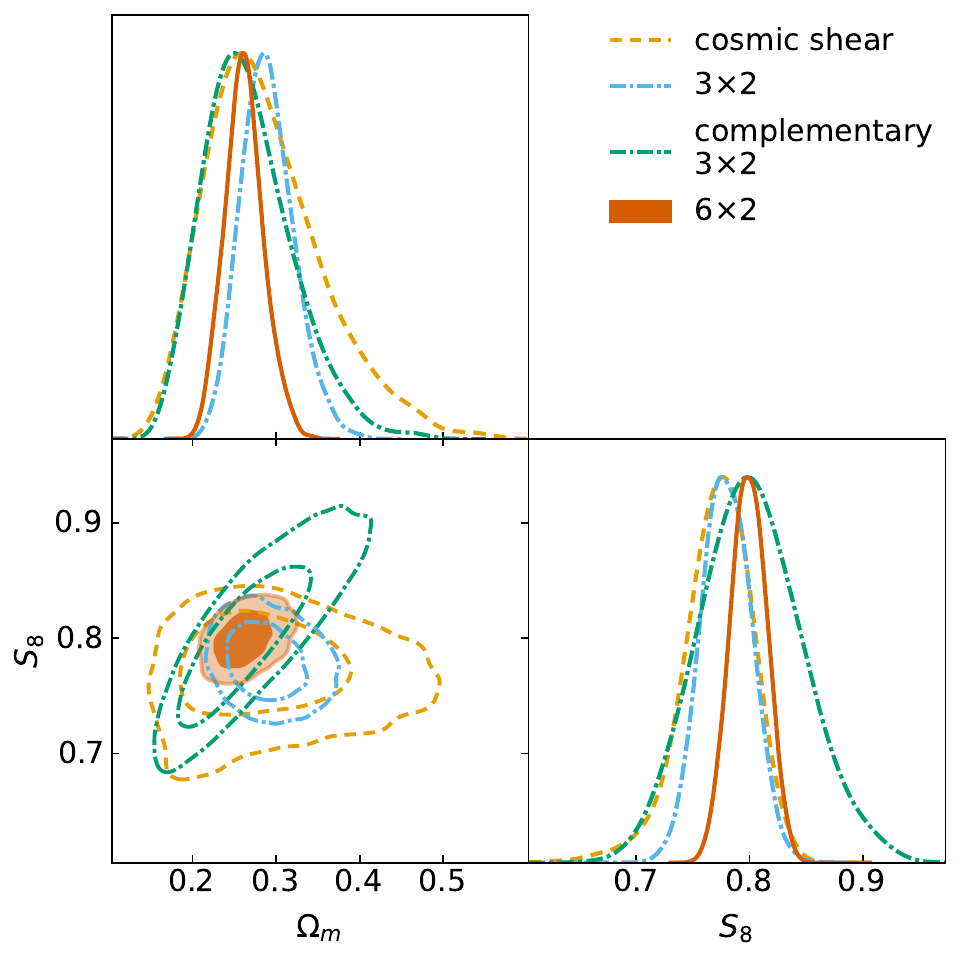}
    \caption{\textbf{Real Analysis:} The results of our baseline 6$\times$2 analysis (solid-filled, vermilion) compared to using subsets of the data vector, i.e., cosmic shear (dashed, orange), 3$\times$2 (dotted-dashed, blue) and c3$\times$2 (dotted-dashed, green). We show the posterior probability in the $\Omega_{m}$ and $S_8$ plane and find the subsets of the data vector in good agreement. The resulting 6$\times$2 analysis prefers slightly higher $S_8$ than the DES Y1 3$\times$2 analysis. }
    \label{fig:result_OmS8_250_Q1info}
\end{figure}

\begin{figure*}
    \centering
     \includegraphics[width=0.325\textwidth]{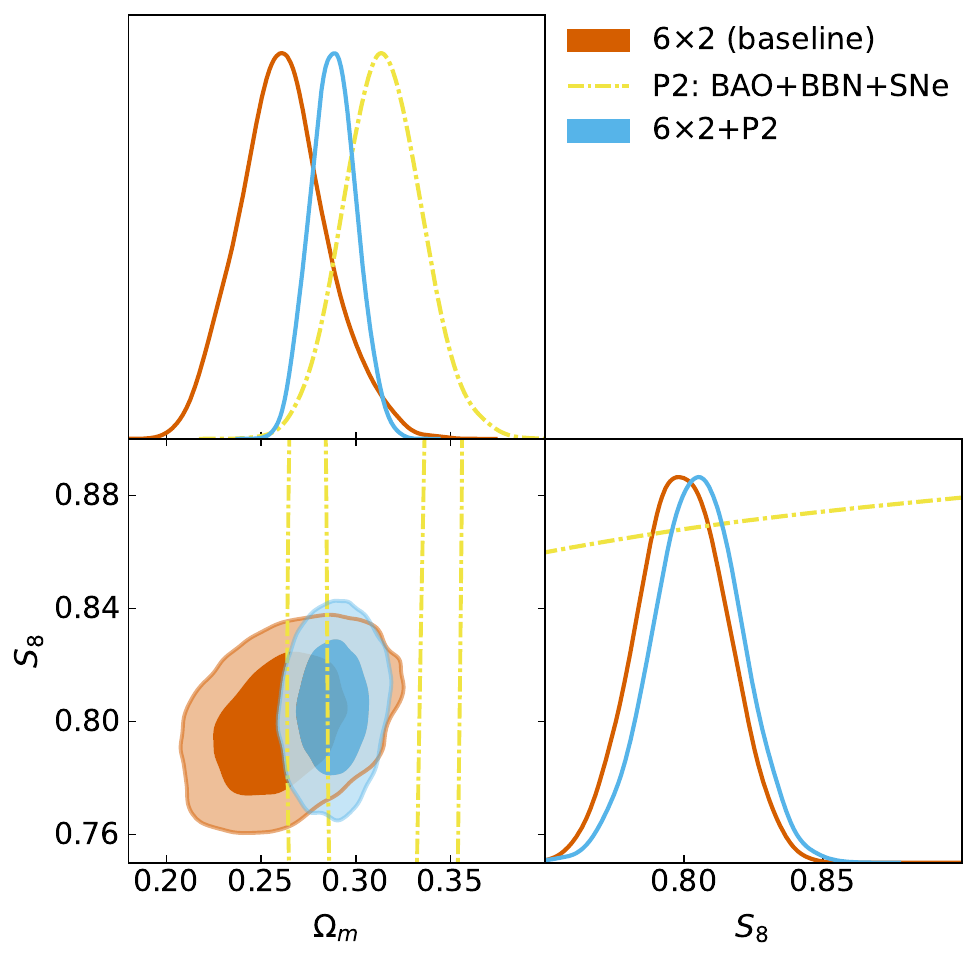}
      \includegraphics[width=0.325\textwidth]{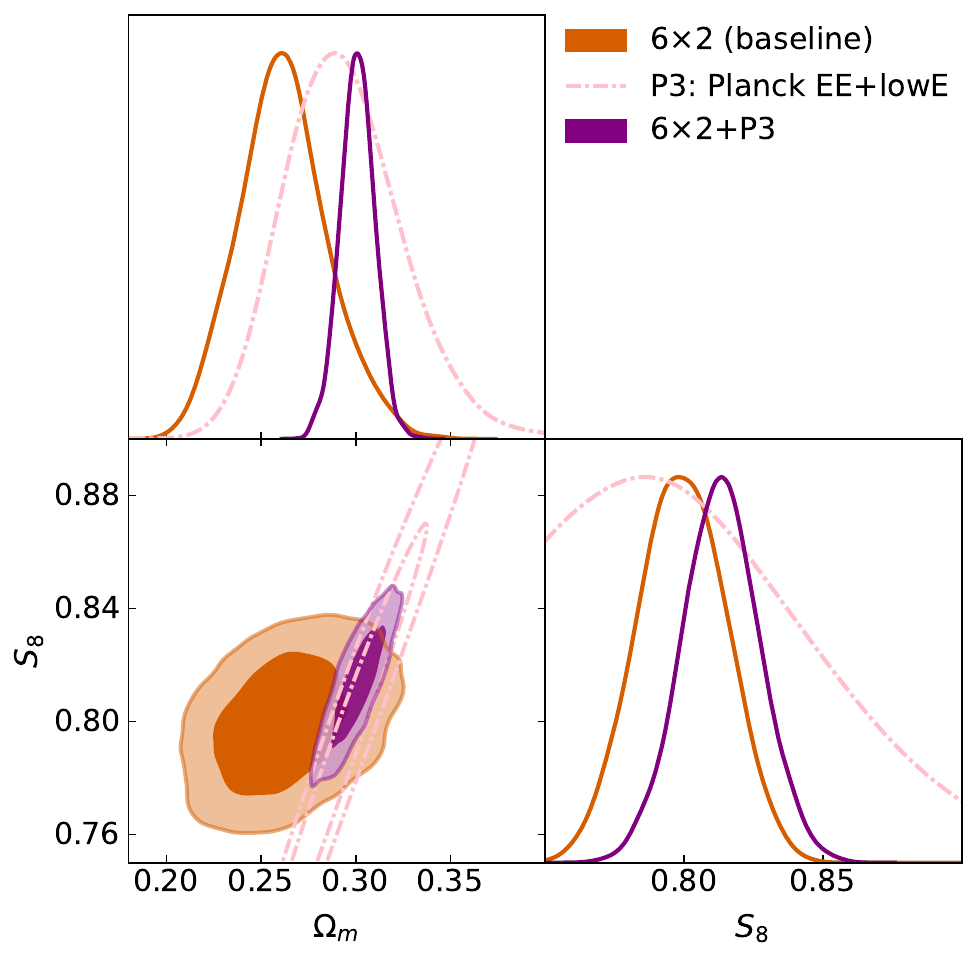}
      \includegraphics[width=0.325\textwidth]{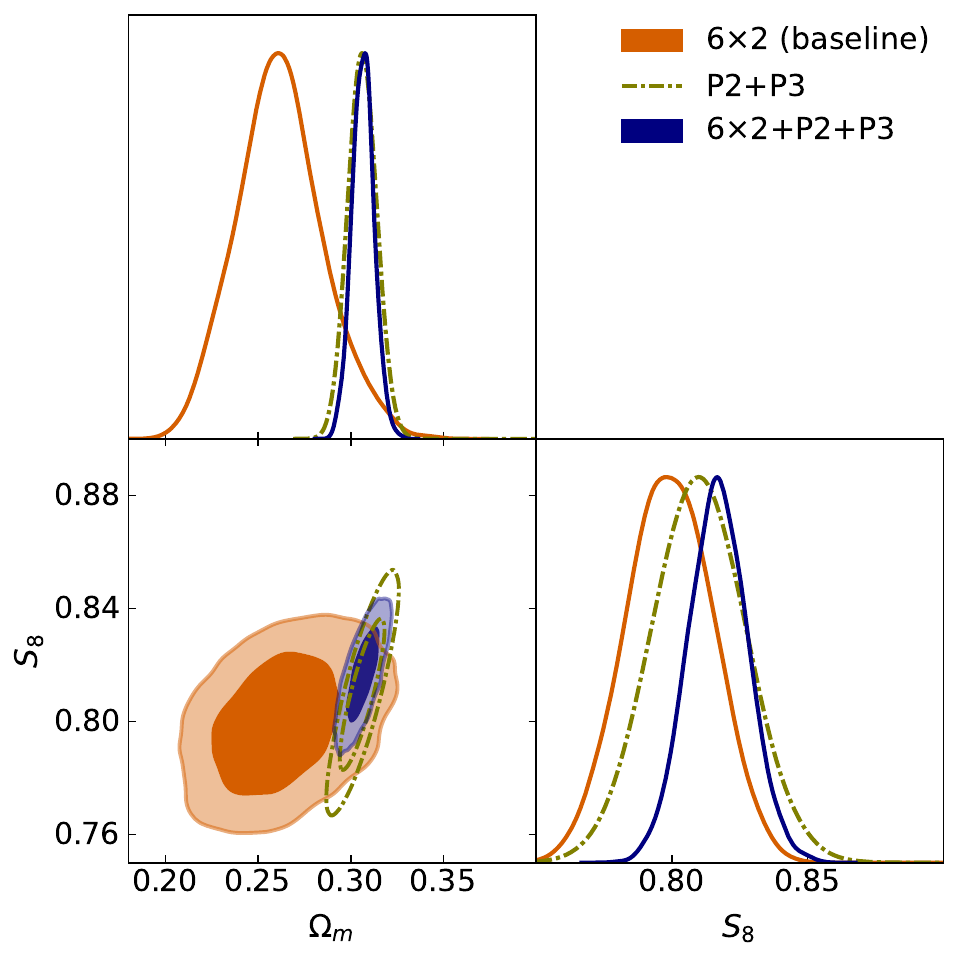}
    \caption{\textbf{Real Analyses with External Probes:} We show $\Omega_{m}$ and $S_8$ posteriors inferred from 6$\times$2 and external data. In each panel, the vermilion solid-filled contour results from our baseline 6$\times$2 analysis, and the dotted-dashed contour corresponds to the external data. The left panel combines 6$\times$2 with BAO+BBN+SNe Ia (P2), the middle panel shows the combination of 6$\times$2 and Planck 2018 EE+lowE (P3), and the right panel shows the result when combining 6$\times$2 with P2 and P3. 
    }
    \label{fig:result_OmS8_250_external_probes}
\end{figure*}
\begin{table*}
    \centering
    \setlength\extrarowheight{2pt}
    \begin{tabular}{lccccccc}
    \toprule
    \multirow{2}{*}{Probes} & \multicolumn{2}{c}{$S_8$} & \multicolumn{2}{c}{$\Omega_{m}$} & \multicolumn{2}{c}{$\sigma_8$} & \multirow{2}{*}{$\Delta\chi^2$}\\
    \cline{2-7}
    & 1D Marg. & MAP & 1D Marg. & MAP & 1D Marg. & MAP & \\
    \tableline
    cosmic shear & $0.771^{+0.034}_{-0.025}$ & 0.797 & $0.288^{+0.048}_{-0.084}$ & 0.301 & $0.804\pm 0.096$ & 0.796 & ... \\
    3$\times$2 & $0.779\pm 0.022$ & 0.792 & $0.290^{+0.028}_{-0.036}$ & 0.287 & $0.797\pm 0.052$ & 0.810 & ...\\
    c3$\times$2 & $0.801\pm 0.045$ & 0.808 & $0.267^{+0.041}_{-0.063}$ & 0.252 & $0.857\pm 0.050$ & 0.881 & ...\\
    6$\times$2 & $0.799\pm 0.016$ & 0.804 & $0.262^{+0.022}_{-0.025}$ & 0.254 & $0.857\pm 0.036$ & 0.874 & ... \\
    \tableline
    6$\times$2 + P2:BAO+BBN+SNe Ia & $0.805\pm0.016$ & 0.805 & $0.288\pm 0.012$ & 0.297 & $0.821\pm 0.022$ & 0.809 & 1.00 \\
    6$\times$2 + P3:Planck EE+lowE & $0.813\pm 0.014$ & 0.814 & $0.3009\pm 0.0090$ & 0.3008 & $0.8121\pm0.0077$ & 0.8124 & 3.70 \\
    6$\times$2 + P2 + P3 & $0.817\pm 0.011$ & 0.825 & $0.3067\pm 0.0059$ & 0.3080 & $0.8081\pm 0.0074$ & 0.8145 & 5.56\\
    \tableline
    \end{tabular}
    \caption{Summary of the cosmology results for our 6$\times$2 analysis and in combination with external probes. We show the 1D marginalized constraints (1D Marg.) on $S_8$, $\Omega_{m}$, and $\sigma_8$, and also their maximum a posteriori (MAP) values. We compare the best-fitting $\chi^2$ of the 6$\times$2-only data vector between analyses with and without external probes in the last column.}
    \label{tab:cosmology_results_summary}
\end{table*}

\section{Data Analysis I: Cosmology Constraints}
\label{sec:result_LCDM}
\subsection{Results for the 6$\times$2 Data Vector without External Priors}
We first study the results from our 6$\times$2 baseline analysis using DES Y1 and Planck CMB lensing information without external priors. 

Fig.~\ref{fig:result_OmS8_250_Q1info} shows four contours, i.e., cosmic shear (dashed, orange), 3$\times$2 (dotted-dashed, blue) and c3$\times$2 (dotted-dashed, green) and the full 6$\times$2 (solid-filled, vermilion). Baryonic uncertainties are marginalized using one PC with the informative prior on the amplitude $Q_1\in[0,\,4]$. As mentioned before, we fix $\sum m_\nu = 0.06$ eV in our baseline analysis. However, in Section \ref{sec:robustness}, we also consider analyses that vary $\sum m_\nu$ using a wide or an informative prior and show that our results are robust against these choices.

We find excellent agreement between the subsets of our data vector (cosmic shear, 3$\times$2, c3$\times$2) and subsequently combine them into a 6$\times$2 analysis. We note that c3$\times$2 shows a nearly orthogonal degeneracy in $\Omega_{m}$-$S_8$ compared to 3$\times$2. A principal component analysis in posterior probability space indicates that 3$\times$2 probes $\Omega_{m}^{0.61}\sigma_8$ best while c3$\times$2 is most sensitive to $\Omega_{m}^{0.30}\sigma_8$. These values agree perfectly with the simulated likelihood analyses we conducted before analyzing the real data (see Appendix \ref{sec:syn_like}). 

Combined into a 6$\times$2 analysis, this breaking of degeneracy in the $S_8$-$\Omega_{m}$ parameter space leads to significant gains in cosmological constraining power. The exact values for the 1D marginalized constraints in $S_8$, $\Omega_{m}$, and $\sigma_8$ and their maximum a posteriori estimate are shown in Table~\ref{tab:cosmology_results_summary}. Specifically, we find $S_8=0.779\pm 0.022$ for 3$\times$2, $S_8=0.801\pm 0.045$ for c3$\times$2 and $S_8=0.799\pm 0.016$ for the combination.

Compared with the DES Y1 3$\times$2 analyses of~\citetalias{hem21}, who measure $S_8=0.788^{+0.018}_{-0.021}$, our 3$\times$2 value is slightly lower. We attribute this difference to small changes in our analysis setup, e.g., we fix the sum of the neutrino mass whereas \citetalias{hem21} varies it, our covariance includes curved sky corrections instead of the flat-sky approximation, and we vary parameters related to shear calibration uncertainties in the analysis rather than including them in the covariance matrix. Also, our covariance matrix is evaluated at a slightly different cosmology than~\citetalias{hem21}.

Our 6$\times$2 analysis prefers higher $S_8$ and lower $\Omega_{m}$ compared to 3$\times$2, which is mainly driven by the Planck full sky CMB lensing measurement within the c3$\times$2 data vector. 

\subsection{Analyses of the 6$\times$2 Data Vector with External Priors}
Next we consider combinations of our 6$\times$2 analysis with external data. We closely follow \cite{ZSM+23} and \citetalias{hem21} in defining the following three priors: 

\paragraph{Primary CMB temperature and polarization information (P1)} The first prior is obtained from the primary CMB temperature and polarization measurements. We use Planck 2018 \texttt{plik} high-$\ell$ TTTEEE spectra truncated after the first peak ($\ell\in[36,\,395]$), as well as low-$\ell$ EE polarization spectrum \citep{P18A5}. The \texttt{plik} nuisance parameters are marginalized assuming the default \texttt{plik} priors. The scale cut imposed on the power spectra removes the impact of integrated Sachs-Wolfe effect and CMB lensing. 

We consider P1 an interesting prior since inflationary parameters $A_s$ and $n_s$ are well constrained by the first peak, but it is insensitive to the non-linear matter power spectrum, which we hope to constrain with our 6$\times$2 analysis. Its error bar on $S_8$ is also larger than the full Planck TTTEEE+lowE result by a factor of $\sim 3$, which should alleviate the anticipated tension between P1 and our low-$z$ 6$\times$2 measurements. 

\paragraph{Geometry information (P2)} The second prior is derived from  BAO and SNe Ia measurements that are sensitive to the geometry of the Universe, as well as $\Omega_{b}$ information from BBN.
We use the Pantheon sample~\citep{SJR+18}, which contains 1048 SNe Ia over the redshift range $0.01 < z < 2.3$. The BBN measurement is obtained from the primordial deuterium-to-hydrogen ratio measured in~\cite{CPNJ16}, and is implemented as a Gaussian prior on $\Omega_{b}h^2\sim \mathcal{N}(0.02208,\,0.00052)$. For the BAO data, we use the SDSS DR7 main galaxy sample~\citep[$z_\mathrm{eff}=0.15$, see][]{RSH+15}, 6dF galaxy survey~\citep[$z_\mathrm{eff}=0.106$, see][]{BBC+11}, and SDSS BOSS DR12 low-$z$ + CMASS combined sample~\citep[$z_\mathrm{eff}=0.38,\,0.51,\,0.61$, see][]{AAB+17}. This combination of priors is powerful in constraining $H_0$ and $\Omega_{b}$, but it is not sensitive to $A_s$, $n_s$, and $S_8$ (see the left panel in Fig.~\ref{fig:result_OmS8_250_external_probes}). 

\paragraph{Primary CMB polarization information (P3)} We also consider the polarization-only CMB anisotropy, i.e. the Planck 2018 \texttt{plik} EE and lowE~\citep{P18A6}. 
As already discussed in \citetalias{hem21}, this prior is an excellent choice to combine with low-$z$ probes that show $S_8$ tension with the full Planck results since the primary driver of this tension is the TT power spectrum. 

\paragraph{Consistency of priors and 6$\times$2}
For the purpose of this paper, we define the threshold to combine our 6$\times$2 analysis with any given prior similar to the internal consistency threshold (see Section \ref{sec:blinding}): We combine if the 1D marginalized $S_8$ are consistent with each other within $1\sigma$, otherwise not. We note that this $1\sigma$ threshold is arbitrary, and exceeding it does not imply that we consider these probes to be in significant tension. We have chosen this relatively stringent $1\sigma$ criterion since we are interested in constraints on baryonic physics and do not want minor differences in cosmology to dilute said constraints. 

\begin{table}
    \centering
    \begin{tabular}{lccc}
    \toprule
    Probe & $S_8$ (1D Marg.) & $S_8$ (MAP) & Combine \\
    \tableline 
    P1 & $0.883\pm0.053$ & $0.878$ & N \\
    P2 & ... &  ...  & Y \\
    P3 & $0.800\pm0.050$ & $0.789$ & Y \\
    \tableline
    6$\times$2 & $0.799 \pm 0.016$ & $0.804$ & ... \\
    \tableline
    \end{tabular}
    \caption{We show the 1D marginalized $S_8$ constraints and the MAP values of our priors and our 6$\times$2 analysis. Note that the external probes included in P2 are geometric only and have no constraining power on $S_8$. The last column shows our decision to combine the prior with the 6$\times$2 analysis.}
    \label{tab:consistency}
\end{table}

We summarize the 1D marginalized $S_8$ constraints and MAP values in Table~\ref{tab:consistency} comparing 6$\times$2 with the external priors. P1 fails the combining criteria while P2 and P3 are consistent with the 6$\times$2 result at $\leq 0.5\sigma$ significance in 1D $S_8$.

\begin{figure}
    \includegraphics[width=0.99\linewidth]{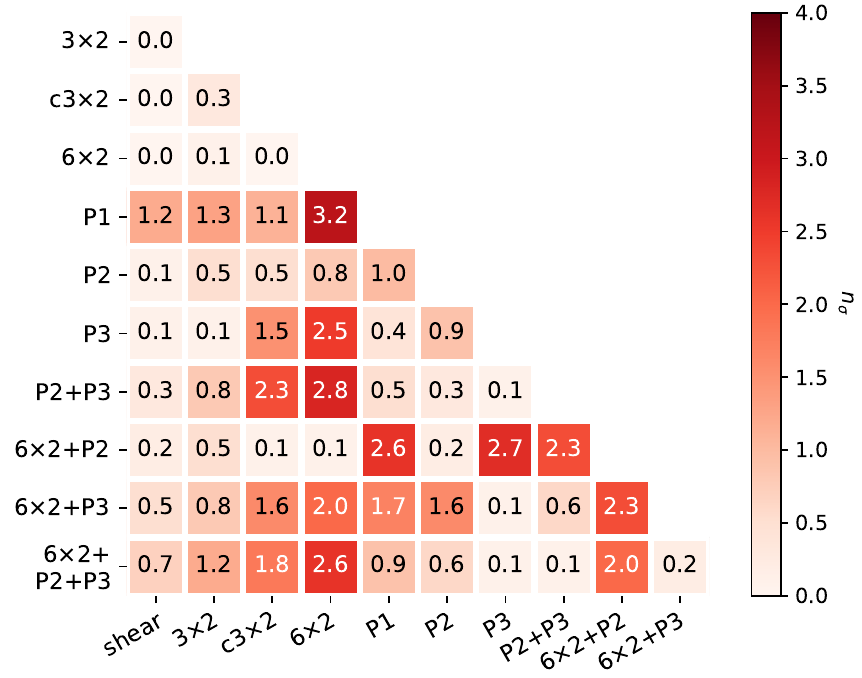}
    \caption{Tensions among different partitions of 6$\times$2, external priors, and 6$\times$2 in combination with external priors, counted in $n_\sigma$ using the parameter difference method (see Section~\ref{sec:robustness}). Since P1 fails the combining criterion, we do not include 6$\times$2+P1 results here.}
    \label{fig:tension_result}
\end{figure}

In addition, we also compute the consistency/tensions among different subsets of the 6$\times$2 data vector, external probes, and 6$\times$2 in combination with external probes using the parameter difference probability as computed in~\citet{RD21,LRC+21,ZSM+23}, which is briefly explained below.

For two chains $\bm{\theta}_1$, $\bm{\theta}_2$, and the corresponding posteriors $P_1(\bm{\theta}_1)$, $P_2(\bm{\theta}_2)$, we evaluate the probability density of $\Delta\bm{\theta}=\bm{\theta}_1-\bm{\theta}_2$ by integrating over the joint distribution $P(\Delta\bm{\theta}, \bm{\theta}_1)=P_1(\bm{\theta}_1)P_2(\bm{\theta}_1-\Delta\bm{\theta})$,
\begin{equation}
\label{eq:parameter_difference_posterior}
    P(\Delta\bm{\theta}) = \int_{V_\Pi}P_1(\bm{\theta}_1)P_2(\bm{\theta}_1-\Delta\bm{\theta})\mathrm{d}\bm{\theta}_1\,.
\end{equation}
$V_\Pi$ is the region of the parameter space where the prior is non-zero. 
The quantity in equation (\ref{eq:parameter_difference_posterior}) is referred to as the parameter difference posterior; integrating it over the range above the iso-contour of no-shift $P(\Delta\bm{\theta})>P(\bm{0})$ gives the probability of a parameter shift
\begin{equation}
    \Delta = \int_{P(\Delta\bm{\theta})>P(\bm{0})}P(\Delta\bm{\theta})\mathrm{d}\Delta\bm{\theta}\,.
\end{equation}
Approximating $\Delta$ as a Gaussian variable, we can quantify the tension in terms of the number of standard deviations,
\begin{equation}
    n_\sigma=\sqrt{2}\mathrm{Erf}^{-1}(\Delta)\,.
\end{equation}
We evaluate $n_\sigma$ using a Masked Autoregressive Flow~\citep{RD21,PPM17} and show the $n_\sigma$ evaluations in Fig.~\ref{fig:tension_result}. Different subsets of 6$\times$2 are highly consistent with each other. We also confirm our previous statement based on the 1D constraints that 6$\times$2 is in agreement ($n_{\sigma}\leq 3$) with external priors P2 and P3, but not P1. 

Consequently, we only combine our 6$\times$2 analysis with P2 and P3 individually and then with the combination P2+P3. Results in the 2D $\Omega_{m}$-$S_8$ parameter space are illustrated in Fig.~\ref{fig:result_OmS8_250_external_probes} while the 1D marginalized results and the MAP constraints are shown in Table~\ref{tab:cosmology_results_summary}. We can see that all external priors  
shift our 6$\times$2 analysis to slightly higher $S_8$ values, the most constraining analysis yields $S_8=0.817\pm 0.011$.

\begin{figure}
\includegraphics[width=0.99\linewidth]{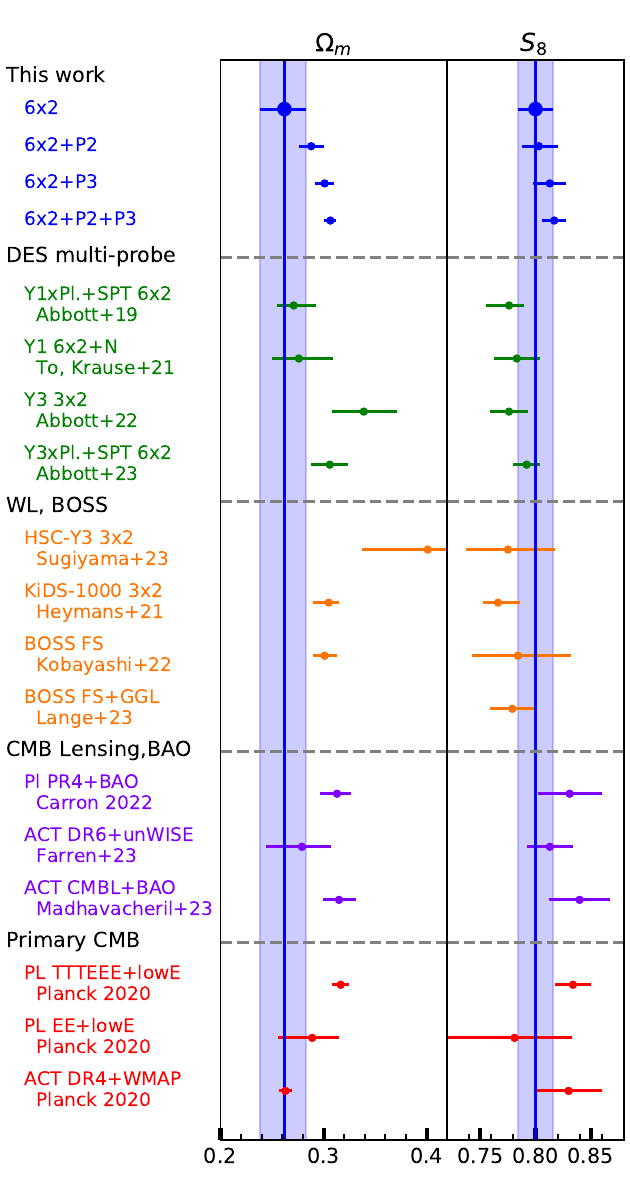}
    \caption{$\Omega_{m}$ and $S_8$ constraints of 6$\times$2 in combinations with external probes Planck 2018 EE+lowE and/or BAO+BBN+SNe Ia, as well as results from other papers, including DES Y1/Y3-related analyses (green), weak lensing and 3D clustering analyses from other stage-IV surveys (HSC Y3, KiDS 1000, BOSS, orange), CMB lensing with BAO and galaxy clustering (violet), and primary CMB power spectrum (red).}
    \label{fig:results_compile2}
\end{figure}

\paragraph{Comparison with other results in the literature} 
In Fig.~\ref{fig:results_compile2}, we compare our constraints with selected results in the literature: ~\cite{DESY1_5x2pt_results,P18A6,KiDS1000_3x2pt,TKR+21,KNTM22,CML22,DES_Y3_3x2pt,DES_Y3_6x2pt_III_23,LHL+23,HSC_Y3_SMM+23,ACT_DR6_CMBL,FKM+23}. We note that our 6$\times$2 constraint is consistent with other results obtained from DES Y1 data although our result prefers a higher $S_8$. DES Y3-based results generally prefer higher $\Omega_{m}$ than Y1, but when including CMB lensing, the DES Y3+Planck/SPT CMB lensing 6$\times$2 is still consistent with our result.
Compared to the Planck TTTEEE+lowE result, the $S_8$ are not in significant tension ($\sim 1.5\sigma$), but the discrepancy in $\Omega_{m}$ is more significant ($\sim 2.4\sigma$). 

Overall, we do not consider these discrepancies to indicate any meaningful tension that indicates new physics. Our main conclusion is that more data is needed. 

\section{Data Analysis II: Baryonic Physics constraints}
\label{sec:result_Q1}
\begin{figure*}
    \centering
    \includegraphics[width=0.46\linewidth]{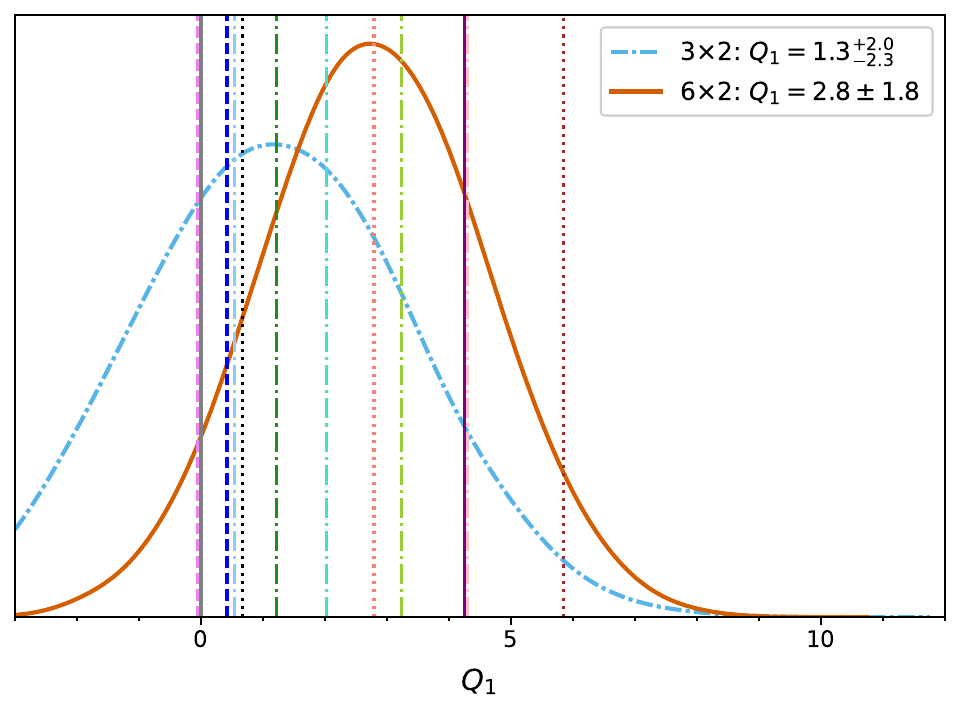}
    \includegraphics[width=0.45\linewidth]{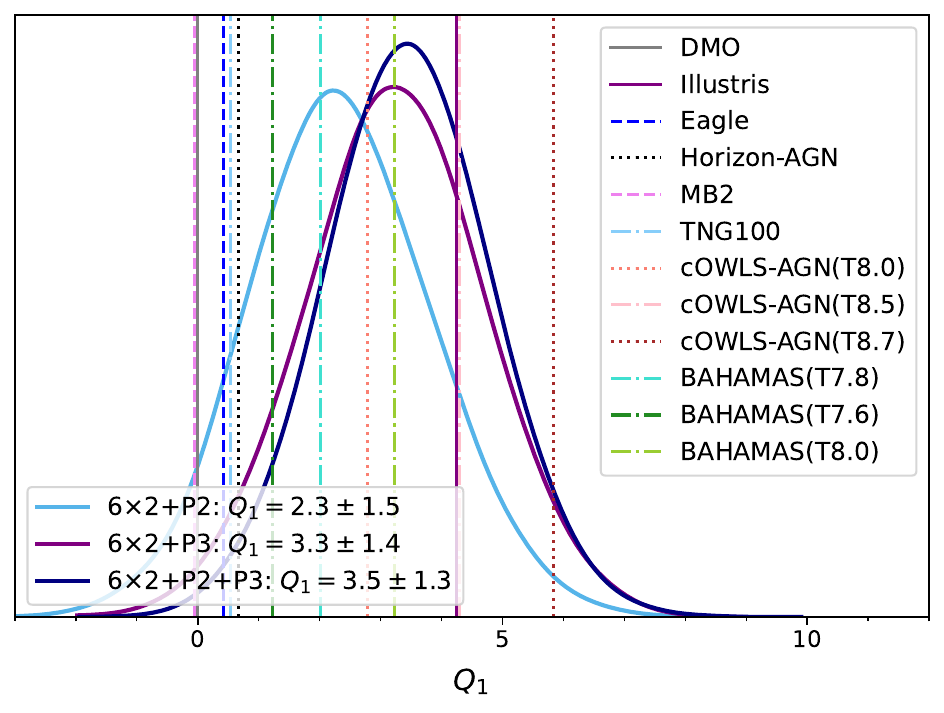}
    \caption{Constraints on $Q_1$ from DES Y1+Planck lensing information (left panel) and in combination with external probes (right panel). In both panels, we show the expected strength $Q_1$ of the hydrodynamical simulations used in this work in vertical non-solid lines. On the left panel, the sky blue dotted-dashed and the vermilion solid lines are the results from 3$\times$2 and 6$\times$2 analyses. 
    On the right panel, the blue, purple, and dark blue solid lines result from 6$\times$2 in combination with BAO+BBN+SNe Ia, Planck 2018 EE+lowE, and both external probes.}
    \label{fig:Q1_results}
\end{figure*}

\begin{figure*}
    \centering
    \includegraphics[width=\linewidth]{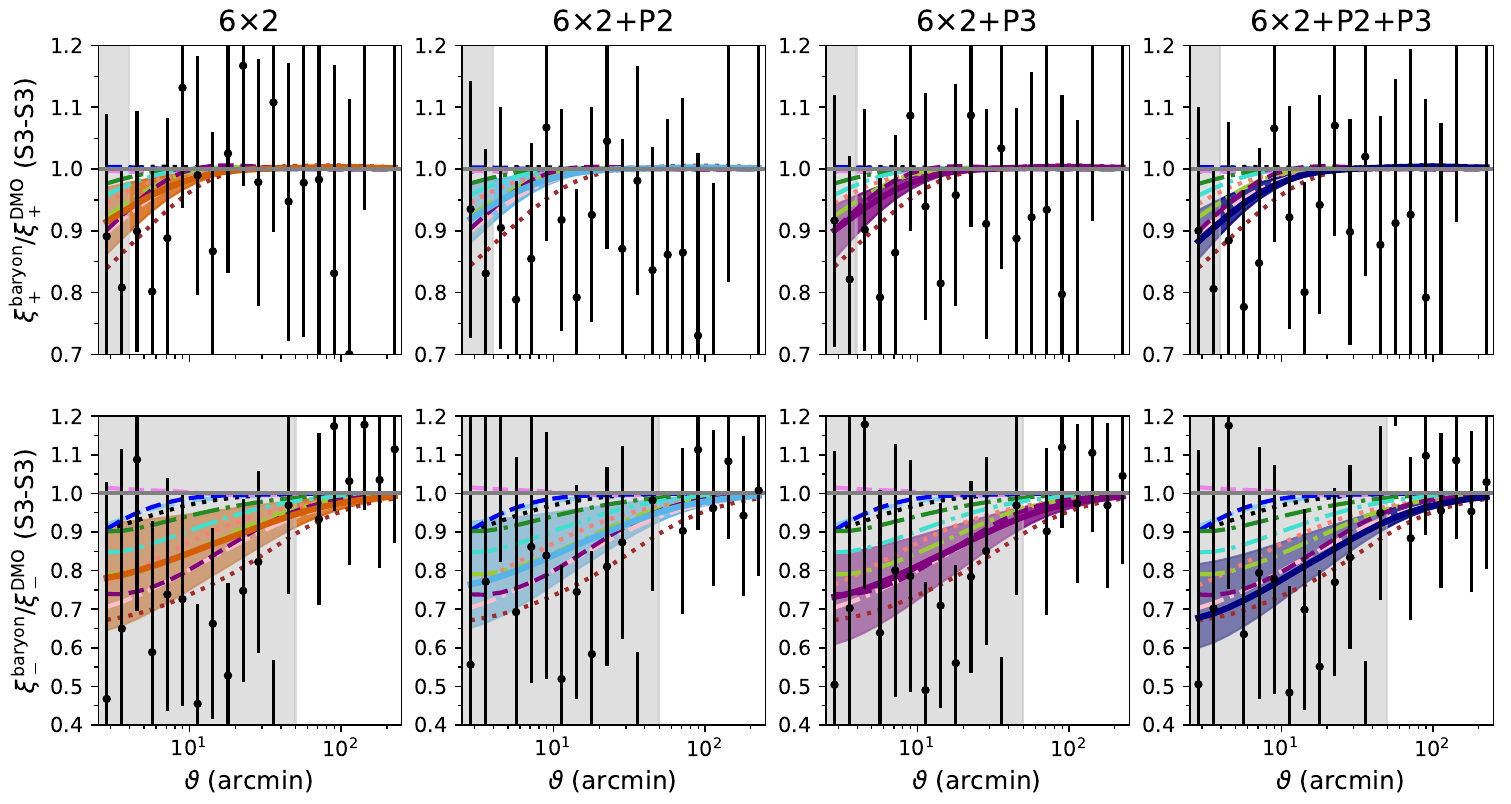}
    \caption{Illustration of the strength of baryonic effects on $\xi_\pm^{33}(\vartheta)$ allowed by 6$\times$2 (vermilion shaded), 6$\times$2+BAO+BBN+SNe Ia (blue shaded), 
    6$\times$2+Planck 2018 EE+lowE (purple shaded),
    and 6$\times$2 in combination with both (deep-blue shaded). 
    The upper (lower) row shows the ratio between $\xi_+^{33}(\vartheta)$ ($\xi_-^{33}(\vartheta)$) with and without baryonic effects. We show the predicted baryonic effects of all the hydrodynamical simulations used in this work as non-solid lines with the same colors as in Fig.~\ref{fig:Q1_results}. The gray-shaded regions are data points discarded in standard DES Y1 scale-cuts.}
    \label{fig:Q1_impact_illustration}
\end{figure*}

\begin{table}
    \centering
    \begin{tabular}{lccccc}
    \toprule
    \multirow{2}{*}{Probes} &  \multicolumn{2}{c}{$Q_1$} & \multicolumn{2}{c}{$S_8$} & \multirow{2}{*}{$\Delta\chi^2$}\\
     \cline{2-5}
     & 1D Marg. & MAP & 1D Marg. & MAP \\
     \tableline
     6$\times$2 & $2.8\pm 1.8$ & 2.8 & $0.804\pm 0.019$ & 0.810 & ...\\
     \,+P2 & $2.3\pm1.5$ & 2.6 & $0.811\pm 0.017$ & 0.807 & 2.73\\
     \,+P3 & $3.2\pm1.4$ & 3.2& $0.815\pm 0.016$ & 0.815 & 3.04\\
     \,+Both& $3.5\pm 1.3$ & 3.9 & $0.820\pm 0.011$ & 0.828 & 2.59\\
    \tableline
    \end{tabular}
    \caption{Constraints on $Q_1$ and $S_8$ from 6$\times$2 and in combination with external probes. All the analyses use the wide $Q_1$ prior [-3, 12]. In the last column, we show the $\Delta\chi^2$ between 6$\times$2 analyses with and without external priors.}
    \label{tab:Q1_results_summary}
\end{table}
We study the baryonic feedback strength $Q_1$ probed by the small-scale cosmic shear component $\hat{\xi}_\pm^{ab}(\vartheta)$ of our data vector. For all the analyses in this section, we use the wide prior on the PC's amplitude $Q_1\in[-3,\,12]$ instead of the informative prior. Similar to the cosmological analyses in the last section, we cut $\hat{\xi}_\pm^{ab}(\vartheta)$ at $2\farcm 5$ and fix $\sum m_\nu=0.06$ eV.

The results on $Q_1$ from the 3$\times$2 (dotted-dashed, blue) and 6$\times$2 (solid-filled, vermilion) analysis are shown in the left panel of Fig.~\ref{fig:Q1_results}. As discussed before, going from 3$\times$2 to 6$\times$2 increases the preferred value of $S_8$ due to the corresponding c3$\times$2 preference. Since $S_8$ and $Q_1$ are positively correlated, the $Q_1$ inferred from 6$\times$2 is also increased. This effect becomes less significant when combining 6$\times$2 with P2 and more significant when combining with P3 (see right panel of Fig.~\ref{fig:Q1_results} and also Table~\ref{tab:Q1_results_summary}).

In addition to our constraints on $Q_1$ from data, Fig.~\ref{fig:Q1_results} shows the $Q_1$ values for a range of hydro-simulations. We find that our constraints show noticeable tension with the most extreme baryonic scenario (cOWLS T8.7): $2.27\sigma$ with 3$\times$2, $1.68\sigma$ with 6$\times$2, $2.36\sigma$ with 6$\times$2+P2, $1.89\sigma$ with 6$\times$2+P3, and $1.8\sigma$ with 6$\times$2+P2+P3. We notice the competing effects of parameter shifts and improved constraining power when including CMB lensing and primary CMB information. The error bars on $Q_1$ are substantially tightened, but the $Q_1$ values are shifted higher in part due to the parameter degeneracy with $S_8$, which reduces the tension with cOWLS. 

For comparison, \citetalias{hem21} find a 2.8$\sigma$ difference between cOWLS and their 3$\times$2+P3 measurement. We do not directly reproduce this measurement but consider our 2.27$\sigma$ result when using 3$\times$2 only to indicate that both papers are consistent.

Overall, we find excellent agreement with the BAHAMAS simulations. Depending on the probe combination considered, the three BAHAMAS scenarios are very close to the peak of the 1D-marginalized posteriors. For example, we compute a 0.03$\sigma$ difference between BAHAMAS (T7.6) and 3$\times$2, and we find similarly low values between BAHAMAS (T7.8) and 6$\times$2+P2, and BAHAMAS (T8.0) and 6$\times$2+P3.

We illustrate how our constraints on $Q_1$ translate into uncertainties in the DES Y1 cosmic shear data vector space in Fig.~\ref{fig:Q1_impact_illustration}.  
In the four panels (from left to right), we show the $\xi^{33}_\pm(\vartheta)$ data points and the best-fitting model and $1\sigma$ uncertainty regions for 6$\times$2, 6$\times$2+P2, 6$\times$2+P3, and 6$\times$2+P2+P3. We also add the model vectors for several baryonic simulations to the panels and indicate the DES Y1 scale cuts from \citet{DES_Y1_3x2pt} in gray.

The upper panels correspond to $\xi^{33}_+(\vartheta)$ and the lower to $\xi^{33}_-(\vartheta)$ data points. The scale cuts in $\xi_-(\vartheta)$ exclude more data points, which can be explained by the increased sensitivity of $\xi_-(\vartheta)$ to small scales/large Fourier modes $\ell$ and consequently also to large $k$ modes. $\xi_+(\vartheta)$ is a filtered version of the shear power spectrum with the Bessel function $J_0(\ell \vartheta)$ (c.f. equation \ref{eqn:dv_xipm}) and due to the functional form of $J_0(\ell \vartheta)$ it will always be sensitive to large scales/small $\ell$ even if $\vartheta$ is small. This fact is also reflected in the smaller range of uncertainty allowed by baryonic physics at given scales in $\xi_+(\vartheta)$ compared to $\xi_-(\vartheta)$. We conclude that $\xi_-(\vartheta)$ is more sensitive to baryonic physics scenarios since small $\vartheta$ are less correlated with large $\vartheta$ and we point out that designing an optimal estimator to measure baryonic physics on small scales is an interesting future concept to explore. 

In any case, we find a clear preference for suppression when going to lower scales in both shear correlation functions, but we also find that the constraining power of the data (shape noise is a concern on these scales) is not yet at the level to exclude any of the hydro-simulations at a meaningful confidence level. DES Y3/Y6 and ultimately LSST Y1 in combination with ACT, SPT, and future CMB experiments, will be exciting to analyze in this context. 

\section{Robustness Tests}
\label{sec:robustness}
We perform several robustness tests of our results.

\paragraph{CMB lensing convergence map reconstruction}
\label{sec:L-dep_multi_bias}

During the CMB lensing convergence reconstruction, the gradient-component $\hat{g}_{LM}$ is extracted from the deflection estimate and is used to calculate the lensing potential $\hat{\phi}$~\citepalias{P18A8}
\begin{equation}
    \hat{\phi}=\frac{1}{\mathcal{R}_L^\phi}\left(\hat{g}_{LM}-\langle\hat{g}_{LM}\rangle_\mathrm{MC}\right)\,,
\end{equation}
where a mean-field $\langle\hat{g}_{LM}\rangle_\mathrm{MC}$ is subtracted and then normalized by an isotropic lensing potential response $\mathcal{R}_L^\phi$~\citep{OH03}.
$\mathcal{R}_L^\phi$ is calculated for the full sky assuming isotropic effective beams and noise levels. This is accurate to sub-percent levels on all but the largest scales for Planck lensing analyses. When cross-correlating the Planck lensing map with other galaxy surveys, however, a scale-dependent bias can occur since the anisotropic beam and noise properties are locally coupled to the reconstructed $\hat{\kappa}$ field. An empirical re-calibration of the lensing response, or MC norm, is needed~\citep[e.g., see Appendix I in][]{FKM+23}.

This bias can be calibrated experimentally via Monte Carlo simulations of the $\hat{\kappa}$ response to the mask used in the CMB lensing reconstruction~\citep{CML22,C23,QSM+23}. To estimate this bias in the context of our analyses, we repeat our measurement on the 300 FFP10 lensing simulations~\citep{P18A3}. Each simulation is a lensed primary CMB map generated from different initial conditions but represents the same fiducial lensing power spectrum $C^{\kappa\kappa}_L$. We measure the pixel-space 2PCF between the input fiducial lensing convergence masked by DES Y1 footprint $\kappa^\mathrm{fid}_\mathrm{DES\,Y1}$ and either the input fiducial lensing convergence masked by Planck footprint ($\kappa^\mathrm{fid}_\mathrm{Planck}$) or the reconstructed lensing convergence ($\hat{\kappa}^\mathrm{rec}_\mathrm{Planck}$). All the maps are smoothed and $L$-cut in the same way as in Section~\ref{sec:estimators}. 
\begin{figure}
  \includegraphics[width=\linewidth]{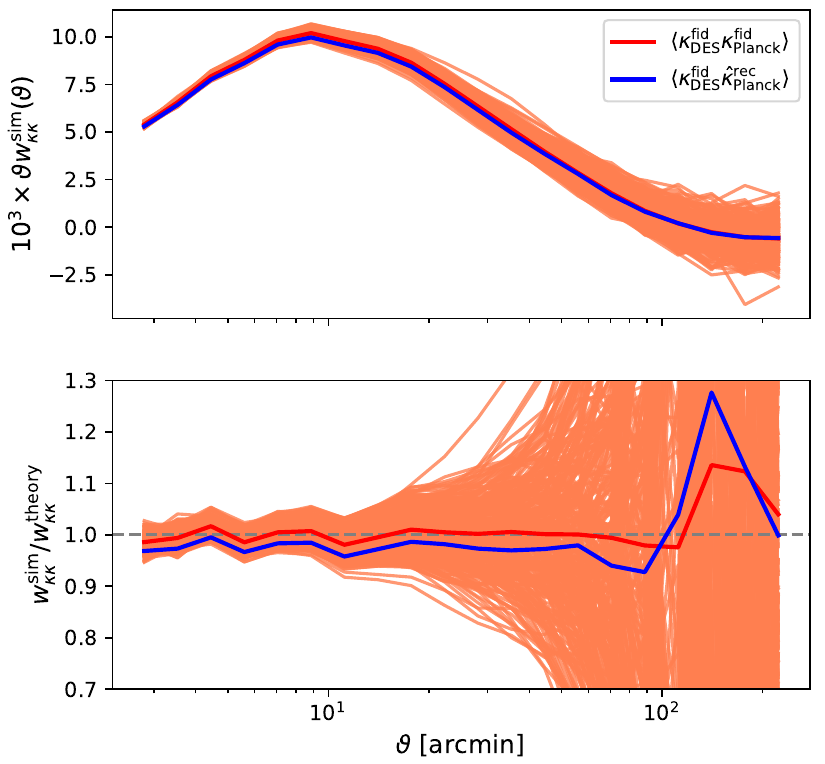}
  \caption{The pixel space 2PCF of CMB lensing convergence field measured from the FFP10 simulations (upper panel) and the theory-normalized values (lower panel). The orange transparent lines are 2PCF between the input fiducial $\kappa$ realization trimmed to DES Y1 footprint and Planck footprint, with the weighted average shown in solid red. The solid blue lines are the mean 2PCF between $\kappa$ trimmed to DES Y1 footprint and reconstructed convergence $\hat{\kappa}$ trimmed to Planck footprint.}
  \label{fi:CMBbias}
\end{figure}

\begin{figure}
  \includegraphics[width=\linewidth]{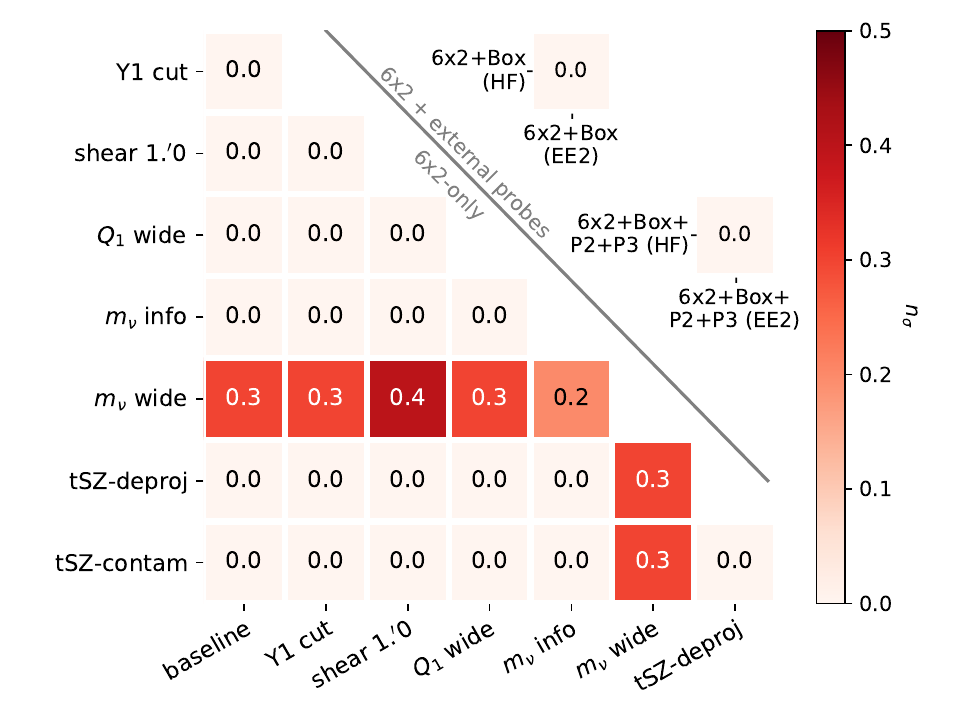}
  \caption{Tension quantified by $n_\sigma$ using parameter difference method among different 6$\times$2 analysis choices. Tension blocks on the lower triangle are 6$\times$2 analyses with different cosmic shear scale cuts (Y1 cut, $1\farcm 0$ cut), $Q_1$ priors, neutrino mass priors, and tSZ treatments. The upper triangle shows the impact of different non-linear matter power spectrum modeling in two external prior settings.
  }
  \label{fi:robust_tension}
\end{figure}

The difference between $w^{\kappa^\mathrm{fid}_\mathrm{DES\,Y1}\hat{\kappa}^\mathrm{rec}_\mathrm{Planck}}(\vartheta)$ and $w^{\kappa^\mathrm{fid}_\mathrm{DES\,Y1}\kappa^\mathrm{fid}_\mathrm{Planck}}(\vartheta)$ serves as a metric for the impact of this scale-dependent bias. 
The results are shown in Fig.~\ref{fi:CMBbias}, where the top panel shows the two ensemble-averaged 2PCFs and the bottom panel shows the same but normalized with the theoretical 2PCF $w^{\kappa\kappa}(\vartheta)$ (see Section~\ref{sec:mv} for definitions of related terms)
\begin{equation}
w^{\kappa\kappa}(\vartheta)\equiv \sum_{L}\frac{2L+1}{4\pi}\overline{P_L}F^2(L)C^{\kappa\kappa}_L \,.
\end{equation}
We see that on small scales, $\hat{\kappa}$ generally is biased low by $\sim 3\%$, while on large scales $\hat{\kappa}$ is biased high.

We apply the multiplicative bias correction factor
\begin{equation}
w^{\kappa^\mathrm{fid}_\mathrm{DES\,Y1}\kappa^\mathrm{fid}_\mathrm{Planck}}(\vartheta)/w^{\kappa^\mathrm{fid}_\mathrm{DES\,Y1}\hat{\kappa}^\mathrm{rec}_\mathrm{Planck}}(\vartheta)
\end{equation}
to our baseline data vector and find this correction to be negligible ($\Delta\chi^2\sim 1.8$). This is also supported by Fig.~\ref{fi:CMBbias} since the envelope generated by the transparent orange lines indicates that cosmic variance is much larger than the difference between the two correlation functions, especially at large scales that survive the scale-cut.

\paragraph{Impact of non-linear power spectrum modeling}
We also explore the robustness of our results when choosing different non-linear matter power spectrum modeling techniques. Specifically, we run 6$\times$2 analyses with \texttt{Halofit} and the \texttt{EuclidEmulator2} separately. We note that the \texttt{EuclidEmulator2} only gives results within a narrower parameter space. Therefore, we run comparison chains assuming two scenarios: 1) 6$\times$2 with a prior that matches the \texttt{EuclidEmulator2} parameter box, and 2) 6$\times$2+P2+P3 where the \texttt{EuclidEmulator2} parameter box is automatically satisfied given the constraining power. 

Again, we quantify the tension between \texttt{Halofit} (HF) and \texttt{EuclidEmulator2} (EE2) within these two scenarios in the upper triangle of Fig.~\ref{fi:robust_tension}. We conclude that non-linear matter power spectrum modeling is not a concern in this work. 

\paragraph{Other analysis choices}
We also explore robustness against a few alternative analysis choices, e.g., sampling the dimension of neutrino mass with wide or informative prior, changing the cosmic shear scale cuts to $1\farcm0$, and changing the prior for the baryonic PC amplitude (see Fig.~\ref{fi:robust_tension}). In all cases, we find negligible tensions and conclude that our results are robust against these variations. 

Propagating the different tSZ data vectors (see Section \ref{sec:optimal_measurements}) through our fiducial 6$\times$2 analysis (assuming informative prior on $Q_1$) we only find negligible differences as indicated by the $n_\sigma$ metric in Fig.~\ref{fi:robust_tension}. We conclude that given our scale-cuts, different tSZ mitigation techniques do not influence our results.

\section{Conclusions}
\label{sec:conclusions}

In this paper, we have built a catalog-to-cosmology pipeline to jointly analyze six two-point statistics measured from photometric LSS and CMB data, namely cosmic shear, galaxy clustering, CMB lensing, and their cross-correlations. 

Our pipeline analytically computes covariances for these measurements that take all cross-covariance terms into account, including mixed Fourier and real-space terms.   
We use our pipeline to conduct a 6$\times$2 analysis of the well-studied DES Y1 and the Planck PR3 data going beyond previous analyses in various aspects: 1) we measure the DES Y1 cosmic shear 2PCF down to $0\farcm25$ and explore possible information gain, 2) we extend CMB cross-correlation measurements to a band surrounding the DES footprint (a $\sim 25\%$ gain in pair counts), and 3) we include the modeling of baryonic physics in our analysis that allows us to include small-scale cosmological information from cosmic shear. 

We validate our pipeline and our analysis choices through an extensive suite of simulated analyses and obtain competitive results on the matter content and matter clustering in our Universe from our 6$\times$2 analysis: $S_8=0.799\pm 0.016$ and $\Omega_{m}=0.262^{+0.022}_{-0.025}$. 

We also combine our 6$\times$2 analysis with independent data from Planck EE+lowE and BAO+BBN+SNe Ia 
and infer $S_8=0.817\pm 0.011$ and $\Omega_{m}=0.3067\pm0.0059$ as our most constraining result. We note that although our 6$\times$2 $S_8$ constraint is located between low-$z$ LSS probes and high-$z$ primary CMB results, our constraints prefer a much lower $\Omega_{m}$ compared to primary CMB results. Our results are consistent with DES Y1 and Planck/SPT constraints~\citep{DESY1xPlanckxSPT_6x2pt} but show slightly lower $\Omega_{m}$ than the DES Y3 analysis~\citep{DES_Y3_3x2pt,DES_Y3_6x2pt_III_23}.

We also constrain baryonic physics in the Universe closely following and extending the work of \citetalias{hem21}. We model baryonic physics through a PCA of existing suites of hydro-simulations. 

The resulting PCs are limited to the physics range spanned by the simulations and do not enable constraints on ``first-principle'' baryonic physics parameters, e.g., the density profile of gas or stellar component, ejection mass fraction and radius, and typical halo mass to retain half of its gas~\citep{ST15,STS+19,AAH+20,AAH+21,GAS+23}. The PCA method considers baryonic physics parameters as linear combinations of the relevant physics across the scales of the summary statistics. 

We use our measurements to constrain baryonic physics quantified by the amplitude of the first PC, $Q_1$. Our 6$\times$2 analysis obtains $Q_1=2.8\pm 1.8$. 
A tighter constraint $Q_1=3.5\pm 1.3$ is derived when combined with Planck EE+lowE and BAO+BBN+SNe Ia. 
These measurements differ from the strongest feedback scenario considered in this work, cOWLS-AGN (T8.7), at $1.7\sigma$ and $1.8\sigma$, respectively. 

Building sophisticated pipelines for the joint analyses of CMB and LSS is an important research topic in cosmological data analysis. Constant refinement on the modeling side is required to meet the quality of upcoming data. In the near future, DES Year 6 and LSST Year 1 data will cover a significant area of the sky (5,000 deg$^2$ and $\sim$12,000 deg$^2$, respectively) overlapping with ACT, SPT, and later SO and S4. Space missions like the recently launched Euclid satellite and the future Roman space telescope will provide exquisite shape and photometry catalogs that will further improve cross-correlation measurements of weak lensing and CMB lensing (and kSZ, tSZ, and CIB).

These data, with their exquisite quality, will require the community to develop models for systematics and statistical uncertainties at the level of accuracy of these data. In that sense, we note that although some features of our pipeline (e.g., the covariance cross-terms and the increased pair counts) are negligible for this DES Y1 plus Planck 2018 analysis, they will likely become important for future analyses when the footprint size of the LSS and CMB components are comparable. On the modeling side, it is also critical to build better models of baryonic physics and better priors for these models. This requires suites of numerical simulations that span a sufficiently large volume with sufficiently high resolution ~\citep[e.g.,][]{VAG+21,AZC+21,SMK+23} over a sufficiently large parameter space (ideally for cosmology and baryonic physics). Results from this paper and future results from our pipeline that constrain the range of allowed physics models can help guide the design of these simulation campaigns such that they are maximally beneficial for future surveys.

\begin{acknowledgments}
We thank Mathew Madhavacheril, Gerrit Farren, Alex Krolewski, Simone Ferraro, and Martin White for helpful discussions on technical details in this work, especially the scale-dependent bias of the reconstructed Planck CMB lensing convergence. 

This work is supported by the Department of Energy Cosmic Frontier program, grant DESC0020215. Simulations in this paper use High Performance Computing (HPC) resources supported by the University of Arizona TRIF, UITS, and RDI and maintained by the UA Research Technologies department.
\end{acknowledgments}

%

\vspace{5mm}

\software{\texttt{CoCoA} \citep{ZSM+23},
\texttt{healpy/HEALPix} \citep{Zonca2019,2005ApJ...622..759G},
\texttt{Getdist} \citep{L19},
\texttt{Cobaya} \citep{TL21,TL19},
\texttt{TreeCorr} \citep{JBJ04}}



\appendix

\section{Details of covariance matrix modeling}
\label{sec:appd_a}

\subsection{Covariance Matrix Terms Related to CMB Lensing}
\label{sec:new_covmat_blocks}
We closely follow \cite{KE17,FES+22} in deriving the Fourier space covariance matrix related to CMB lensing convergence. The Gaussian covariance between two probes $C_{XY}^{ab}(\ell_1)$ and $C_{PQ}^{cd}(\ell_2)$ reads
\begin{equation}
\begin{aligned}
    \mathrm{Cov}^\mathrm{G}\left(C_{XY}^{ab}(\ell_1),\,C_{PQ}^{cd}(\ell_2)\right)=\frac{4\pi\delta_{\ell_1\ell_2}}{\Omega_s(2\ell_1+1)}\times&\left[\,\left(C_{XP}^{ac}(\ell_1)+\delta^{ac}\delta_{XP}N_{X}^a(\ell_1)\right)\,\left(C_{YQ}^{bd}(\ell_2)+\delta^{bd}\delta_{YQ}N_{Y}^b(\ell_2)\right)+\right.\\
    &\left.\left(C_{XQ}^{ad}(\ell_1)+\delta^{ad}\delta_{XQ}N_{X}^a(\ell_1)\right)\,\left(C_{YP}^{bc}(\ell_2)+\delta^{bc}\delta_{YP}N_{Y}^b(\ell_2)\right)\right]\,,
\end{aligned}
\end{equation}
where $\delta^{ab}$, $\delta_{XY}$ are the Kronecker delta functions, $\Omega_s$ is the survey area. $N_{X}^a(\ell_1)$ is the Fourier space noise. For the lens and source galaxy sample, the noise is white, i.e., $N_{\delta_g}^a(\ell_1)=1/\bar{n}_\mathrm{lens}^a$, $N_{\kappa_g}^a(\ell_1)=2\sigma_\epsilon^2/\bar{n}_\mathrm{src}^a$. 
For the reconstructed CMB lensing convergence, the noise power spectrum $N_{\kappa}(\ell_1)$ is scale-dependent. We adopt the MV lensing reconstruction noise power spectrum from the Planck PR3. The non-Gaussian covariance matrix, in the absence of survey geometry effects, is 
\begin{equation}
\label{eqn:cov_ng_0}
\begin{aligned}
    \mathrm{Cov}^\mathrm{NG,0}\left(C_{XY}^{ab}(\ell_1),\,C_{PQ}^{cd}(\ell_2)\right)=&\frac{1}{\Omega_s}\int_{\left|\bm{l}\right|\in\ell_1}\frac{\mathrm{d}^2\bm{l}}{A(\ell_1)}\int_{\left|\bm{l}^\prime\right|\in\ell_2}\frac{\mathrm{d}^2\bm{l}^\prime}{A(\ell_2)}\int\mathrm{d}\chi\frac{q_{X}^a(\chi)q_{Y}^b(\chi)q_{P}^c(\chi)q_{Q}^d(\chi)}{\chi^6}\times\\
    &T_{XYPQ}^{abcd}\left(\frac{\bm{l}}{\chi}, \frac{-\bm{l}}{\chi}, \frac{\bm{l}^\prime}{\chi},\frac{-\bm{l}^\prime}{\chi};z(\chi)\right)\,,
\end{aligned}
\end{equation}
where $A(\ell_1)$ is the area of the annulus of bin $\ell_1$, $T_{XYPQ}^{abcd}$ is the trispectrum of the tracers. To account for the response of the summary statistics to long wavelength modes that exceed the survey footprint, we compute the super sample covariance component as
\begin{equation}
\label{eqn:cov_ng_ssc}
    \begin{aligned}
        \mathrm{Cov}^\mathrm{SSC}\left(C_{XY}^{ab}(\ell_1),\,C_{PQ}^{cd}(\ell_2)\right)=\int \mathrm{d}\chi \frac{q_{X}^{a}(\chi)q_{Y}^{b}(\chi)q_{P}^{c}(\chi)q_{Q}^{d}(\chi)}{\chi^4}\frac{\partial P_{XY}(\ell_1/\chi;z(\chi))}{\partial \delta_b}\frac{\partial P_{PQ}(\ell_2/\chi;z(\chi))}{\partial \delta_b}\sigma_b^2\,,
    \end{aligned}
\end{equation}
where $\sigma_b^2$ is the variance of the background modes $\delta_b$ over the survey window with Fourier space window function $\widetilde{W}_s$,
\begin{equation}
    \begin{aligned}
        \sigma_b^2\equiv\int \frac{\mathrm{d}^2k_\perp}{(2\pi)^2}P_\mathrm{lin}(k_\perp,z)\left|\frac{\widetilde{W}_s(k_\perp)}{\Omega_s}\right|^2\,.
    \end{aligned}
\end{equation}

\subsection{Survey Geometry Correction}
\label{sec:footprint_correction}

Consider a scenario where two tracer fields $X(\bm{x})$ and $Y(\bm{x})$ are measured in different footprints with window function $W_X(\bm{x})$ and $W_Y(\bm{x})$, respectively, and assume that the former footprint is fully embedded in the latter, as sketched in Fig.~\ref{fig:footprint} (i.e., the deep-blue DES Y1 footprint for $X(\bm{x})$/$W_X(\bm{x})$ and the green Planck footprint for $Y(\bm{x})$/$W_Y(\bm{x})$ ). The window functions are defined such that
\begin{equation}
    W_i(\bm{x}) = \left\lbrace \begin{array}{cl}
    1~, & \bm{x} {\rm \ within\ footprint\ }i \\
    0~, & {\rm else}
    \end{array}\right.~,~~(i=X,Y)
\end{equation}

The Gaussian, connected non-Gaussian, and super-sample components of the Fourier space covariances when cross-correlating surveys with different footprints have been derived in \eg, Appendix G of \cite{vUJJ+18}.

In real space, the pure noise terms in the Gaussian auto-covariances of 2PCFs (such as galaxy clustering, galaxy-galaxy lensing, and cosmic shear) measured in the same footprint are proportional to the inverse of the expected numbers of random pairs $N_{p}(\vartheta)$ within each angular separation bin. 
Without the boundary effect, $N_{p}(\vartheta)$ is proportional to the area of the angular bin. 
\cite{TKC+18} have shown that for a finite survey, the boundary effect reduces the expected number of random pairs as approximated by the area scaling, hence increasing the error bar \cite[also see Appendix C of][]{FAC+21}. We incorporate this boundary effect for the DES Y1 6$\times$2 covariance, and show that this boundary effect vanishes for the auto-covariances of the cross-survey probes.

The expected $N_{p}(\vartheta)$ on the 2D sky within angular separation bin $[\vartheta-\Delta\vartheta/2,\vartheta+\Delta\vartheta/2)$ is 
\begin{equation}
\begin{aligned}
N_{p}(\vartheta)&\equiv\sum_{ij}W_{X,i}W_{Y,j}\Delta_\vartheta(ij)=
\bar{n}_X\bar{n}_Y\int_{\hat{\bm{n}}_X} \int_{\hat{\bm{n}}_Y}\,\left(1+w(\hat{\bm{n}}_X\cdot\hat{\bm{n}}_Y)\right)\,W_{X}(\hat{\bm{n}}_X)W_{Y}(\hat{\bm{n}}_Y)\,\delta(\hat{\bm{n}}_X\cdot\hat{\bm{n}}_Y-\mathrm{cos}\vartheta)\,\mathrm{sin}\vartheta\Delta\vartheta\,,
\end{aligned}
\end{equation}
here we approximate the summation over galaxy catalog or \texttt{HEALPix} by an integral over the $S^2$ sphere $\sum_i = \int_{\hat{\bm{n}}}\,n(\hat{\bm{n}})$, where $\hat{\bm{n}}$ is the unit normal vector on the $S^2$ sphere, $n(\hat{\bm{n}})$ is the number density of the tracer, and $\int_{\hat{\bm{n}}}$ is the abbreviation of $\int \mathrm{d}^2\hat{\bm{n}}$. 
$w(\hat{\bm{n}}_X\cdot\hat{\bm{n}}_Y)\equiv\langle n_X(\hat{\bm{n}}_X)n_{Y}(\hat{\bm{n}}_Y)\rangle/(\bar{n}_X\bar{n}_Y)-1$ is the angular galaxy 2PCF. 
We ignore the source galaxy clustering in this work. We also assume that the angular bin is narrow enough such that we can approximate $\Delta_\vartheta(\hat{\bm{n}}_X\cdot \hat{\bm{n}}_Y)$ as $\delta(\hat{\bm{n}}_X\cdot \hat{\bm{n}}_Y - \mathrm{cos}\vartheta)\,\mathrm{sin}\vartheta\Delta\vartheta$. Note that 
\begin{equation}
    \delta(\hat{\bm{n}}_X\cdot \hat{\bm{n}}_Y - \mathrm{cos}\vartheta) = \sum_{\ell=0}^{\infty} \frac{2\ell+1}{2}P_\ell(\mathrm{cos}\vartheta)P_\ell(\hat{\bm{n}}_X\cdot \hat{\bm{n}}_Y)=\sum_{\ell,m=0}^{\infty}2\pi\,P_\ell(\mathrm{cos}\vartheta)Y_{\ell m}(\hat{\bm{n}}_X)Y_{\ell m}^*(\hat{\bm{n}}_Y)\,,
\end{equation}
thus we have
\begin{equation}
\begin{aligned}
    \frac{N_{p}(\vartheta)}{\bar{n}_X\bar{n}_Y\,2\pi\mathrm{sin}\vartheta\Delta\vartheta} &=\int_{\hat{\bm{n}}_X} \int_{\hat{\bm{n}}_Y}\, W_{X}(\hat{\bm{n}}_X)W_{Y}(\hat{\bm{n}}_Y)\sum_{\ell,m=0}^{\infty}\,P_\ell(\mathrm{cos}\vartheta)Y_{\ell m}(\hat{\bm{n}}_X)Y_{\ell m}^*(\hat{\bm{n}}_Y)=
    4\pi\,\xi^{W_XW_Y}(\vartheta)\,,
\end{aligned}
\end{equation}
where we have defined
\begin{subequations}
    \begin{equation}
        \widetilde{W}_{X,\ell m} = \int_{\hat{\bm{n}}_X} W_X(\hat{\bm{n}}_X) Y_{\ell m}(\hat{\bm{n}}_X)\,,
    \end{equation}
    \begin{equation}
        C_\ell^{W_X W_Y} = \frac{1}{2\ell+1}\sum_{m} \widetilde{W}_{X,\ell m}\widetilde{W}_{Y,\ell m}^*\,,
    \end{equation}
    \begin{equation}
        \xi^{W_X W_Y}(\vartheta) = \sum_\ell \frac{2\ell+1}{4\pi} C_\ell^{W_X W_Y} P_\ell(\cos\vartheta)\,.
    \end{equation}
\end{subequations}
Note that by definition, $\xi^{W_XW_Y}(\vartheta)$ is normalized such that $\xi^{W_X W_Y}(0)=\Omega_{X\cap Y}/(4\pi)=f_\mathrm{sky}$, where $\Omega_{X\cap Y}$ is the area of the intersection of footprints $X$ and $Y$. We define $\bar{\xi}^{W_X W_Y}(\vartheta)\equiv \xi^{W_X W_Y}(\vartheta)/f_\mathrm{sky}$.
If the two surveys have the same footprint, we have $\bar{\xi}^{W_X W_Y}(\vartheta)=\bar{\xi}^W(\vartheta)$.

If we neglect the boundary effect, then for every pair with one point $\hat{\bm{n}}_X$ in footprint $X$, the other point $\hat{\bm{n}}_Y$ is assumed to be in footprint $Y$ as well, $W(\hat{\bm{n}}_X)W(\hat{\bm{n}}_Y)\simeq W(\hat{\bm{n}}_X)$, \ie, no missing pair due to the survey boundary. This approximation leads to
\begin{equation}
    \begin{aligned}
        N_{p}(\vartheta) &=\bar{n}_X\bar{n}_Y\int_{\hat{\bm{n}}_X}\int_{\hat{\bm{n}}_Y}\, W_{X}(\hat{\bm{n}}_X)\,\mathrm{sin}\vartheta\Delta\vartheta\sum_{\ell}\frac{2\ell+1}{2}\,P_\ell(\mathrm{cos}\vartheta)P_\ell(\hat{\bm{n}}_X\cdot\hat{\bm{n}}_Y)=\bar{n}_X\bar{n}_Y\, 2\pi \mathrm{sin}\vartheta\Delta\vartheta \,\Omega_{X\cap Y}\,,
    \end{aligned}
\end{equation}

In the scenario where footprint $X$ is fully embedded in footprint $Y$ and the angular scale $\vartheta$ considered is much smaller than the scale of the outer region of footprint $Y$, we always have $W_X(\hat{\bm{n}}_X)W_Y(\hat{\bm{n}}_Y)=W_X(\hat{\bm{n}}_X)$. 
Therefore, for cross-survey 2PCF $C_{XY}(\ell)$, the pure noise term in its Gaussian (auto-)covariance does not need a boundary effect correction, and the usual $1/f_{\rm sky}$ expression is more accurate. Here $f_{\rm sky}$ should take the sky coverage of the inner footprint $f_{\mathrm{sky},X}$. \textit{Thus when considering the survey geometry effect, the $\bar{\xi}^\mathrm{DES\,Y1}(\vartheta)$ correction should be considered in 3$\times$2 but not in $\hat{w}^a_{g\kappa}$ and $\hat{w}^a_{s\kappa}$}.

For completeness, we also consider the Gaussian auto-covariance blocks of $\hat{w}_{g\kappa}^a$ and $\hat{w}_{s\kappa}^a$ if the CMB map has the same footprint as DES Y1. In that case we also have to include the $\bar{\xi}^\mathrm{DES\,Y1}(\vartheta)$ factor in $N_\kappa(\ell)$. However, the scale-dependent nature of $N_\kappa(\ell)$ makes the modeling more complex. One solution is to model the pure shot noise terms $D[\hat{w}_{g\kappa}^a(\vartheta);\,\hat{w}_{g\kappa}^b(\vartheta^\prime)]$ and $D[\hat{w}_{s\kappa}^a(\vartheta);\,\hat{w}_{s\kappa}^b(\vartheta^\prime)]$ in real space. Following~\cite{SWK+02,JSE08}, we have 
\begin{subequations}\label{eqn:append_NN_gksk}
    \begin{equation}\label{eqn:append_NN_sk}
        D\left[\hat{w}_{sk}^a(\vartheta); \hat{w}_{sk}^b(\vartheta^\prime)\right]=\frac{\delta_{ab}\bar{n}_\mathrm{src}^2\bar{n}_\kappa^2}{N_{p}(\vartheta)N_{p}(\vartheta^\prime)}
        \int_{\bm{\theta}}\int_{\bm{\theta^\prime}}\Delta_{\vartheta}(\theta)\Delta_{\vartheta^\prime}(\theta^\prime)\frac{\sigma_\epsilon^2}{2\bar{n}_\mathrm{src}} N(\left|\bm{\theta}-\bm{\theta^\prime}\right|)\,\zeta_{skk}^W(\theta,\theta^\prime,\Delta\psi)\,\mathrm{cos}2\Delta\psi\,,
    \end{equation}
    \begin{equation}\label{eqn:append_NN_gk}
        D\left[\hat{w}_{gk}^a(\vartheta); \hat{w}_{gk}^b(\vartheta^\prime)\right]=\frac{\delta_{ab}\bar{n}_\mathrm{lens}^2\bar{n}_\kappa^2}{N_{p}(\vartheta)N_{p}(\vartheta^\prime)}
        \int_{\bm{\theta}}\int_{\bm{\theta^\prime}}\Delta_{\vartheta}(\theta)\Delta_{\vartheta^\prime}(\theta^\prime)\frac{1}{\bar{n}_\mathrm{lens}} N(\left|\bm{\theta}-\bm{\theta^\prime}\right|)\,\zeta_{gkk}^W(\theta,\theta^\prime,\Delta\psi)\,,
    \end{equation}
\end{subequations}
where $\Delta\psi$ is the angle between $\bm{\theta}$ and $\bm{\theta}^\prime$, $N(\theta)$ is the 2PCF of CMB lensing reconstruction noise in configuration space, $\bar{n}_\kappa$ is the number density of \texttt{HEALPix} pixels, and $\zeta_{gkk}^W$/$\zeta_{skk}^W$ is the survey mask three-point correlation function, e.g.
\begin{equation}
    \zeta_{gkk}^W(\theta,\theta^\prime,\Delta\psi) = \int_{\bm{\theta}^{\prime\prime}}W_g(\bm{\theta}^{\prime\prime})W_{\kappa}(\bm{\theta}^{\prime\prime}+\bm{\theta})W_{\kappa}(\bm{\theta}^{\prime\prime}+\bm{\theta}^{\prime})\,.
\end{equation}
To further simplify, note that the unsmoothed lensing reconstruction noise is dominated by small-scale noise; most of the covariance is coming from $\theta\approx \theta^\prime$ and $\Delta\psi\ll 1$. Also, consider the Cauchy-Schwarz inequality
\begin{equation}
    \left|\int f(\bm{x})g(\bm{x})\mathrm{d}\bm{x}\right|^2 \leq \int \left|f(\bm{x})\right|^2\mathrm{d}\bm{x} \int \left|g(\bm{y})\right|^2\mathrm{d}\bm{y}\,,
\end{equation}
the two sides are equal if and only if $f(\bm{x})$ and $g(\bm{x})$ are linearly dependent.
We take $f(\bm{x})=W^\mathrm{DES\,Y1}(\bm{x})W^\mathrm{Planck}(\bm{x}+\bm{\theta})$ and $g(\bm{x})=W^\mathrm{DES\,Y1}(\bm{x})W^\mathrm{Planck}(\bm{x}+\bm{\theta}^\prime)$, and note that for binary survey mask $W$, $f(\bm{x})=f(\bm{x})^2$, $g(\bm{x})=g(\bm{x})^2$, then we have 
\begin{equation}
    \zeta^W_{gkk}(\bm{\theta},\bm{\theta}^\prime) < \sqrt{\xi^W_{gk}(\bm{\theta})\xi^W_{gk}(\bm{\theta}^\prime)}\,,
\end{equation}
since $f(\bm{x})$ and $g(\bm{x})$ are linear independent. Then equations~(\ref{eqn:append_NN_gksk}) reduce to 
\begin{equation}
    \begin{aligned}
    D\left[\hat{w}_{gk}(\vartheta); \hat{w}_{gk}(\vartheta^\prime)\right]&< \frac{D^F\left[\hat{w}_{gk}(\vartheta); \hat{w}_{gk}(\vartheta^\prime)\right]}{\sqrt{\xi^W_{gk}(\vartheta)\xi^W_{gk}(\vartheta^\prime)}/f_\mathrm{sky}};\,
    D\left[\hat{w}_{sk}(\vartheta); \hat{w}_{sk}(\vartheta^\prime)\right]&< \frac{D^F\left[\hat{w}_{sk}(\vartheta); \hat{w}_{sk}(\vartheta^\prime)\right]}{\sqrt{\xi^W_{sk}(\vartheta)\xi^W_{sk}(\vartheta^\prime)}/f_\mathrm{sky}}\,,
    \end{aligned}
\end{equation}
where $D^F$ is the pure-noise term without edge correction being considered. We adopt the right-hand-side expressions in our covariance matrix modeling, which slightly overestimates the real covariance matrix. We note that this amplification should not be large, since when $\theta\approx\theta^\prime$ and $\Delta\psi\ll 1$, $f(\bm{x})$ and $g(\bm{x})$ are close to linearly dependent. 

The survey geometry corrections are easier for non-Gaussian components. \cite{TH13,KE17} derive non-Gaussian covariance assuming the same footprint for all tracers. Following similar derivation but adopting different footprints, we only have to do the following replacements in equation~(\ref{eqn:cov_ng_0}) and (\ref{eqn:cov_ng_ssc})
\begin{equation}
    \frac{1}{\Omega_s} \rightarrow \frac{\Omega_s^{X\cap Y\cap P\cap Q}}{\Omega_s^{X\cap Y}\Omega_s^{P\cap Q}};\, \sigma_b^2\rightarrow \int \frac{\mathrm{d}^2k_\perp}{(2\pi)^2}P_\mathrm{lin}(k_\perp,z)\frac{\widetilde{W}_{X\cap Y}(k_\perp)}{\Omega_s^{X\cap Y}}\frac{\widetilde{W}^*_{P\cap Q}(k_\perp)}{\Omega_s^{P\cap Q}}\,.
\end{equation}
The $\Omega_s^{X\cap Y}$ and $\Omega_s^{P\cap Q}$ factors in the denominators come from the modified power spectrum estimator, $\Omega_s^{X\cap Y\cap P\cap Q}$ comes from the convolution between the trispectrum and survey window functions.

\section{Synthetic Likelihood Analysis}
\label{sec:syn_like}
\begin{figure}
    \centering
    \includegraphics[width=0.45\linewidth]{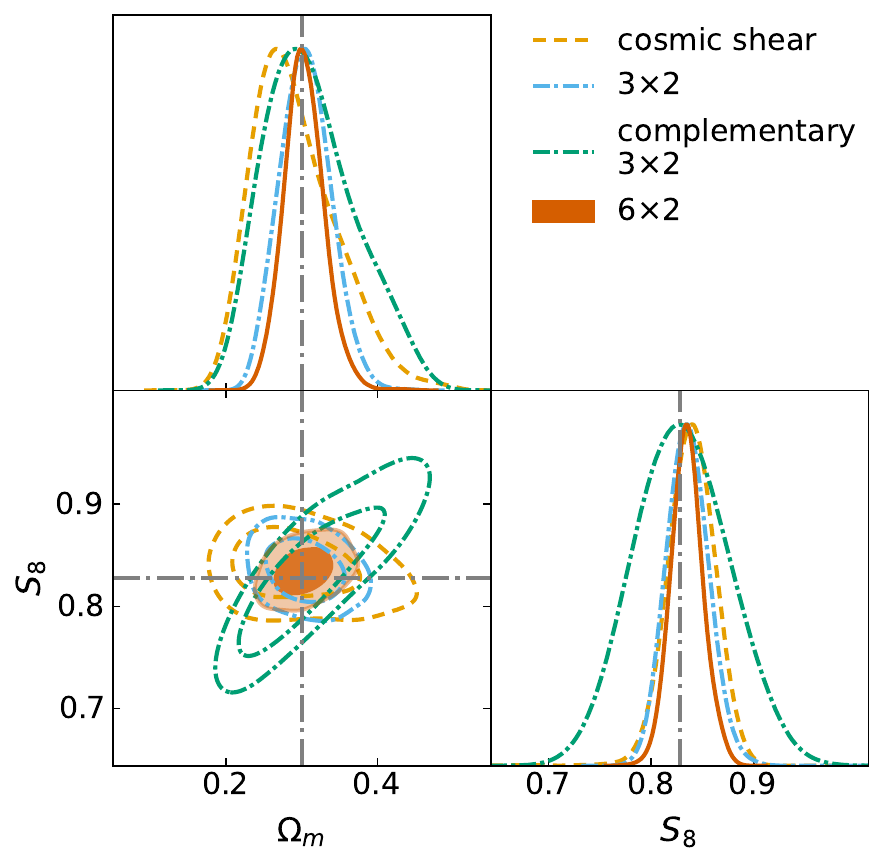}
    \includegraphics[width=0.5\linewidth]{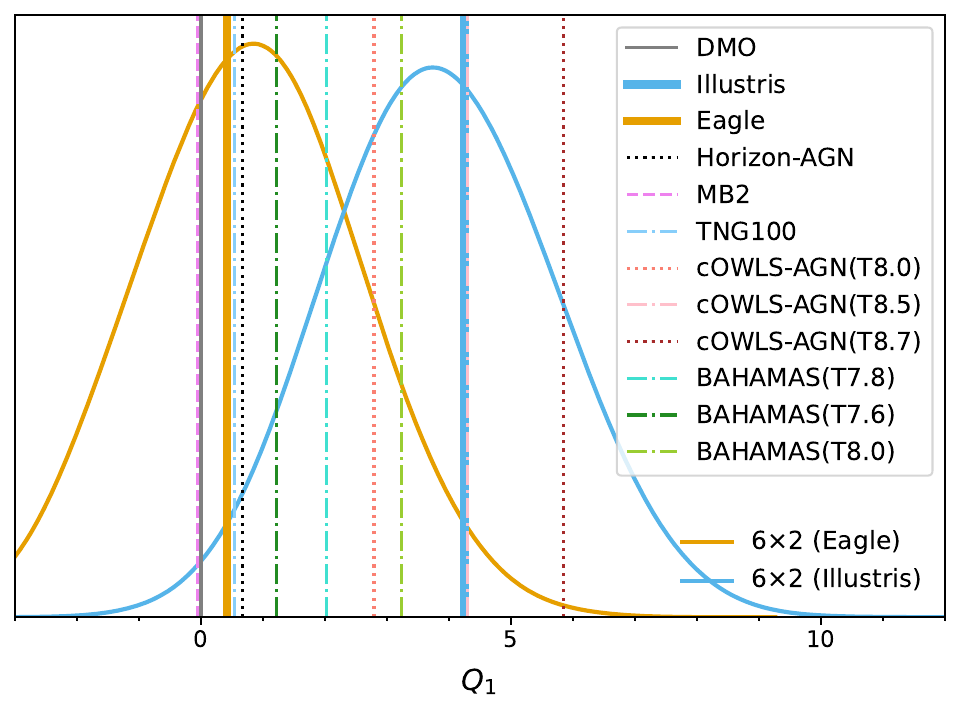}
    \caption{\textbf{Simulated Analysis:} The left panel shows the $\Omega_{m}$-$S_8$ posteriors for synthetic analyses based on Eagle-like baryonic effects contamination. Analyses are performed by marginalizing over $Q_1\in[0,\,4]$. The right panel shows the 1D marginalized posterior of the amplitude $Q_1$ of the first PC of baryonic feedback. The orange and blue curves show the posteriors of Eagle-like and Illustris-like contamination in synthetic data vectors, and the thick vertical solid lines mark their fiducial $Q_1$ values. We also show the fiducial $Q_1$ values for other hydrodynamical simulations. These simulated analyses are forecasts for the results on real data presented in Fig. \ref{fig:result_OmS8_250_Q1info} and Fig. \ref{fig:Q1_results}.}
    \label{fig:probe_combine_Q1info}
\end{figure}

In this section, we present selected simulated likelihood analyses that use synthetic data vectors computed at the fiducial Planck 2018 $\Lambda$CDM cosmology \citep{P18A6} as inputs. 

We show the multi-probe constraining power on $\Omega_{m}$-$S_8$ in the left panel of Fig.~\ref{fig:probe_combine_Q1info}, where we assume an Eagle-like baryonic contamination and an informative prior on $Q_1$. This figure forecasts the 6$\times$2 data analysis presented in Fig. \ref{fig:result_OmS8_250_Q1info}. 

We see that the standard 3$\times$2 analysis based on DES Y1-like data only and the complementary 3$\times$2 (c3$\times$2) analysis that includes auto and cross-correlations with CMB lensing have significantly different parameter degeneracies. 
A principal component analysis in posterior probability space indicates that 3$\times$2 probes $\Omega_{m}^{0.58}\sigma_8$ best while c3$\times$2 is most sensitive to $\Omega_{m}^{0.31}\sigma_8$.

The combined 6$\times$2 analysis is predicted to give 1D marginalized $\Delta S_8=0.007\pm0.016$ when choosing $\vartheta_\mathrm{min}=2\farcm 5$ in ${\hat{\xi}^{ab}_\pm}$. This corresponds to an 79\% increase in the $\Omega_{m}$-$S_8$ figure-of-merit (FoM) from 3$\times$2 to 6$\times$2. Replacing the informative prior on $Q_1$ with the wide prior $Q_1\in[-3,\,12]$, we obtain $\Delta S_8=0.02\pm 0.018$ and the FoM is increased by 70\% compared to 3$\times$2 with informative prior on $Q_1$.

To study the constraining power on $Q_1$ we run 6$\times$2 chains sampling $Q_1$ with wide prior $Q_1\in[-3,\,12]$ and fix $Q_2\equiv 0$. We cut $\hat{\xi}_\pm^{ab}(\vartheta)$ to $\vartheta_\mathrm{min}=2\farcm 5$. Our synthetic input data vectors are derived from Eagle-like and Illustris-like baryon contamination scenarios. 

The $Q_1$ constraints of our simulated analyses are shown in the right panel of Fig.~\ref{fig:probe_combine_Q1info}, which should be compared to the real data analyses presented in Fig. \ref{fig:Q1_results}. We report $\Delta Q_1^\mathrm{Eagle}=0.48\pm 1.7$ and $\Delta Q_1^\mathrm{Illustris}=-0.35\pm 1.8$, which is an excellent forecast of the uncertainties we see when using real data. 

Compared to the 3$\times$2 analyses in~\citetalias{hem21} where $\sigma(Q_1)\sim 2.5$, our uncertainties decreased by $\sim 28\%$ due to the additional constraining power on cosmology and lens/source galaxy properties from CMB lensing and its cross-correlations with DES Y1 galaxies. However, biases of order $\sim 0.3\sigma$ in $Q_1$ can occur due to the degeneracy between $Q_1$ and other cosmological parameters like $S_8$ and $n_s$.

\bibliography{main}{}

\begin{thebibliography}{}
\expandafter\ifx\csname natexlab\endcsname\relax\def\natexlab#1{#1}\fi
\providecommand{\url}[1]{\href{#1}{#1}}
\providecommand{\dodoi}[1]{doi:~\href{http://doi.org/#1}{\nolinkurl{#1}}}
\providecommand{\doeprint}[1]{\href{http://ascl.net/#1}{\nolinkurl{http://ascl.net/#1}}}
\providecommand{\doarXiv}[1]{\href{https://arxiv.org/abs/#1}{\nolinkurl{https://arxiv.org/abs/#1}}}

\bibitem[{{Abbott} {et~al.}(2018){Abbott}, {Abdalla}, {Alarcon}, {Aleksi{\'c}}, {Allam}, {Allen}, {Amara}, {Annis}, {Asorey}, {Avila}, {Bacon}, {Balbinot}, {Banerji}, {Banik}, {Barkhouse}, {Baumer}, {Baxter}, {Bechtol}, {Becker}, {Benoit-L{\'e}vy}, {Benson}, {Bernstein}, {Bertin}, {Blazek}, {Bridle}, {Brooks}, {Brout}, {Buckley-Geer}, {Burke}, {Busha}, {Campos}, {Capozzi}, {Carnero Rosell}, {Carrasco Kind}, {Carretero}, {Castander}, {Cawthon}, {Chang}, {Chen}, {Childress}, {Choi}, {Conselice}, {Crittenden}, {Crocce}, {Cunha}, {D'Andrea}, {da Costa}, {Das}, {Davis}, {Davis}, {De Vicente}, {DePoy}, {DeRose}, {Desai}, {Diehl}, {Dietrich}, {Dodelson}, {Doel}, {Drlica-Wagner}, {Eifler}, {Elliott}, {Elsner}, {Elvin-Poole}, {Estrada}, {Evrard}, {Fang}, {Fernandez}, {Fert{\'e}}, {Finley}, {Flaugher}, {Fosalba}, {Friedrich}, {Frieman}, {Garc{\'\i}a-Bellido}, {Garcia-Fernandez}, {Gatti}, {Gaztanaga}, {Gerdes}, {Giannantonio}, {Gill}, {Glazebrook}, {Goldstein}, {Gruen}, {Gruendl}, {Gschwend}, {Gutierrez}, {Hamilton},
  {Hartley}, {Hinton}, {Honscheid}, {Hoyle}, {Huterer}, {Jain}, {James}, {Jarvis}, {Jeltema}, {Johnson}, {Johnson}, {Kacprzak}, {Kent}, {Kim}, {King}, {Kirk}, {Kokron}, {Kovacs}, {Krause}, {Krawiec}, {Kremin}, {Kuehn}, {Kuhlmann}, {Kuropatkin}, {Lacasa}, {Lahav}, {Li}, {Liddle}, {Lidman}, {Lima}, {Lin}, {MacCrann}, {Maia}, {Makler}, {Manera}, {March}, {Marshall}, {Martini}, {McMahon}, {Melchior}, {Menanteau}, {Miquel}, {Miranda}, {Mudd}, {Muir}, {M{\"o}ller}, {Neilsen}, {Nichol}, {Nord}, {Nugent}, {Ogando}, {Palmese}, {Peacock}, {Peiris}, {Peoples}, {Percival}, {Petravick}, {Plazas}, {Porredon}, {Prat}, {Pujol}, {Rau}, {Refregier}, {Ricker}, {Roe}, {Rollins}, {Romer}, {Roodman}, {Rosenfeld}, {Ross}, {Rozo}, {Rykoff}, {Sako}, {Salvador}, {Samuroff}, {S{\'a}nchez}, {Sanchez}, {Santiago}, {Scarpine}, {Schindler}, {Scolnic}, {Secco}, {Serrano}, {Sevilla-Noarbe}, {Sheldon}, {Smith}, {Smith}, {Smith}, {Soares-Santos}, {Sobreira}, {Suchyta}, {Tarle}, {Thomas}, {Troxel}, {Tucker}, {Tucker}, {Uddin}, {Varga},
  {Vielzeuf}, {Vikram}, {Vivas}, {Walker}, {Wang}, {Wechsler}, {Weller}, {Wester}, {Wolf}, {Yanny}, {Yuan}, {Zenteno}, {Zhang}, {Zhang}, {Zuntz}, \& {Dark Energy Survey Collaboration}}]{DES_Y1_3x2pt}
{Abbott}, T.~M.~C., {Abdalla}, F.~B., {Alarcon}, A., {et~al.} 2018, \prd, 98, 043526, \dodoi{10.1103/PhysRevD.98.043526}

\bibitem[{{Abbott} {et~al.}(2019{\natexlab{a}}){Abbott}, {Abdalla}, {Alarcon}, {Allam}, {Annis}, {Avila}, {Aylor}, {Banerji}, {Banik}, {Baxter}, {Bechtol}, {Becker}, {Benson}, {Bernstein}, {Bertin}, {Bianchini}, {Blazek}, {Bleem}, {Bridle}, {Brooks}, {Buckley-Geer}, {Burke}, {Carlstrom}, {Carnero Rosell}, {Carrasco Kind}, {Carretero}, {Castander}, {Cawthon}, {Chang}, {Chang}, {Cho}, {Choi}, {Chown}, {Crawford}, {Crites}, {Crocce}, {Cunha}, {D'Andrea}, {da Costa}, {Davis}, {de Haan}, {DeRose}, {Desai}, {De Vicente}, {Diehl}, {Dietrich}, {Dobbs}, {Dodelson}, {Doel}, {Drlica-Wagner}, {Eifler}, {Elvin-Poole}, {Everett}, {Flaugher}, {Fosalba}, {Friedrich}, {Frieman}, {Garc{\'\i}a-Bellido}, {Gatti}, {Gaztanaga}, {George}, {Gerdes}, {Giannantonio}, {Gruen}, {Gruendl}, {Gschwend}, {Gutierrez}, {Halverson}, {Harrington}, {Hartley}, {Holder}, {Hollowood}, {Holzapfel}, {Honscheid}, {Hou}, {Hoyle}, {Hrubes}, {Huterer}, {Jain}, {James}, {Jarvis}, {Jeltema}, {Johnson}, {Johnson}, {Kent}, {Kirk}, {Knox}, {Kokron}, {Krause},
  {Kuehn}, {Lahav}, {Lee}, {Leitch}, {Li}, {Lima}, {Lin}, {Luong-Van}, {MacCrann}, {Maia}, {Manzotti}, {Marrone}, {Marshall}, {Martini}, {McMahon}, {Menanteau}, {Meyer}, {Miquel}, {Mocanu}, {Mohr}, {Muir}, {Natoli}, {Nicola}, {Nord}, {Omori}, {Padin}, {Pandey}, {Plazas}, {Porredon}, {Prat}, {Pryke}, {Rau}, {Reichardt}, {Rollins}, {Romer}, {Roodman}, {Ross}, {Rozo}, {Ruhl}, {Rykoff}, {Samuroff}, {S{\'a}nchez}, {Sanchez}, {Sayre}, {Scarpine}, {Schaffer}, {Secco}, {Serrano}, {Sevilla-Noarbe}, {Sheldon}, {Shirokoff}, {Simard}, {Smith}, {Soares-Santos}, {Sobreira}, {Staniszewski}, {Stark}, {Story}, {Suchyta}, {Swanson}, {Tarle}, {Thomas}, {Troxel}, {Tucker}, {Vanderlinde}, {Vieira}, {Vielzeuf}, {Vikram}, {Walker}, {Wechsler}, {Weller}, {Williamson}, {Wu}, {Yanny}, {Zahn}, {Zhang}, {Zuntz}, {DES}, \& {SPT Collaborations}}]{DESY1xPlanckxSPT_6x2pt}
---. 2019{\natexlab{a}}, \prd, 100, 023541, \dodoi{10.1103/PhysRevD.100.023541}

\bibitem[{{Abbott} {et~al.}(2019{\natexlab{b}}){Abbott}, {Abdalla}, {Alarcon}, {Allam}, {Annis}, {Avila}, {Aylor}, {Banerji}, {Banik}, {Baxter}, {Bechtol}, {Becker}, {Benson}, {Bernstein}, {Bertin}, {Bianchini}, {Blazek}, {Bleem}, {Bridle}, {Brooks}, {Buckley-Geer}, {Burke}, {Carlstrom}, {Carnero Rosell}, {Carrasco Kind}, {Carretero}, {Castander}, {Cawthon}, {Chang}, {Chang}, {Cho}, {Choi}, {Chown}, {Crawford}, {Crites}, {Crocce}, {Cunha}, {D'Andrea}, {da Costa}, {Davis}, {de Haan}, {DeRose}, {Desai}, {De Vicente}, {Diehl}, {Dietrich}, {Dobbs}, {Dodelson}, {Doel}, {Drlica-Wagner}, {Eifler}, {Elvin-Poole}, {Everett}, {Flaugher}, {Fosalba}, {Friedrich}, {Frieman}, {Garc{\'\i}a-Bellido}, {Gatti}, {Gaztanaga}, {George}, {Gerdes}, {Giannantonio}, {Gruen}, {Gruendl}, {Gschwend}, {Gutierrez}, {Halverson}, {Harrington}, {Hartley}, {Holder}, {Hollowood}, {Holzapfel}, {Honscheid}, {Hou}, {Hoyle}, {Hrubes}, {Huterer}, {Jain}, {James}, {Jarvis}, {Jeltema}, {Johnson}, {Johnson}, {Kent}, {Kirk}, {Knox}, {Kokron}, {Krause},
  {Kuehn}, {Lahav}, {Lee}, {Leitch}, {Li}, {Lima}, {Lin}, {Luong-Van}, {MacCrann}, {Maia}, {Manzotti}, {Marrone}, {Marshall}, {Martini}, {McMahon}, {Menanteau}, {Meyer}, {Miquel}, {Mocanu}, {Mohr}, {Muir}, {Natoli}, {Nicola}, {Nord}, {Omori}, {Padin}, {Pandey}, {Plazas}, {Porredon}, {Prat}, {Pryke}, {Rau}, {Reichardt}, {Rollins}, {Romer}, {Roodman}, {Ross}, {Rozo}, {Ruhl}, {Rykoff}, {Samuroff}, {S{\'a}nchez}, {Sanchez}, {Sayre}, {Scarpine}, {Schaffer}, {Secco}, {Serrano}, {Sevilla-Noarbe}, {Sheldon}, {Shirokoff}, {Simard}, {Smith}, {Soares-Santos}, {Sobreira}, {Staniszewski}, {Stark}, {Story}, {Suchyta}, {Swanson}, {Tarle}, {Thomas}, {Troxel}, {Tucker}, {Vanderlinde}, {Vieira}, {Vielzeuf}, {Vikram}, {Walker}, {Wechsler}, {Weller}, {Williamson}, {Wu}, {Yanny}, {Zahn}, {Zhang}, {Zuntz}, {DES}, \& {SPT Collaborations}}]{DESY1_5x2pt_results}
---. 2019{\natexlab{b}}, \prd, 100, 023541, \dodoi{10.1103/PhysRevD.100.023541}

\bibitem[{{Abbott} {et~al.}(2022){Abbott}, {Aguena}, {Alarcon}, {Allam}, {Alves}, {Amon}, {Andrade-Oliveira}, {Annis}, {Avila}, {Bacon}, {Baxter}, {Bechtol}, {Becker}, {Bernstein}, {Bhargava}, {Birrer}, {Blazek}, {Brandao-Souza}, {Bridle}, {Brooks}, {Buckley-Geer}, {Burke}, {Camacho}, {Campos}, {Carnero Rosell}, {Carrasco Kind}, {Carretero}, {Castander}, {Cawthon}, {Chang}, {Chen}, {Chen}, {Choi}, {Conselice}, {Cordero}, {Costanzi}, {Crocce}, {da Costa}, {da Silva Pereira}, {Davis}, {Davis}, {De Vicente}, {DeRose}, {Desai}, {Di Valentino}, {Diehl}, {Dietrich}, {Dodelson}, {Doel}, {Doux}, {Drlica-Wagner}, {Eckert}, {Eifler}, {Elsner}, {Elvin-Poole}, {Everett}, {Evrard}, {Fang}, {Farahi}, {Fernandez}, {Ferrero}, {Fert{\'e}}, {Fosalba}, {Friedrich}, {Frieman}, {Garc{\'\i}a-Bellido}, {Gatti}, {Gaztanaga}, {Gerdes}, {Giannantonio}, {Giannini}, {Gruen}, {Gruendl}, {Gschwend}, {Gutierrez}, {Harrison}, {Hartley}, {Herner}, {Hinton}, {Hollowood}, {Honscheid}, {Hoyle}, {Huff}, {Huterer}, {Jain}, {James}, {Jarvis},
  {Jeffrey}, {Jeltema}, {Kovacs}, {Krause}, {Kron}, {Kuehn}, {Kuropatkin}, {Lahav}, {Leget}, {Lemos}, {Liddle}, {Lidman}, {Lima}, {Lin}, {MacCrann}, {Maia}, {Marshall}, {Martini}, {McCullough}, {Melchior}, {Mena-Fern{\'a}ndez}, {Menanteau}, {Miquel}, {Mohr}, {Morgan}, {Muir}, {Myles}, {Nadathur}, {Navarro-Alsina}, {Nichol}, {Ogando}, {Omori}, {Palmese}, {Pandey}, {Park}, {Paz-Chinch{\'o}n}, {Petravick}, {Pieres}, {Plazas Malag{\'o}n}, {Porredon}, {Prat}, {Raveri}, {Rodriguez-Monroy}, {Rollins}, {Romer}, {Roodman}, {Rosenfeld}, {Ross}, {Rykoff}, {Samuroff}, {S{\'a}nchez}, {Sanchez}, {Sanchez}, {Sanchez Cid}, {Scarpine}, {Schubnell}, {Scolnic}, {Secco}, {Serrano}, {Sevilla-Noarbe}, {Sheldon}, {Shin}, {Smith}, {Soares-Santos}, {Suchyta}, {Swanson}, {Tabbutt}, {Tarle}, {Thomas}, {To}, {Troja}, {Troxel}, {Tucker}, {Tutusaus}, {Varga}, {Walker}, {Weaverdyck}, {Wechsler}, {Weller}, {Yanny}, {Yin}, {Zhang}, {Zuntz}, \& {DES Collaboration}}]{DES_Y3_3x2pt}
{Abbott}, T.~M.~C., {Aguena}, M., {Alarcon}, A., {et~al.} 2022, \prd, 105, 023520, \dodoi{10.1103/PhysRevD.105.023520}

\bibitem[{{Abbott} {et~al.}(2023{\natexlab{a}}){Abbott}, {Aguena}, {Alarcon}, {Alves}, {Amon}, {Andrade-Oliveira}, {Asgari}, {Avila}, {Bacon}, {Bechtol}, {Becker}, {Bernstein}, {Bertin}, {Bilicki}, {Blazek}, {Bocquet}, {Brooks}, {Burger}, {Burke}, {Camacho}, {Campos}, {Carnero Rosell}, {Carrasco Kind}, {Carretero}, {Castander}, {Cawthon}, {Chang}, {Chen}, {Choi}, {Conselice}, {Cordero}, {da Costa}, {Pereira}, {Dalal}, {Davis}, {deJong}, {DeRose}, {Desai}, {Diehl}, {Dodelson}, {Doel}, {Doux}, {Drlica-Wagner}, {Dvornik}, {Eckert}, {Eifler}, {Elvin-Poole}, {Everett}, {Fang}, {Ferrero}, {Fert{\'e}}, {Flaugher}, {Friedrich}, {Frieman}, {Garc{\'\i}a-Bellido}, {Gatti}, {Giannini}, {Giblin}, {Gruen}, {Gruendl}, {Gutierrez}, {Harrison}, {Hartley}, {Herner}, {Heymans}, {Hildebrandt}, {Hinton}, {Hoekstra}, {Hollowood}, {Honscheid}, {Huang}, {Huff}, {Huterer}, {James}, {Jarvis}, {Jeffrey}, {Jeltema}, {Joachimi}, {Joudaki}, {Kannawadi}, {Krause}, {Kuehn}, {Kuijken}, {Kuropatkin}, {Leget}, {Lemos}, {Li}, {Li}, {Liddle},
  {Lima}, {Lin}, {Lin}, {MacCrann}, {Mahony}, {Marshall}, {McCullough}, {Mena-Fern{\'a}ndez}, {Menanteau}, {Miquel}, {Mohr}, {Muir}, {Myles}, {Napolitano}, {Navarro-Alsina}, {Ogando}, {Palmese}, {Pandey}, {Park}, {Paterno}, {Peacock}, {Petravick}, {Pieres}, {Plazas Malag{\'o}n}, {Porredon}, {Prat}, {Radovich}, {Raveri}, {Reischke}, {Rollins}, {Romer}, {Roodman}, {Rykoff}, {Samuroff}, {S{\'a}nchez}, {Sanchez}, {Sanchez}, {Schneider}, {Secco}, {Sevilla-Noarbe}, {Shan}, {Sheldon}, {Shin}, {Sif{\'o}n}, {Smith}, {Soares-Santos}, {St{\"o}lzner}, {Suchyta}, {Swanson}, {Tarle}, {Thomas}, {To}, {Troxel}, {Tr{\"o}ster}, {Tutusaus}, {van den Busch}, {Varga}, {Walker}, {Weaverdyck}, {Wechsler}, {Weller}, {Wiseman}, {Wright}, {Yanny}, {Yin}, {Yoon}, {Zhang}, \& {Zuntz}}]{DESY3_KiDS1000_shear}
---. 2023{\natexlab{a}}, arXiv e-prints, arXiv:2305.17173.
\newblock \doarXiv{2305.17173}

\bibitem[{{Abbott} {et~al.}(2023{\natexlab{b}}){Abbott}, {Aguena}, {Alarcon}, {Alves}, {Amon}, {Andrade-Oliveira}, {Annis}, {Ansarinejad}, {Avila}, {Bacon}, {Baxter}, {Bechtol}, {Becker}, {Benson}, {Bernstein}, {Bertin}, {Blazek}, {Bleem}, {Bocquet}, {Brooks}, {Buckley-Geer}, {Burke}, {Camacho}, {Campos}, {Carlstrom}, {Carnero Rosell}, {Carrasco Kind}, {Carretero}, {Cawthon}, {Chang}, {Chang}, {Chen}, {Choi}, {Chown}, {Conselice}, {Cordero}, {Costanzi}, {Crawford}, {Crites}, {Crocce}, {da Costa}, {Davis}, {Davis}, {de Haan}, {De Vicente}, {DeRose}, {Desai}, {Diehl}, {Dobbs}, {Dodelson}, {Doel}, {Doux}, {Drlica-Wagner}, {Eckert}, {Eifler}, {Elsner}, {Elvin-Poole}, {Everett}, {Everett}, {Fang}, {Ferrero}, {Fert{\'e}}, {Flaugher}, {Fosalba}, {Friedrich}, {Frieman}, {Garc{\'\i}a-Bellido}, {Gatti}, {George}, {Giannantonio}, {Giannini}, {Gruen}, {Gruendl}, {Gschwend}, {Gutierrez}, {Halverson}, {Harrison}, {Herner}, {Hinton}, {Holder}, {Hollowood}, {Holzapfel}, {Honscheid}, {Hrubes}, {Huang}, {Huff}, {Huterer},
  {Jain}, {James}, {Jarvis}, {Jeltema}, {Kent}, {Knox}, {Kovacs}, {Krause}, {Kuehn}, {Kuropatkin}, {Lahav}, {Lee}, {Leget}, {Lemos}, {Liddle}, {Lidman}, {Luong-Van}, {McMahon}, {MacCrann}, {March}, {Marshall}, {Martini}, {McCullough}, {Melchior}, {Menanteau}, {Meyer}, {Miquel}, {Mocanu}, {Mohr}, {Morgan}, {Muir}, {Myles}, {Natoli}, {Navarro-Alsina}, {Nichol}, {Omori}, {Padin}, {Pandey}, {Park}, {Paz-Chinch{\'o}n}, {Pereira}, {Pieres}, {Plazas Malag{\'o}n}, {Porredon}, {Prat}, {Pryke}, {Raveri}, {Reichardt}, {Rollins}, {Romer}, {Roodman}, {Rosenfeld}, {Ross}, {Ruhl}, {Rykoff}, {S{\'a}nchez}, {Sanchez}, {Sanchez}, {Schaffer}, {Secco}, {Sevilla-Noarbe}, {Sheldon}, {Shin}, {Shirokoff}, {Smith}, {Staniszewski}, {Stark}, {Suchyta}, {Swanson}, {Tarle}, {To}, {Troxel}, {Tutusaus}, {Varga}, {Vieira}, {Weaverdyck}, {Wechsler}, {Weller}, {Williamson}, {Wu}, {Yanny}, {Yin}, {Zhang}, {Zuntz}, {DES}, \& {SPT Collaborations}}]{DES_Y3_6x2pt_III_23}
---. 2023{\natexlab{b}}, \prd, 107, 023531, \dodoi{10.1103/PhysRevD.107.023531}

\bibitem[{{Alam} {et~al.}(2017){Alam}, {Ata}, {Bailey}, {Beutler}, {Bizyaev}, {Blazek}, {Bolton}, {Brownstein}, {Burden}, {Chuang}, {Comparat}, {Cuesta}, {Dawson}, {Eisenstein}, {Escoffier}, {Gil-Mar{\'\i}n}, {Grieb}, {Hand}, {Ho}, {Kinemuchi}, {Kirkby}, {Kitaura}, {Malanushenko}, {Malanushenko}, {Maraston}, {McBride}, {Nichol}, {Olmstead}, {Oravetz}, {Padmanabhan}, {Palanque-Delabrouille}, {Pan}, {Pellejero-Ibanez}, {Percival}, {Petitjean}, {Prada}, {Price-Whelan}, {Reid}, {Rodr{\'\i}guez-Torres}, {Roe}, {Ross}, {Ross}, {Rossi}, {Rubi{\~n}o-Mart{\'\i}n}, {Saito}, {Salazar-Albornoz}, {Samushia}, {S{\'a}nchez}, {Satpathy}, {Schlegel}, {Schneider}, {Sc{\'o}ccola}, {Seo}, {Sheldon}, {Simmons}, {Slosar}, {Strauss}, {Swanson}, {Thomas}, {Tinker}, {Tojeiro}, {Maga{\~n}a}, {Vazquez}, {Verde}, {Wake}, {Wang}, {Weinberg}, {White}, {Wood-Vasey}, {Y{\`e}che}, {Zehavi}, {Zhai}, \& {Zhao}}]{AAB+17}
{Alam}, S., {Ata}, M., {Bailey}, S., {et~al.} 2017, \mnras, 470, 2617, \dodoi{10.1093/mnras/stx721}

\bibitem[{{Angulo} {et~al.}(2021){Angulo}, {Zennaro}, {Contreras}, {Aric{\`o}}, {Pellejero-Iba{\~n}ez}, \& {St{\"u}cker}}]{AZC+21}
{Angulo}, R.~E., {Zennaro}, M., {Contreras}, S., {et~al.} 2021, \mnras, 507, 5869, \dodoi{10.1093/mnras/stab2018}

\bibitem[{{Aric{\`o}} {et~al.}(2021{\natexlab{a}}){Aric{\`o}}, {Angulo}, {Contreras}, {Ondaro-Mallea}, {Pellejero-Iba{\~n}ez}, \& {Zennaro}}]{AAC+21}
{Aric{\`o}}, G., {Angulo}, R.~E., {Contreras}, S., {et~al.} 2021{\natexlab{a}}, \mnras, 506, 4070, \dodoi{10.1093/mnras/stab1911}

\bibitem[{{Aric{\`o}} {et~al.}(2021{\natexlab{b}}){Aric{\`o}}, {Angulo}, {Hern{\'a}ndez-Monteagudo}, {Contreras}, \& {Zennaro}}]{AAH+21}
{Aric{\`o}}, G., {Angulo}, R.~E., {Hern{\'a}ndez-Monteagudo}, C., {Contreras}, S., \& {Zennaro}, M. 2021{\natexlab{b}}, \mnras, 503, 3596, \dodoi{10.1093/mnras/stab699}

\bibitem[{{Aric{\`o}} {et~al.}(2020){Aric{\`o}}, {Angulo}, {Hern{\'a}ndez-Monteagudo}, {Contreras}, {Zennaro}, {Pellejero-Iba{\~n}ez}, \& {Rosas-Guevara}}]{AAH+20}
{Aric{\`o}}, G., {Angulo}, R.~E., {Hern{\'a}ndez-Monteagudo}, C., {et~al.} 2020, \mnras, 495, 4800, \dodoi{10.1093/mnras/staa1478}

\bibitem[{{Aric{\`o}} {et~al.}(2023){Aric{\`o}}, {Angulo}, {Zennaro}, {Contreras}, {Chen}, \& {Hern{\'a}ndez-Monteagudo}}]{AAZ+23}
{Aric{\`o}}, G., {Angulo}, R.~E., {Zennaro}, M., {et~al.} 2023, \aap, 678, A109, \dodoi{10.1051/0004-6361/202346539}

\bibitem[{{Asgari} {et~al.}(2020){Asgari}, {Tr{\"o}ster}, {Heymans}, {Hildebrandt}, {van den Busch}, {Wright}, {Choi}, {Erben}, {Joachimi}, {Joudaki}, {Kannawadi}, {Kuijken}, {Lin}, {Schneider}, \& {Zuntz}}]{ATH+20}
{Asgari}, M., {Tr{\"o}ster}, T., {Heymans}, C., {et~al.} 2020, \aap, 634, A127, \dodoi{10.1051/0004-6361/201936512}

\bibitem[{{Asgari} {et~al.}(2021){Asgari}, {Lin}, {Joachimi}, {Giblin}, {Heymans}, {Hildebrandt}, {Kannawadi}, {St{\"o}lzner}, {Tr{\"o}ster}, {van den Busch}, {Wright}, {Bilicki}, {Blake}, {de Jong}, {Dvornik}, {Erben}, {Getman}, {Hoekstra}, {K{\"o}hlinger}, {Kuijken}, {Miller}, {Radovich}, {Schneider}, {Shan}, \& {Valentijn}}]{KiDS_shear20}
{Asgari}, M., {Lin}, C.-A., {Joachimi}, B., {et~al.} 2021, \aap, 645, A104, \dodoi{10.1051/0004-6361/202039070}

\bibitem[{{Baxter} {et~al.}(2019){Baxter}, {Omori}, {Chang}, {Giannantonio}, {Kirk}, {Krause}, {Blazek}, {Bleem}, {Choi}, {Crawford}, {Dodelson}, {Eifler}, {Friedrich}, {Gruen}, {Holder}, {Jain}, {Jarvis}, {MacCrann}, {Nicola}, {Pandey}, {Prat}, {Reichardt}, {Samuroff}, {S{\'a}nchez}, {Secco}, {Sheldon}, {Troxel}, {Zuntz}, {Abbott}, {Abdalla}, {Annis}, {Avila}, {Bechtol}, {Benson}, {Bertin}, {Brooks}, {Buckley-Geer}, {Burke}, {Carnero Rosell}, {Carrasco Kind}, {Carretero}, {Castander}, {Cawthon}, {Cunha}, {D'Andrea}, {da Costa}, {Davis}, {De Vicente}, {DePoy}, {Diehl}, {Doel}, {Estrada}, {Evrard}, {Flaugher}, {Fosalba}, {Frieman}, {Garc{\'\i}a-Bellido}, {Gaztanaga}, {Gerdes}, {Gruendl}, {Gschwend}, {Gutierrez}, {Hartley}, {Hollowood}, {Hoyle}, {James}, {Kent}, {Kuehn}, {Kuropatkin}, {Lahav}, {Lima}, {Maia}, {March}, {Marshall}, {Melchior}, {Menanteau}, {Miquel}, {Plazas}, {Roodman}, {Rykoff}, {Sanchez}, {Schindler}, {Schubnell}, {Sevilla-Noarbe}, {Smith}, {Smith}, {Soares-Santos}, {Sobreira}, {Suchyta},
  {Swanson}, {Tarle}, {Walker}, {Wu}, {Weller}, {DES}, \& {SPT Collaborations}}]{DESY1_5x2pt_method}
{Baxter}, E.~J., {Omori}, Y., {Chang}, C., {et~al.} 2019, \prd, 99, 023508, \dodoi{10.1103/PhysRevD.99.023508}

\bibitem[{{Beutler} {et~al.}(2011){Beutler}, {Blake}, {Colless}, {Jones}, {Staveley-Smith}, {Campbell}, {Parker}, {Saunders}, \& {Watson}}]{BBC+11}
{Beutler}, F., {Blake}, C., {Colless}, M., {et~al.} 2011, \mnras, 416, 3017, \dodoi{10.1111/j.1365-2966.2011.19250.x}

\bibitem[{{Bridle} \& {King}(2007)}]{brk07}
{Bridle}, S., \& {King}, L. 2007, New Journal of Physics, 9, 444, \dodoi{10.1088/1367-2630/9/12/444}

\bibitem[{{Carron}(2023)}]{C23}
{Carron}, J. 2023, \jcap, 2023, 057, \dodoi{10.1088/1475-7516/2023/02/057}

\bibitem[{{Carron} {et~al.}(2022){Carron}, {Mirmelstein}, \& {Lewis}}]{CML22}
{Carron}, J., {Mirmelstein}, M., \& {Lewis}, A. 2022, \jcap, 2022, 039, \dodoi{10.1088/1475-7516/2022/09/039}

\bibitem[{{Chen} {et~al.}(2023){Chen}, {Aric{\`o}}, {Huterer}, {Angulo}, {Weaverdyck}, {Friedrich}, {Secco}, {Hern{\'a}ndez-Monteagudo}, {Alarcon}, {Alves}, {Amon}, {Andrade-Oliveira}, {Baxter}, {Bechtol}, {Becker}, {Bernstein}, {Blazek}, {Brandao-Souza}, {Bridle}, {Camacho}, {Campos}, {Carnero Rosell}, {Carrasco Kind}, {Cawthon}, {Chang}, {Chen}, {Chintalapati}, {Choi}, {Cordero}, {Crocce}, {Pereira}, {Davis}, {DeRose}, {Di Valentino}, {Diehl}, {Dodelson}, {Doux}, {Drlica-Wagner}, {Eckert}, {Eifler}, {Elsner}, {Elvin-Poole}, {Everett}, {Fang}, {Fert{\'e}}, {Fosalba}, {Gatti}, {Gaztanaga}, {Giannini}, {Gruen}, {Gruendl}, {Harrison}, {Hartley}, {Herner}, {Hoffmann}, {Huang}, {Huff}, {Jain}, {Jarvis}, {Jeffrey}, {Kacprzak}, {Krause}, {Kuropatkin}, {Leget}, {Lemos}, {Liddle}, {MacCrann}, {McCullough}, {Muir}, {Myles}, {Navarro-Alsina}, {Omori}, {Pandey}, {Park}, {Porredon}, {Prat}, {Raveri}, {Refregier}, {Rollins}, {Roodman}, {Rosenfeld}, {Ross}, {Rykoff}, {Samuroff}, {S{\'a}nchez}, {Sanchez}, {Sevilla-Noarbe},
  {Sheldon}, {Shin}, {Troja}, {Troxel}, {Tutusaus}, {Varga}, {Wechsler}, {Yanny}, {Yin}, {Zhang}, {Zuntz}, {Aguena}, {Annis}, {Bacon}, {Bertin}, {Bocquet}, {Brooks}, {Burke}, {Carretero}, {Conselice}, {Costanzi}, {da Costa}, {De Vicente}, {Desai}, {Doel}, {Ferrero}, {Flaugher}, {Frieman}, {Garc{\'\i}a-Bellido}, {Gerdes}, {Giannantonio}, {Gschwend}, {Gutierrez}, {Hinton}, {Hollowood}, {Honscheid}, {James}, {Kuehn}, {Lahav}, {March}, {Marshall}, {Melchior}, {Menanteau}, {Miquel}, {Mohr}, {Morgan}, {Paz-Chinch{\'o}n}, {Pieres}, {Sanchez}, {Smith}, {Suchyta}, {Swanson}, {Tarle}, {Thomas}, {To}, \& {DES Collaboration}}]{CAA+23}
{Chen}, A., {Aric{\`o}}, G., {Huterer}, D., {et~al.} 2023, \mnras, 518, 5340, \dodoi{10.1093/mnras/stac3213}

\bibitem[{{Chisari} {et~al.}(2018){Chisari}, {Richardson}, {Devriendt}, {Dubois}, {Schneider}, {Le Brun}, {Beckmann}, {Peirani}, {Slyz}, \& {Pichon}}]{CRD+18}
{Chisari}, N.~E., {Richardson}, M.~L.~A., {Devriendt}, J., {et~al.} 2018, \mnras, 480, 3962, \dodoi{10.1093/mnras/sty2093}

\bibitem[{{Chisari} {et~al.}(2019){Chisari}, {Mead}, {Joudaki}, {Ferreira}, {Schneider}, {Mohr}, {Tr{\"o}ster}, {Alonso}, {McCarthy}, {Martin-Alvarez}, {Devriendt}, {Slyz}, \& {van Daalen}}]{CMJ+19}
{Chisari}, N.~E., {Mead}, A.~J., {Joudaki}, S., {et~al.} 2019, The Open Journal of Astrophysics, 2, 4, \dodoi{10.21105/astro.1905.06082}

\bibitem[{{Cooke} {et~al.}(2016){Cooke}, {Pettini}, {Nollett}, \& {Jorgenson}}]{CPNJ16}
{Cooke}, R.~J., {Pettini}, M., {Nollett}, K.~M., \& {Jorgenson}, R. 2016, \apj, 830, 148, \dodoi{10.3847/0004-637X/830/2/148}

\bibitem[{{Dai} {et~al.}(2018){Dai}, {Feng}, \& {Seljak}}]{DFS18}
{Dai}, B., {Feng}, Y., \& {Seljak}, U. 2018, \jcap, 2018, 009, \dodoi{10.1088/1475-7516/2018/11/009}

\bibitem[{{Dalal} {et~al.}(2023){Dalal}, {Li}, {Nicola}, {Zuntz}, {Strauss}, {Sugiyama}, {Zhang}, {Rau}, {Mandelbaum}, {Takada}, {More}, {Miyatake}, {Kannawadi}, {Shirasaki}, {Taniguchi}, {Takahashi}, {Osato}, {Hamana}, {Oguri}, {Nishizawa}, {Plazas Malag{\'o}n}, {Sunayama}, {Alonso}, {Slosar}, {Armstrong}, {Bosch}, {Komiyama}, {Lupton}, {Lust}, {MacArthur}, {Miyazaki}, {Murayama}, {Nishimichi}, {Okura}, {Price}, {Tait}, {Tanaka}, \& {Wang}}]{HSC_Y3_DLN+23}
{Dalal}, R., {Li}, X., {Nicola}, A., {et~al.} 2023, arXiv e-prints, arXiv:2304.00701, \dodoi{10.48550/arXiv.2304.00701}

\bibitem[{{Delgado} {et~al.}(2023){Delgado}, {Angl{\'e}s-Alc{\'a}zar}, {Thiele}, {Pandey}, {Lehman}, {Somerville}, {Ntampaka}, {Genel}, {Villaescusa-Navarro}, \& {Hernquist}}]{DAT+23}
{Delgado}, A.~M., {Angl{\'e}s-Alc{\'a}zar}, D., {Thiele}, L., {et~al.} 2023, \mnras, \dodoi{10.1093/mnras/stad2992}

\bibitem[{{Dubois} {et~al.}(2014){Dubois}, {Pichon}, {Welker}, {Le Borgne}, {Devriendt}, {Laigle}, {Codis}, {Pogosyan}, {Arnouts}, {Benabed}, {Bertin}, {Blaizot}, {Bouchet}, {Cardoso}, {Colombi}, {de Lapparent}, {Desjacques}, {Gavazzi}, {Kassin}, {Kimm}, {McCracken}, {Milliard}, {Peirani}, {Prunet}, {Rouberol}, {Silk}, {Slyz}, {Sousbie}, {Teyssier}, {Tresse}, {Treyer}, {Vibert}, \& {Volonteri}}]{DPW+14}
{Dubois}, Y., {Pichon}, C., {Welker}, C., {et~al.} 2014, \mnras, 444, 1453, \dodoi{10.1093/mnras/stu1227}

\bibitem[{{Eifler} {et~al.}(2015){Eifler}, {Krause}, {Dodelson}, {Zentner}, {Hearin}, \& {Gnedin}}]{ekd15}
{Eifler}, T., {Krause}, E., {Dodelson}, S., {et~al.} 2015, \mnras, 454, 2451, \dodoi{10.1093/mnras/stv2000}

\bibitem[{{Eifler} {et~al.}(2014){Eifler}, {Krause}, {Schneider}, \& {Honscheid}}]{EKS+14}
{Eifler}, T., {Krause}, E., {Schneider}, P., \& {Honscheid}, K. 2014, \mnras, 440, 1379, \dodoi{10.1093/mnras/stu251}

\bibitem[{{Elvin-Poole} {et~al.}(2018){Elvin-Poole}, {Crocce}, {Ross}, {Giannantonio}, {Rozo}, {Rykoff}, {Avila}, {Banik}, {Blazek}, {Bridle}, {Cawthon}, {Drlica-Wagner}, {Friedrich}, {Kokron}, {Krause}, {MacCrann}, {Prat}, {S{\'a}nchez}, {Secco}, {Sevilla-Noarbe}, {Troxel}, {Abbott}, {Abdalla}, {Allam}, {Annis}, {Asorey}, {Bechtol}, {Becker}, {Benoit-L{\'e}vy}, {Bernstein}, {Bertin}, {Brooks}, {Buckley-Geer}, {Burke}, {Carnero Rosell}, {Carollo}, {Carrasco Kind}, {Carretero}, {Castander}, {Cunha}, {D'Andrea}, {da Costa}, {Davis}, {Davis}, {Desai}, {Diehl}, {Dietrich}, {Dodelson}, {Doel}, {Eifler}, {Evrard}, {Fernandez}, {Flaugher}, {Fosalba}, {Frieman}, {Garc{\'\i}a-Bellido}, {Gaztanaga}, {Gerdes}, {Glazebrook}, {Gruen}, {Gruendl}, {Gschwend}, {Gutierrez}, {Hartley}, {Hinton}, {Honscheid}, {Hoormann}, {Jain}, {James}, {Jarvis}, {Jeltema}, {Johnson}, {Johnson}, {King}, {Kuehn}, {Kuhlmann}, {Kuropatkin}, {Lahav}, {Lewis}, {Li}, {Lidman}, {Lima}, {Lin}, {Macaulay}, {March}, {Marshall}, {Martini}, {Melchior},
  {Menanteau}, {Miquel}, {Mohr}, {M{\"o}ller}, {Nichol}, {Nord}, {O'Neill}, {Percival}, {Petravick}, {Plazas}, {Romer}, {Sako}, {Sanchez}, {Scarpine}, {Schindler}, {Schubnell}, {Sheldon}, {Smith}, {Smith}, {Soares-Santos}, {Sobreira}, {Sommer}, {Suchyta}, {Swanson}, {Tarle}, {Thomas}, {Tucker}, {Tucker}, {Uddin}, {Vikram}, {Walker}, {Wechsler}, {Weller}, {Wester}, {Wolf}, {Yuan}, {Zhang}, {Zuntz}, \& {DES Collaboration}}]{EP+_redmagic}
{Elvin-Poole}, J., {Crocce}, M., {Ross}, A.~J., {et~al.} 2018, \prd, 98, 042006, \dodoi{10.1103/PhysRevD.98.042006}

\bibitem[{{Euclid Collaboration} {et~al.}(2021){Euclid Collaboration}, {Knabenhans}, {Stadel}, {Potter}, {Dakin}, {Hannestad}, {Tram}, {Marelli}, {Schneider}, {Teyssier}, {Fosalba}, {Andreon}, {Auricchio}, {Baccigalupi}, {Balaguera-Antol{\'\i}nez}, {Baldi}, {Bardelli}, {Battaglia}, {Bender}, {Biviano}, {Bodendorf}, {Bozzo}, {Branchini}, {Brescia}, {Burigana}, {Cabanac}, {Camera}, {Capobianco}, {Cappi}, {Carbone}, {Carretero}, {Carvalho}, {Casas}, {Casas}, {Castellano}, {Castignani}, {Cavuoti}, {Cledassou}, {Colodro-Conde}, {Congedo}, {Conselice}, {Conversi}, {Copin}, {Corcione}, {Coupon}, {Courtois}, {Da Silva}, {de la Torre}, {Di Ferdinando}, {Duncan}, {Dupac}, {Fabbian}, {Farrens}, {Ferreira}, {Finelli}, {Frailis}, {Franceschi}, {Galeotta}, {Garilli}, {Giocoli}, {Gozaliasl}, {Graci{\'a}-Carpio}, {Grupp}, {Guzzo}, {Holmes}, {Hormuth}, {Israel}, {Jahnke}, {Keihanen}, {Kermiche}, {Kirkpatrick}, {Kubik}, {Kunz}, {Kurki-Suonio}, {Ligori}, {Lilje}, {Lloro}, {Maino}, {Marggraf}, {Markovic}, {Martinet}, {Marulli},
  {Massey}, {Mauri}, {Maurogordato}, {Medinaceli}, {Meneghetti}, {Metcalf}, {Meylan}, {Moresco}, {Morin}, {Moscardini}, {Munari}, {Neissner}, {Niemi}, {Padilla}, {Paltani}, {Pasian}, {Patrizii}, {Pettorino}, {Pires}, {Polenta}, {Poncet}, {Raison}, {Renzi}, {Rhodes}, {Riccio}, {Romelli}, {Roncarelli}, {Saglia}, {S{\'a}nchez}, {Sapone}, {Schneider}, {Scottez}, {Secroun}, {Serrano}, {Sirignano}, {Sirri}, {Stanco}, {Sureau}, {Tallada Cresp{\'\i}}, {Taylor}, {Tenti}, {Tereno}, {Toledo-Moreo}, {Torradeflot}, {Valenziano}, {Valiviita}, {Vassallo}, {Viel}, {Wang}, {Welikala}, {Whittaker}, {Zacchei}, \& {Zucca}}]{EuclidEmu2}
{Euclid Collaboration}, {Knabenhans}, M., {Stadel}, J., {et~al.} 2021, \mnras, 505, 2840, \dodoi{10.1093/mnras/stab1366}

\bibitem[{{Fang} {et~al.}(2022){Fang}, {Eifler}, {Schaan}, {Huang}, {Krause}, \& {Ferraro}}]{FES+22}
{Fang}, X., {Eifler}, T., {Schaan}, E., {et~al.} 2022, \mnras, 509, 5721, \dodoi{10.1093/mnras/stab3410}

\bibitem[{{Fang} {et~al.}(2020){Fang}, {Krause}, {Eifler}, \& {MacCrann}}]{FKEM20}
{Fang}, X., {Krause}, E., {Eifler}, T., \& {MacCrann}, N. 2020, \jcap, 2020, 010, \dodoi{10.1088/1475-7516/2020/05/010}

\bibitem[{{Fang} {et~al.}(2023){Fang}, {Krause}, {Eifler}, {Ferraro}, {Benabed}, {Pranjal R.}, {Ay{\c{c}}oberry}, {Dubois}, \& {Miranda}}]{FKE+23}
{Fang}, X., {Krause}, E., {Eifler}, T., {et~al.} 2023, arXiv e-prints, arXiv:2308.01856, \dodoi{10.48550/arXiv.2308.01856}

\bibitem[{{Farren} {et~al.}(2023){Farren}, {Krolewski}, {MacCrann}, {Ferraro}, {Abril-Cabezas}, {An}, {Atkins}, {Battaglia}, {Bond}, {Calabrese}, {Choi}, {Darwish}, {Devlin}, {Duivenvoorden}, {Dunkley}, {Hill}, {Hilton}, {Huffenberger}, {Kim}, {Louis}, {Madhavacheril}, {Marques}, {Moodley}, {Page}, {Partridge}, {Qu}, {Sehgal}, {Sherwin}, {Sif{\'o}n}, {Staggs}, {Van Engelen}, {Vargas}, {Wenzl}, {White}, \& {Wollack}}]{FKM+23}
{Farren}, G.~S., {Krolewski}, A., {MacCrann}, N., {et~al.} 2023, arXiv e-prints, arXiv:2309.05659, \dodoi{10.48550/arXiv.2309.05659}

\bibitem[{{Fedeli}(2014)}]{F14}
{Fedeli}, C. 2014, \jcap, 2014, 028, \dodoi{10.1088/1475-7516/2014/04/028}

\bibitem[{{Ferlito} {et~al.}(2023){Ferlito}, {Springel}, {Davies}, {Hern{\'a}ndez-Aguayo}, {Pakmor}, {Barrera}, {White}, {Delgado}, {Hadzhiyska}, {Hernquist}, {Kannan}, {Bose}, \& {Frenk}}]{FSD+23}
{Ferlito}, F., {Springel}, V., {Davies}, C.~T., {et~al.} 2023, \mnras, 524, 5591, \dodoi{10.1093/mnras/stad2205}

\bibitem[{{Friedrich} {et~al.}(2021){Friedrich}, {Andrade-Oliveira}, {Camacho}, {Alves}, {Rosenfeld}, {Sanchez}, {Fang}, {Eifler}, {Krause}, {Chang}, {Omori}, {Amon}, {Baxter}, {Elvin-Poole}, {Huterer}, {Porredon}, {Prat}, {Terra}, {Troja}, {Alarcon}, {Bechtol}, {Bernstein}, {Buchs}, {Campos}, {Carnero Rosell}, {Carrasco Kind}, {Cawthon}, {Choi}, {Cordero}, {Crocce}, {Davis}, {DeRose}, {Diehl}, {Dodelson}, {Doux}, {Drlica-Wagner}, {Elsner}, {Everett}, {Fosalba}, {Gatti}, {Giannini}, {Gruen}, {Gruendl}, {Harrison}, {Hartley}, {Jain}, {Jarvis}, {MacCrann}, {McCullough}, {Muir}, {Myles}, {Pandey}, {Raveri}, {Roodman}, {Rodriguez-Monroy}, {Rykoff}, {Samuroff}, {S{\'a}nchez}, {Secco}, {Sevilla-Noarbe}, {Sheldon}, {Troxel}, {Weaverdyck}, {Yanny}, {Aguena}, {Avila}, {Bacon}, {Bertin}, {Bhargava}, {Brooks}, {Burke}, {Carretero}, {Costanzi}, {da Costa}, {Pereira}, {De Vicente}, {Desai}, {Evrard}, {Ferrero}, {Frieman}, {Garc{\'\i}a-Bellido}, {Gaztanaga}, {Gerdes}, {Giannantonio}, {Gschwend}, {Gutierrez}, {Hinton},
  {Hollowood}, {Honscheid}, {James}, {Kuehn}, {Lahav}, {Lima}, {Maia}, {Menanteau}, {Miquel}, {Morgan}, {Palmese}, {Paz-Chinch{\'o}n}, {Plazas}, {Sanchez}, {Scarpine}, {Serrano}, {Soares-Santos}, {Smith}, {Suchyta}, {Tarle}, {Thomas}, {To}, {Varga}, {Weller}, {Wilkinson}, {Wilkinson}, \& {DES Collaboration}}]{FAC+21}
{Friedrich}, O., {Andrade-Oliveira}, F., {Camacho}, H., {et~al.} 2021, \mnras, 508, 3125, \dodoi{10.1093/mnras/stab2384}

\bibitem[{{Gatti} {et~al.}(2018){Gatti}, {Vielzeuf}, {Davis}, {Cawthon}, {Rau}, {DeRose}, {De Vicente}, {Alarcon}, {Rozo}, {Gaztanaga}, {Hoyle}, {Miquel}, {Bernstein}, {Bonnett}, {Carnero Rosell}, {Castander}, {Chang}, {da Costa}, {Gruen}, {Gschwend}, {Hartley}, {Lin}, {MacCrann}, {Maia}, {Ogando}, {Roodman}, {Sevilla-Noarbe}, {Troxel}, {Wechsler}, {Asorey}, {Davis}, {Glazebrook}, {Hinton}, {Lewis}, {Lidman}, {Macaulay}, {M{\"o}ller}, {O'Neill}, {Sommer}, {Uddin}, {Yuan}, {Zhang}, {Abbott}, {Allam}, {Annis}, {Bechtol}, {Brooks}, {Burke}, {Carollo}, {Carrasco Kind}, {Carretero}, {Cunha}, {D'Andrea}, {DePoy}, {Desai}, {Eifler}, {Evrard}, {Flaugher}, {Fosalba}, {Frieman}, {Garc{\'\i}a-Bellido}, {Gerdes}, {Goldstein}, {Gruendl}, {Gutierrez}, {Honscheid}, {Hoormann}, {Jain}, {James}, {Jarvis}, {Jeltema}, {Johnson}, {Johnson}, {Krause}, {Kuehn}, {Kuhlmann}, {Kuropatkin}, {Li}, {Lima}, {Marshall}, {Melchior}, {Menanteau}, {Nichol}, {Nord}, {Plazas}, {Reil}, {Rykoff}, {Sako}, {Sanchez}, {Scarpine}, {Schubnell},
  {Sheldon}, {Smith}, {Smith}, {Soares-Santos}, {Sobreira}, {Suchyta}, {Swanson}, {Tarle}, {Thomas}, {Tucker}, {Tucker}, {Vikram}, {Walker}, {Weller}, {Wester}, \& {Wolf}}]{GVD+18}
{Gatti}, M., {Vielzeuf}, P., {Davis}, C., {et~al.} 2018, \mnras, 477, 1664, \dodoi{10.1093/mnras/sty466}

\bibitem[{{Genel} {et~al.}(2014){Genel}, {Vogelsberger}, {Springel}, {Sijacki}, {Nelson}, {Snyder}, {Rodriguez-Gomez}, {Torrey}, \& {Hernquist}}]{GVS+14}
{Genel}, S., {Vogelsberger}, M., {Springel}, V., {et~al.} 2014, \mnras, 445, 175, \dodoi{10.1093/mnras/stu1654}

\bibitem[{{Giri} \& {Schneider}(2021)}]{GS21}
{Giri}, S.~K., \& {Schneider}, A. 2021, \jcap, 2021, 046, \dodoi{10.1088/1475-7516/2021/12/046}

\bibitem[{{G{\'o}rski} {et~al.}(2005){G{\'o}rski}, {Hivon}, {Banday}, {Wandelt}, {Hansen}, {Reinecke}, \& {Bartelmann}}]{2005ApJ...622..759G}
{G{\'o}rski}, K.~M., {Hivon}, E., {Banday}, A.~J., {et~al.} 2005, \apj, 622, 759, \dodoi{10.1086/427976}

\bibitem[{{Grandis} {et~al.}(2023){Grandis}, {Arico'}, {Schneider}, \& {Linke}}]{GAS+23}
{Grandis}, S., {Arico'}, G., {Schneider}, A., \& {Linke}, L. 2023, arXiv e-prints, arXiv:2309.02920, \dodoi{10.48550/arXiv.2309.02920}

\bibitem[{{Hadzhiyska} {et~al.}(2023){Hadzhiyska}, {Ferraro}, {Pakmor}, {Bose}, {Delgado}, {Hern{\'a}ndez-Aguayo}, {Kannan}, {Springel}, {White}, \& {Hernquist}}]{HFP+23}
{Hadzhiyska}, B., {Ferraro}, S., {Pakmor}, R., {et~al.} 2023, \mnras, 526, 369, \dodoi{10.1093/mnras/stad2751}

\bibitem[{{Haider} {et~al.}(2016){Haider}, {Steinhauser}, {Vogelsberger}, {Genel}, {Springel}, {Torrey}, \& {Hernquist}}]{HSV+16}
{Haider}, M., {Steinhauser}, D., {Vogelsberger}, M., {et~al.} 2016, \mnras, 457, 3024, \dodoi{10.1093/mnras/stw077}

\bibitem[{{Hamana} {et~al.}(2020){Hamana}, {Shirasaki}, {Miyazaki}, {Hikage}, {Oguri}, {More}, {Armstrong}, {Leauthaud}, {Mandelbaum}, {Miyatake}, {Nishizawa}, {Simet}, {Takada}, {Aihara}, {Bosch}, {Komiyama}, {Lupton}, {Murayama}, {Strauss}, \& {Tanaka}}]{HSC_Y1_2PCF}
{Hamana}, T., {Shirasaki}, M., {Miyazaki}, S., {et~al.} 2020, \pasj, 72, 16, \dodoi{10.1093/pasj/psz138}

\bibitem[{{Heymans} {et~al.}(2021){Heymans}, {Tr{\"o}ster}, {Asgari}, {Blake}, {Hildebrandt}, {Joachimi}, {Kuijken}, {Lin}, {S{\'a}nchez}, {van den Busch}, {Wright}, {Amon}, {Bilicki}, {de Jong}, {Crocce}, {Dvornik}, {Erben}, {Fortuna}, {Getman}, {Giblin}, {Glazebrook}, {Hoekstra}, {Joudaki}, {Kannawadi}, {K{\"o}hlinger}, {Lidman}, {Miller}, {Napolitano}, {Parkinson}, {Schneider}, {Shan}, {Valentijn}, {Verdoes Kleijn}, \& {Wolf}}]{KiDS1000_3x2pt}
{Heymans}, C., {Tr{\"o}ster}, T., {Asgari}, M., {et~al.} 2021, \aap, 646, A140, \dodoi{10.1051/0004-6361/202039063}

\bibitem[{{Hirata} \& {Seljak}(2004)}]{hs04}
{Hirata}, C.~M., \& {Seljak}, U. 2004, \prd, 70, 063526, \dodoi{10.1103/PhysRevD.70.063526}

\bibitem[{{Hoyle} {et~al.}(2018){Hoyle}, {Gruen}, {Bernstein}, {Rau}, {De Vicente}, {Hartley}, {Gaztanaga}, {DeRose}, {Troxel}, {Davis}, {Alarcon}, {MacCrann}, {Prat}, {S{\'a}nchez}, {Sheldon}, {Wechsler}, {Asorey}, {Becker}, {Bonnett}, {Carnero Rosell}, {Carollo}, {Carrasco Kind}, {Castander}, {Cawthon}, {Chang}, {Childress}, {Davis}, {Drlica-Wagner}, {Gatti}, {Glazebrook}, {Gschwend}, {Hinton}, {Hoormann}, {Kim}, {King}, {Kuehn}, {Lewis}, {Lidman}, {Lin}, {Macaulay}, {Maia}, {Martini}, {Mudd}, {M{\"o}ller}, {Nichol}, {Ogando}, {Rollins}, {Roodman}, {Ross}, {Rozo}, {Rykoff}, {Samuroff}, {Sevilla-Noarbe}, {Sharp}, {Sommer}, {Tucker}, {Uddin}, {Varga}, {Vielzeuf}, {Yuan}, {Zhang}, {Abbott}, {Abdalla}, {Allam}, {Annis}, {Bechtol}, {Benoit-L{\'e}vy}, {Bertin}, {Brooks}, {Buckley-Geer}, {Burke}, {Busha}, {Capozzi}, {Carretero}, {Crocce}, {D'Andrea}, {da Costa}, {DePoy}, {Desai}, {Diehl}, {Doel}, {Eifler}, {Estrada}, {Evrard}, {Fernandez}, {Flaugher}, {Fosalba}, {Frieman}, {Garc{\'\i}a-Bellido}, {Gerdes},
  {Giannantonio}, {Goldstein}, {Gruendl}, {Gutierrez}, {Honscheid}, {James}, {Jarvis}, {Jeltema}, {Johnson}, {Johnson}, {Kirk}, {Krause}, {Kuhlmann}, {Kuropatkin}, {Lahav}, {Li}, {Lima}, {March}, {Marshall}, {Melchior}, {Menanteau}, {Miquel}, {Nord}, {O'Neill}, {Plazas}, {Romer}, {Sako}, {Sanchez}, {Santiago}, {Scarpine}, {Schindler}, {Schubnell}, {Smith}, {Smith}, {Soares-Santos}, {Sobreira}, {Suchyta}, {Swanson}, {Tarle}, {Thomas}, {Tucker}, {Vikram}, {Walker}, {Weller}, {Wester}, {Wolf}, {Yanny}, {Zuntz}, \& {DES Collaboration}}]{HGB+18}
{Hoyle}, B., {Gruen}, D., {Bernstein}, G.~M., {et~al.} 2018, \mnras, 478, 592, \dodoi{10.1093/mnras/sty957}

\bibitem[{{Huang} {et~al.}(2019){Huang}, {Eifler}, {Mandelbaum}, \& {Dodelson}}]{hem19}
{Huang}, H.-J., {Eifler}, T., {Mandelbaum}, R., \& {Dodelson}, S. 2019, \mnras, 488, 1652, \dodoi{10.1093/mnras/stz1714}

\bibitem[{{Huang} {et~al.}(2021){Huang}, {Eifler}, {Mandelbaum}, {Bernstein}, {Chen}, {Choi}, {Garc{\'\i}a-Bellido}, {Huterer}, {Krause}, {Rozo}, {Singh}, {Bridle}, {DeRose}, {Elvin-Poole}, {Fang}, {Friedrich}, {Gatti}, {Gaztanaga}, {Gruen}, {Hartley}, {Hoyle}, {Jarvis}, {MacCrann}, {Miranda}, {Rau}, {Prat}, {S{\'a}nchez}, {Samuroff}, {Troxel}, {Zuntz}, {Abbott}, {Aguena}, {Annis}, {Avila}, {Becker}, {Bertin}, {Brooks}, {Burke}, {Carnero Rosell}, {Carrasco Kind}, {Carretero}, {Castander}, {da Costa}, {De Vicente}, {Dietrich}, {Doel}, {Everett}, {Flaugher}, {Fosalba}, {Frieman}, {Gruendl}, {Gutierrez}, {Hinton}, {Honscheid}, {James}, {Kuehn}, {Lahav}, {Lima}, {Maia}, {Marshall}, {Menanteau}, {Miquel}, {Paz-Chinch{\'o}n}, {Malag{\'o}n}, {Romer}, {Roodman}, {Sanchez}, {Scarpine}, {Serrano}, {Sevilla}, {Smith}, {Soares-Santos}, {Suchyta}, {Swanson}, {Tarle}, {Thomas}, {Weller}, \& {DES Collaboration}}]{hem21}
{Huang}, H.-J., {Eifler}, T., {Mandelbaum}, R., {et~al.} 2021, \mnras, 502, 6010, \dodoi{10.1093/mnras/stab357}

\bibitem[{{Huff} \& {Mandelbaum}(2017)}]{HM17}
{Huff}, E., \& {Mandelbaum}, R. 2017, arXiv e-prints, arXiv:1702.02600.
\newblock \doarXiv{1702.02600}

\bibitem[{{Jarvis} {et~al.}(2004){Jarvis}, {Bernstein}, \& {Jain}}]{JBJ04}
{Jarvis}, M., {Bernstein}, G., \& {Jain}, B. 2004, \mnras, 352, 338, \dodoi{10.1111/j.1365-2966.2004.07926.x}

\bibitem[{{Jing} {et~al.}(2006){Jing}, {Zhang}, {Lin}, {Gao}, \& {Springel}}]{JZL+06}
{Jing}, Y.~P., {Zhang}, P., {Lin}, W.~P., {Gao}, L., \& {Springel}, V. 2006, \apjl, 640, L119, \dodoi{10.1086/503547}

\bibitem[{{Joachimi} {et~al.}(2008){Joachimi}, {Schneider}, \& {Eifler}}]{JSE08}
{Joachimi}, B., {Schneider}, P., \& {Eifler}, T. 2008, \aap, 477, 43, \dodoi{10.1051/0004-6361:20078400}

\bibitem[{{Joachimi} {et~al.}(2021){Joachimi}, {Lin}, {Asgari}, {Tr{\"o}ster}, {Heymans}, {Hildebrandt}, {K{\"o}hlinger}, {S{\'a}nchez}, {Wright}, {Bilicki}, {Blake}, {van den Busch}, {Crocce}, {Dvornik}, {Erben}, {Getman}, {Giblin}, {Hoekstra}, {Kannawadi}, {Kuijken}, {Napolitano}, {Schneider}, {Scoccimarro}, {Sellentin}, {Shan}, {von Wietersheim-Kramsta}, \& {Zuntz}}]{JLA+21}
{Joachimi}, B., {Lin}, C.~A., {Asgari}, M., {et~al.} 2021, \aap, 646, A129, \dodoi{10.1051/0004-6361/202038831}

\bibitem[{{Khandai} {et~al.}(2015){Khandai}, {Di Matteo}, {Croft}, {Wilkins}, {Feng}, {Tucker}, {DeGraf}, \& {Liu}}]{KMC+15}
{Khandai}, N., {Di Matteo}, T., {Croft}, R., {et~al.} 2015, \mnras, 450, 1349, \dodoi{10.1093/mnras/stv627}

\bibitem[{{Kobayashi} {et~al.}(2022){Kobayashi}, {Nishimichi}, {Takada}, \& {Miyatake}}]{KNTM22}
{Kobayashi}, Y., {Nishimichi}, T., {Takada}, M., \& {Miyatake}, H. 2022, \prd, 105, 083517, \dodoi{10.1103/PhysRevD.105.083517}

\bibitem[{{Krause} \& {Eifler}(2017)}]{KE17}
{Krause}, E., \& {Eifler}, T. 2017, \mnras, 470, 2100, \dodoi{10.1093/mnras/stx1261}

\bibitem[{{Krause} {et~al.}(2016){Krause}, {Eifler}, \& {Blazek}}]{Krause16_IA}
{Krause}, E., {Eifler}, T., \& {Blazek}, J. 2016, \mnras, 456, 207, \dodoi{10.1093/mnras/stv2615}

\bibitem[{{Krause} {et~al.}(2017){Krause}, {Eifler}, {Zuntz}, {Friedrich}, {Troxel}, {Dodelson}, {Blazek}, {Secco}, {MacCrann}, {Baxter}, {Chang}, {Chen}, {Crocce}, {DeRose}, {Ferte}, {Kokron}, {Lacasa}, {Miranda}, {Omori}, {Porredon}, {Rosenfeld}, {Samuroff}, {Wang}, {Wechsler}, {Abbott}, {Abdalla}, {Allam}, {Annis}, {Bechtol}, {Benoit-Levy}, {Bernstein}, {Brooks}, {Burke}, {Capozzi}, {Carrasco Kind}, {Carretero}, {D'Andrea}, {da Costa}, {Davis}, {DePoy}, {Desai}, {Diehl}, {Dietrich}, {Evrard}, {Flaugher}, {Fosalba}, {Frieman}, {Garcia-Bellido}, {Gaztanaga}, {Giannantonio}, {Gruen}, {Gruendl}, {Gschwend}, {Gutierrez}, {Honscheid}, {James}, {Jeltema}, {Kuehn}, {Kuhlmann}, {Lahav}, {Lima}, {Maia}, {March}, {Marshall}, {Martini}, {Menanteau}, {Miquel}, {Nichol}, {Plazas}, {Romer}, {Rykoff}, {Sanchez}, {Scarpine}, {Schindler}, {Schubnell}, {Sevilla-Noarbe}, {Smith}, {Soares-Santos}, {Sobreira}, {Suchyta}, {Swanson}, {Tarle}, {Tucker}, {Vikram}, {Walker}, \& {Weller}}]{DESY1_method}
{Krause}, E., {Eifler}, T.~F., {Zuntz}, J., {et~al.} 2017, arXiv e-prints, arXiv:1706.09359.
\newblock \doarXiv{1706.09359}

\bibitem[{{Lange} {et~al.}(2023){Lange}, {Hearin}, {Leauthaud}, {van den Bosch}, {Xhakaj}, {Guo}, {Wechsler}, \& {DeRose}}]{LHL+23}
{Lange}, J.~U., {Hearin}, A.~P., {Leauthaud}, A., {et~al.} 2023, \mnras, 520, 5373, \dodoi{10.1093/mnras/stad473}

\bibitem[{{Le Brun} {et~al.}(2014){Le Brun}, {McCarthy}, {Schaye}, \& {Ponman}}]{BMSP14}
{Le Brun}, A. M.~C., {McCarthy}, I.~G., {Schaye}, J., \& {Ponman}, T.~J. 2014, \mnras, 441, 1270, \dodoi{10.1093/mnras/stu608}

\bibitem[{{Lemos} {et~al.}(2021){Lemos}, {Raveri}, {Campos}, {Park}, {Chang}, {Weaverdyck}, {Huterer}, {Liddle}, {Blazek}, {Cawthon}, {Choi}, {DeRose}, {Dodelson}, {Doux}, {Gatti}, {Gruen}, {Harrison}, {Krause}, {Lahav}, {MacCrann}, {Muir}, {Prat}, {Rau}, {Rollins}, {Samuroff}, {Zuntz}, {Aguena}, {Allam}, {Annis}, {Avila}, {Bacon}, {Bernstein}, {Bertin}, {Brooks}, {Burke}, {Carnero Rosell}, {Carrasco Kind}, {Carretero}, {Castander}, {Conselice}, {Costanzi}, {Crocce}, {Pereira}, {Davis}, {De Vicente}, {Desai}, {Diehl}, {Doel}, {Eckert}, {Eifler}, {Elvin-Poole}, {Everett}, {Evrard}, {Ferrero}, {Fert{\'e}}, {Flaugher}, {Fosalba}, {Frieman}, {Garc{\'\i}a-Bellido}, {Gaztanaga}, {Gerdes}, {Giannantonio}, {Gruendl}, {Gschwend}, {Gutierrez}, {Hartley}, {Hinton}, {Hollowood}, {Honscheid}, {Hoyle}, {Huff}, {James}, {Jarvis}, {Lima}, {Maia}, {March}, {Marshall}, {Martini}, {Melchior}, {Menanteau}, {Miquel}, {Mohr}, {Morgan}, {Myles}, {Ogando}, {Palmese}, {Pandey}, {Paz-Chinch{\'o}n}, {Plazas Malag{\'o}n},
  {Rodriguez-Monroy}, {Roodman}, {Sanchez}, {Scarpine}, {Schubnell}, {Secco}, {Serrano}, {Sevilla-Noarbe}, {Smith}, {Soares-Santos}, {Suchyta}, {Swanson}, {Tarle}, {Thomas}, {To}, {Troxel}, {Varga}, {Weller}, {Wester}, \& {DES Collaboration}}]{LRC+21}
{Lemos}, P., {Raveri}, M., {Campos}, A., {et~al.} 2021, \mnras, 505, 6179, \dodoi{10.1093/mnras/stab1670}

\bibitem[{{Lewis}(2019)}]{L19}
{Lewis}, A. 2019, arXiv e-prints, arXiv:1910.13970, \dodoi{10.48550/arXiv.1910.13970}

\bibitem[{{Li} {et~al.}(2023){Li}, {Zhang}, {Sugiyama}, {Dalal}, {Rau}, {Mandelbaum}, {Takada}, {More}, {Strauss}, {Miyatake}, {Shirasaki}, {Hamana}, {Oguri}, {Luo}, {Nishizawa}, {Takahashi}, {Nicola}, {Osato}, {Kannawadi}, {Sunayama}, {Armstrong}, {Komiyama}, {Lupton}, {Lust}, {Miyazaki}, {Murayama}, {Nishimichi}, {Okura}, {Price}, {Tait}, {Tanaka}, \& {Wang}}]{HSC_Y3_LZS+23}
{Li}, X., {Zhang}, T., {Sugiyama}, S., {et~al.} 2023, arXiv e-prints, arXiv:2304.00702, \dodoi{10.48550/arXiv.2304.00702}

\bibitem[{{Lin} {et~al.}(2020){Lin}, {Harnois-D{\'e}raps}, {Eifler}, {Pospisil}, {Mandelbaum}, {Lee}, {Singh}, \& {LSST Dark Energy Science Collaboration}}]{likelihood}
{Lin}, C.-H., {Harnois-D{\'e}raps}, J., {Eifler}, T., {et~al.} 2020, \mnras, 499, 2977, \dodoi{10.1093/mnras/staa2948}

\bibitem[{{Madhavacheril} {et~al.}(2023){Madhavacheril}, {Qu}, {Sherwin}, {MacCrann}, {Li}, {Abril-Cabezas}, {Ade}, {Aiola}, {Alford}, {Amiri}, {Amodeo}, {An}, {Atkins}, {Austermann}, {Battaglia}, {Battistelli}, {Beall}, {Bean}, {Beringue}, {Bhandarkar}, {Biermann}, {Bolliet}, {Bond}, {Cai}, {Calabrese}, {Calafut}, {Capalbo}, {Carrero}, {Challinor}, {Chesmore}, {Cho}, {Choi}, {Clark}, {C{\'o}rdova Rosado}, {Cothard}, {Coughlin}, {Coulton}, {Crowley}, {Dalal}, {Darwish}, {Devlin}, {Dicker}, {Doze}, {Duell}, {Duff}, {Duivenvoorden}, {Dunkley}, {D{\"u}nner}, {Fanfani}, {Fankhanel}, {Farren}, {Ferraro}, {Freundt}, {Fuzia}, {Gallardo}, {Garrido}, {Givans}, {Gluscevic}, {Golec}, {Guan}, {Hall}, {Halpern}, {Han}, {Harrison}, {Hasselfield}, {Healy}, {Henderson}, {Hensley}, {Herv{\'\i}as-Caimapo}, {Hill}, {Hilton}, {Hilton}, {Hincks}, {Hlo{\v{z}}ek}, {Ho}, {Huber}, {Hubmayr}, {Huffenberger}, {Hughes}, {Irwin}, {Isopi}, {Jense}, {Keller}, {Kim}, {Knowles}, {Koopman}, {Kosowsky}, {Kramer}, {Kusiak}, {La Posta}, {Lague},
  {Lakey}, {Lee}, {Li}, {Limon}, {Lokken}, {Louis}, {Lungu}, {MacInnis}, {Maldonado}, {Maldonado}, {Mallaby-Kay}, {Marques}, {McMahon}, {Mehta}, {Menanteau}, {Moodley}, {Morris}, {Mroczkowski}, {Naess}, {Namikawa}, {Nati}, {Newburgh}, {Nicola}, {Niemack}, {Nolta}, {Orlowski-Scherer}, {Page}, {Pandey}, {Partridge}, {Prince}, {Puddu}, {Radiconi}, {Robertson}, {Rojas}, {Sakuma}, {Salatino}, {Schaan}, {Schmitt}, {Sehgal}, {Shaikh}, {Sierra}, {Sievers}, {Sif{\'o}n}, {Simon}, {Sonka}, {Spergel}, {Staggs}, {Storer}, {Switzer}, {Tampier}, {Thornton}, {Trac}, {Treu}, {Tucker}, {Ulluom}, {Vale}, {Van Engelen}, {Van Lanen}, {van Marrewijk}, {Vargas}, {Vavagiakis}, {Wagoner}, {Wang}, {Wenzl}, {Wollack}, {Xu}, {Zago}, \& {Zhang}}]{ACT_DR6_CMBL}
{Madhavacheril}, M.~S., {Qu}, F.~J., {Sherwin}, B.~D., {et~al.} 2023, arXiv e-prints, arXiv:2304.05203, \dodoi{10.48550/arXiv.2304.05203}

\bibitem[{{Marinacci} {et~al.}(2018){Marinacci}, {Vogelsberger}, {Pakmor}, {Torrey}, {Springel}, {Hernquist}, {Nelson}, {Weinberger}, {Pillepich}, {Naiman}, \& {Genel}}]{MVP+18}
{Marinacci}, F., {Vogelsberger}, M., {Pakmor}, R., {et~al.} 2018, \mnras, 480, 5113, \dodoi{10.1093/mnras/sty2206}

\bibitem[{{McCarthy} {et~al.}(2017){McCarthy}, {Schaye}, {Bird}, \& {Le Brun}}]{MSB+17}
{McCarthy}, I.~G., {Schaye}, J., {Bird}, S., \& {Le Brun}, A. M.~C. 2017, \mnras, 465, 2936, \dodoi{10.1093/mnras/stw2792}

\bibitem[{{Mead} {et~al.}(2021){Mead}, {Brieden}, {Tr{\"o}ster}, \& {Heymans}}]{MBT+21}
{Mead}, A.~J., {Brieden}, S., {Tr{\"o}ster}, T., \& {Heymans}, C. 2021, \mnras, 502, 1401, \dodoi{10.1093/mnras/stab082}

\bibitem[{{Mead} {et~al.}(2015){Mead}, {Peacock}, {Heymans}, {Joudaki}, \& {Heavens}}]{MPH+15}
{Mead}, A.~J., {Peacock}, J.~A., {Heymans}, C., {Joudaki}, S., \& {Heavens}, A.~F. 2015, \mnras, 454, 1958, \dodoi{10.1093/mnras/stv2036}

\bibitem[{{Mead} {et~al.}(2020){Mead}, {Tr{\"o}ster}, {Heymans}, {Van Waerbeke}, \& {McCarthy}}]{MTH+20}
{Mead}, A.~J., {Tr{\"o}ster}, T., {Heymans}, C., {Van Waerbeke}, L., \& {McCarthy}, I.~G. 2020, \aap, 641, A130, \dodoi{10.1051/0004-6361/202038308}

\bibitem[{{Miyatake} {et~al.}(2023){Miyatake}, {Sugiyama}, {Takada}, {Nishimichi}, {Li}, {Shirasaki}, {More}, {Kobayashi}, {Nishizawa}, {Rau}, {Zhang}, {Takahashi}, {Dalal}, {Mandelbaum}, {Strauss}, {Hamana}, {Oguri}, {Osato}, {Luo}, {Kannawadi}, {Hsieh}, {Armstrong}, {Komiyama}, {Lupton}, {Lust}, {MacArthur}, {Miyazaki}, {Murayama}, {Okura}, {Price}, {Sunayama}, {Tait}, {Tanaka}, \& {Wang}}]{HSC_Y3_MST+23}
{Miyatake}, H., {Sugiyama}, S., {Takada}, M., {et~al.} 2023, arXiv e-prints, arXiv:2304.00704, \dodoi{10.48550/arXiv.2304.00704}

\bibitem[{{Mohammed} {et~al.}(2014){Mohammed}, {Martizzi}, {Teyssier}, \& {Amara}}]{MMT+14}
{Mohammed}, I., {Martizzi}, D., {Teyssier}, R., \& {Amara}, A. 2014, arXiv e-prints, arXiv:1410.6826, \dodoi{10.48550/arXiv.1410.6826}

\bibitem[{{Naiman} {et~al.}(2018){Naiman}, {Pillepich}, {Springel}, {Ramirez-Ruiz}, {Torrey}, {Vogelsberger}, {Pakmor}, {Nelson}, {Marinacci}, {Hernquist}, {Weinberger}, \& {Genel}}]{NPSR+18}
{Naiman}, J.~P., {Pillepich}, A., {Springel}, V., {et~al.} 2018, \mnras, 477, 1206, \dodoi{10.1093/mnras/sty618}

\bibitem[{{Nelson} {et~al.}(2018){Nelson}, {Pillepich}, {Springel}, {Weinberger}, {Hernquist}, {Pakmor}, {Genel}, {Torrey}, {Vogelsberger}, {Kauffmann}, {Marinacci}, \& {Naiman}}]{NPSW+18}
{Nelson}, D., {Pillepich}, A., {Springel}, V., {et~al.} 2018, \mnras, 475, 624, \dodoi{10.1093/mnras/stx3040}

\bibitem[{{Okamoto} \& {Hu}(2003)}]{OH03}
{Okamoto}, T., \& {Hu}, W. 2003, \prd, 67, 083002, \dodoi{10.1103/PhysRevD.67.083002}

\bibitem[{{Omori}(2022)}]{O22}
{Omori}, Y. 2022, arXiv e-prints, arXiv:2212.07420, \dodoi{10.48550/arXiv.2212.07420}

\bibitem[{{Omori} {et~al.}(2019{\natexlab{a}}){Omori}, {Giannantonio}, {Porredon}, {Baxter}, {Chang}, {Crocce}, {Fosalba}, {Alarcon}, {Banik}, {Blazek}, {Bleem}, {Bridle}, {Cawthon}, {Choi}, {Chown}, {Crawford}, {Dodelson}, {Drlica-Wagner}, {Eifler}, {Elvin-Poole}, {Friedrich}, {Gruen}, {Holder}, {Huterer}, {Jain}, {Jarvis}, {Kirk}, {Kokron}, {Krause}, {MacCrann}, {Muir}, {Prat}, {Reichardt}, {Ross}, {Rozo}, {Rykoff}, {S{\'a}nchez}, {Secco}, {Simard}, {Wechsler}, {Zuntz}, {Abbott}, {Abdalla}, {Allam}, {Avila}, {Aylor}, {Benson}, {Bernstein}, {Bertin}, {Bianchini}, {Brooks}, {Buckley-Geer}, {Burke}, {Carlstrom}, {Carnero Rosell}, {Carrasco Kind}, {Carretero}, {Castander}, {Chang}, {Cho}, {Crites}, {Cunha}, {da Costa}, {de Haan}, {Davis}, {De Vicente}, {Desai}, {Diehl}, {Dietrich}, {Dobbs}, {Everett}, {Doel}, {Estrada}, {Flaugher}, {Frieman}, {Garc{\'\i}a-Bellido}, {Gaztanaga}, {Gerdes}, {George}, {Gruendl}, {Gschwend}, {Gutierrez}, {Halverson}, {Harrington}, {Hartley}, {Hollowood}, {Holzapfel}, {Honscheid},
  {Hou}, {Hoyle}, {Hrubes}, {James}, {Jeltema}, {Kuehn}, {Kuropatkin}, {Lee}, {Leitch}, {Lima}, {Luong-Van}, {Manzotti}, {Marrone}, {Marshall}, {McMahon}, {Melchior}, {Menanteau}, {Meyer}, {Miller}, {Miquel}, {Mocanu}, {Mohr}, {Natoli}, {Padin}, {Plazas}, {Pryke}, {Romer}, {Roodman}, {Ruhl}, {Sanchez}, {Scarpine}, {Schaffer}, {Schubnell}, {Serrano}, {Sevilla-Noarbe}, {Shirokoff}, {Smith}, {Soares-Santos}, {Sobreira}, {Staniszewski}, {Stark}, {Story}, {Suchyta}, {Swanson}, {Tarle}, {Thomas}, {Troxel}, {Vanderlinde}, {Vieira}, {Walker}, {Wu}, {Zahn}, {DES Collaboration}, \& {SPT Collaboration}}]{ogp19}
{Omori}, Y., {Giannantonio}, T., {Porredon}, A., {et~al.} 2019{\natexlab{a}}, \prd, 100, 043501, \dodoi{10.1103/PhysRevD.100.043501}

\bibitem[{{Omori} {et~al.}(2019{\natexlab{b}}){Omori}, {Baxter}, {Chang}, {Kirk}, {Alarcon}, {Bernstein}, {Bleem}, {Cawthon}, {Choi}, {Chown}, {Crawford}, {Davis}, {De Vicente}, {DeRose}, {Dodelson}, {Eifler}, {Fosalba}, {Friedrich}, {Gatti}, {Gaztanaga}, {Giannantonio}, {Gruen}, {Hartley}, {Holder}, {Hoyle}, {Huterer}, {Jain}, {Jarvis}, {Krause}, {MacCrann}, {Miquel}, {Prat}, {Rau}, {Reichardt}, {Rozo}, {Samuroff}, {S{\'a}nchez}, {Secco}, {Sheldon}, {Simard}, {Troxel}, {Vielzeuf}, {Wechsler}, {Zuntz}, {Abbott}, {Abdalla}, {Allam}, {Annis}, {Avila}, {Aylor}, {Benson}, {Bertin}, {Bridle}, {Brooks}, {Burke}, {Carlstrom}, {Carnero Rosell}, {Carrasco Kind}, {Carretero}, {Castander}, {Chang}, {Cho}, {Crites}, {Crocce}, {Cunha}, {da Costa}, {de Haan}, {Desai}, {Diehl}, {Dietrich}, {Dobbs}, {Everett}, {Fernandez}, {Flaugher}, {Frieman}, {Garc{\'\i}a-Bellido}, {George}, {Gruendl}, {Gutierrez}, {Halverson}, {Harrington}, {Hollowood}, {Honscheid}, {Holzapfel}, {Hou}, {Hrubes}, {James}, {Jeltema}, {Kuehn}, {Kuropatkin},
  {Lima}, {Lin}, {Lee}, {Leitch}, {Luong-Van}, {Maia}, {Manzotti}, {Marrone}, {Marshall}, {Martini}, {McMahon}, {Melchior}, {Menanteau}, {Meyer}, {Mocanu}, {Mohr}, {Natoli}, {Ogando}, {Padin}, {Plazas}, {Pryke}, {Romer}, {Roodman}, {Ruhl}, {Rykoff}, {Sanchez}, {Scarpine}, {Schaffer}, {Schindler}, {Sevilla-Noarbe}, {Shirokoff}, {Smith}, {Smith}, {Soares-Santos}, {Sobreira}, {Staniszewski}, {Stark}, {Story}, {Suchyta}, {Swanson}, {Tarle}, {Thomas}, {Vanderlinde}, {Vieira}, {Vikram}, {Walker}, {Weller}, {Williamson}, {Wu}, {Zahn}, {DES Collaboration}, \& {SPT Collaboration}}]{obc19}
{Omori}, Y., {Baxter}, E.~J., {Chang}, C., {et~al.} 2019{\natexlab{b}}, \prd, 100, 043517, \dodoi{10.1103/PhysRevD.100.043517}

\bibitem[{{Pandey} {et~al.}(2023){Pandey}, {Lehman}, {Baxter}, {Ni}, {Angl{\'e}s-Alc{\'a}zar}, {Genel}, {Villaescusa-Navarro}, {Delgado}, \& {di Matteo}}]{PLB+23}
{Pandey}, S., {Lehman}, K., {Baxter}, E.~J., {et~al.} 2023, \mnras, 525, 1779, \dodoi{10.1093/mnras/stad2268}

\bibitem[{{Papamakarios} {et~al.}(2017){Papamakarios}, {Pavlakou}, \& {Murray}}]{PPM17}
{Papamakarios}, G., {Pavlakou}, T., \& {Murray}, I. 2017, arXiv e-prints, arXiv:1705.07057, \dodoi{10.48550/arXiv.1705.07057}

\bibitem[{{Pillepich} {et~al.}(2018){Pillepich}, {Nelson}, {Hernquist}, {Springel}, {Pakmor}, {Torrey}, {Weinberger}, {Genel}, {Naiman}, {Marinacci}, \& {Vogelsberger}}]{PNH+18}
{Pillepich}, A., {Nelson}, D., {Hernquist}, L., {et~al.} 2018, \mnras, 475, 648, \dodoi{10.1093/mnras/stx3112}

\bibitem[{{Planck Collaboration} {et~al.}(2016){Planck Collaboration}, {Adam}, {Ade}, {Aghanim}, {Arnaud}, {Ashdown}, {Aumont}, {Baccigalupi}, {Banday}, {Barreiro}, {Bartolo}, {Battaner}, {Benabed}, {Beno{\^\i}t}, {Benoit-L{\'e}vy}, {Bernard}, {Bersanelli}, {Bertincourt}, {Bielewicz}, {Bock}, {Bonavera}, {Bond}, {Borrill}, {Bouchet}, {Boulanger}, {Bucher}, {Burigana}, {Calabrese}, {Cardoso}, {Catalano}, {Challinor}, {Chamballu}, {Chary}, {Chiang}, {Christensen}, {Clements}, {Colombi}, {Colombo}, {Combet}, {Couchot}, {Coulais}, {Crill}, {Curto}, {Cuttaia}, {Danese}, {Davies}, {Davis}, {de Bernardis}, {de Rosa}, {de Zotti}, {Delabrouille}, {Delouis}, {D{\'e}sert}, {Diego}, {Dole}, {Donzelli}, {Dor{\'e}}, {Douspis}, {Ducout}, {Dupac}, {Efstathiou}, {Elsner}, {En{\ss}lin}, {Eriksen}, {Falgarone}, {Fergusson}, {Finelli}, {Forni}, {Frailis}, {Fraisse}, {Franceschi}, {Frejsel}, {Galeotta}, {Galli}, {Ganga}, {Ghosh}, {Giard}, {Giraud-H{\'e}raud}, {Gjerl{\o}w}, {Gonz{\'a}lez-Nuevo}, {G{\'o}rski}, {Gratton},
  {Gruppuso}, {Gudmundsson}, {Hansen}, {Hanson}, {Harrison}, {Henrot-Versill{\'e}}, {Herranz}, {Hildebrandt}, {Hivon}, {Hobson}, {Holmes}, {Hornstrup}, {Hovest}, {Huffenberger}, {Hurier}, {Jaffe}, {Jaffe}, {Jones}, {Juvela}, {Keih{\"a}nen}, {Keskitalo}, {Kisner}, {Kneissl}, {Knoche}, {Kunz}, {Kurki-Suonio}, {Lagache}, {Lamarre}, {Lasenby}, {Lattanzi}, {Lawrence}, {Le Jeune}, {Leahy}, {Lellouch}, {Leonardi}, {Lesgourgues}, {Levrier}, {Liguori}, {Lilje}, {Linden-V{\o}rnle}, {L{\'o}pez-Caniego}, {Lubin}, {Mac{\'\i}as-P{\'e}rez}, {Maggio}, {Maino}, {Mandolesi}, {Mangilli}, {Maris}, {Martin}, {Mart{\'\i}nez-Gonz{\'a}lez}, {Masi}, {Matarrese}, {McGehee}, {Melchiorri}, {Mendes}, {Mennella}, {Migliaccio}, {Mitra}, {Miville-Desch{\^e}nes}, {Moneti}, {Montier}, {Moreno}, {Morgante}, {Mortlock}, {Moss}, {Mottet}, {Munshi}, {Murphy}, {Naselsky}, {Nati}, {Natoli}, {Netterfield}, {N{\o}rgaard-Nielsen}, {Noviello}, {Novikov}, {Novikov}, {Oxborrow}, {Paci}, {Pagano}, {Pajot}, {Paoletti}, {Pasian}, {Patanchon}, {Pearson},
  {Perdereau}, {Perotto}, {Perrotta}, {Pettorino}, {Piacentini}, {Piat}, {Pierpaoli}, {Pietrobon}, {Plaszczynski}, {Pointecouteau}, {Polenta}, {Pratt}, {Pr{\'e}zeau}, {Prunet}, {Puget}, {Rachen}, {Reinecke}, {Remazeilles}, {Renault}, {Renzi}, {Ristorcelli}, {Rocha}, {Rosset}, {Rossetti}, {Roudier}, {Rowan-Robinson}, {Rusholme}, {Sandri}, {Santos}, {Sauv{\'e}}, {Savelainen}, {Savini}, {Scott}, {Seiffert}, {Shellard}, {Spencer}, {Stolyarov}, {Stompor}, {Sudiwala}, {Sutton}, {Suur-Uski}, {Sygnet}, {Tauber}, {Terenzi}, {Toffolatti}, {Tomasi}, {Tristram}, {Tucci}, {Tuovinen}, {Valenziano}, {Valiviita}, {Van Tent}, {Vibert}, {Vielva}, {Villa}, {Wade}, {Wandelt}, {Watson}, {Wehus}, {Yvon}, {Zacchei}, \& {Zonca}}]{P15PA7}
{Planck Collaboration}, {Adam}, R., {Ade}, P.~A.~R., {et~al.} 2016, \aap, 594, A7, \dodoi{10.1051/0004-6361/201525844}

\bibitem[{{Planck Collaboration} {et~al.}(2020{\natexlab{a}}){Planck Collaboration}, {Aghanim}, {Akrami}, {Ashdown}, {Aumont}, {Baccigalupi}, {Ballardini}, {Banday}, {Barreiro}, {Bartolo}, {Basak}, {Benabed}, {Bernard}, {Bersanelli}, {Bielewicz}, {Bock}, {Bond}, {Borrill}, {Bouchet}, {Boulanger}, {Bucher}, {Burigana}, {Calabrese}, {Cardoso}, {Carron}, {Challinor}, {Chiang}, {Colombo}, {Combet}, {Crill}, {Cuttaia}, {de Bernardis}, {de Zotti}, {Delabrouille}, {Di Valentino}, {Diego}, {Dor{\'e}}, {Douspis}, {Ducout}, {Dupac}, {Efstathiou}, {Elsner}, {En{\ss}lin}, {Eriksen}, {Fantaye}, {Fernandez-Cobos}, {Finelli}, {Forastieri}, {Frailis}, {Fraisse}, {Franceschi}, {Frolov}, {Galeotta}, {Galli}, {Ganga}, {G{\'e}nova-Santos}, {Gerbino}, {Ghosh}, {Gonz{\'a}lez-Nuevo}, {G{\'o}rski}, {Gratton}, {Gruppuso}, {Gudmundsson}, {Hamann}, {Handley}, {Hansen}, {Herranz}, {Hivon}, {Huang}, {Jaffe}, {Jones}, {Karakci}, {Keih{\"a}nen}, {Keskitalo}, {Kiiveri}, {Kim}, {Knox}, {Krachmalnicoff}, {Kunz}, {Kurki-Suonio}, {Lagache},
  {Lamarre}, {Lasenby}, {Lattanzi}, {Lawrence}, {Le Jeune}, {Levrier}, {Lewis}, {Liguori}, {Lilje}, {Lindholm}, {L{\'o}pez-Caniego}, {Lubin}, {Ma}, {Mac{\'\i}as-P{\'e}rez}, {Maggio}, {Maino}, {Mandolesi}, {Mangilli}, {Marcos-Caballero}, {Maris}, {Martin}, {Mart{\'\i}nez-Gonz{\'a}lez}, {Matarrese}, {Mauri}, {McEwen}, {Melchiorri}, {Mennella}, {Migliaccio}, {Miville-Desch{\^e}nes}, {Molinari}, {Moneti}, {Montier}, {Morgante}, {Moss}, {Natoli}, {Pagano}, {Paoletti}, {Partridge}, {Patanchon}, {Perrotta}, {Pettorino}, {Piacentini}, {Polastri}, {Polenta}, {Puget}, {Rachen}, {Reinecke}, {Remazeilles}, {Renzi}, {Rocha}, {Rosset}, {Roudier}, {Rubi{\~n}o-Mart{\'\i}n}, {Ruiz-Granados}, {Salvati}, {Sandri}, {Savelainen}, {Scott}, {Sirignano}, {Sunyaev}, {Suur-Uski}, {Tauber}, {Tavagnacco}, {Tenti}, {Toffolatti}, {Tomasi}, {Trombetti}, {Valiviita}, {Van Tent}, {Vielva}, {Villa}, {Vittorio}, {Wandelt}, {Wehus}, {White}, {White}, {Zacchei}, \& {Zonca}}]{P18A8}
{Planck Collaboration}, {Aghanim}, N., {Akrami}, Y., {et~al.} 2020{\natexlab{a}}, \aap, 641, A8, \dodoi{10.1051/0004-6361/201833886}

\bibitem[{{Planck Collaboration} {et~al.}(2020{\natexlab{b}}){Planck Collaboration}, {Akrami}, {Ashdown}, {Aumont}, {Baccigalupi}, {Ballardini}, {Banday}, {Barreiro}, {Bartolo}, {Basak}, {Benabed}, {Bersanelli}, {Bielewicz}, {Bond}, {Borrill}, {Bouchet}, {Boulanger}, {Bucher}, {Burigana}, {Calabrese}, {Cardoso}, {Carron}, {Casaponsa}, {Challinor}, {Colombo}, {Combet}, {Crill}, {Cuttaia}, {de Bernardis}, {de Rosa}, {de Zotti}, {Delabrouille}, {Delouis}, {Di Valentino}, {Dickinson}, {Diego}, {Donzelli}, {Dor{\'e}}, {Ducout}, {Dupac}, {Efstathiou}, {Elsner}, {En{\ss}lin}, {Eriksen}, {Falgarone}, {Fernandez-Cobos}, {Finelli}, {Forastieri}, {Frailis}, {Fraisse}, {Franceschi}, {Frolov}, {Galeotta}, {Galli}, {Ganga}, {G{\'e}nova-Santos}, {Gerbino}, {Ghosh}, {Gonz{\'a}lez-Nuevo}, {G{\'o}rski}, {Gratton}, {Gruppuso}, {Gudmundsson}, {Handley}, {Hansen}, {Helou}, {Herranz}, {Hildebrandt}, {Huang}, {Jaffe}, {Karakci}, {Keih{\"a}nen}, {Keskitalo}, {Kiiveri}, {Kim}, {Kisner}, {Krachmalnicoff}, {Kunz}, {Kurki-Suonio},
  {Lagache}, {Lamarre}, {Lasenby}, {Lattanzi}, {Lawrence}, {Le Jeune}, {Levrier}, {Liguori}, {Lilje}, {Lindholm}, {L{\'o}pez-Caniego}, {Lubin}, {Ma}, {Mac{\'\i}as-P{\'e}rez}, {Maggio}, {Maino}, {Mandolesi}, {Mangilli}, {Marcos-Caballero}, {Maris}, {Martin}, {Mart{\'\i}nez-Gonz{\'a}lez}, {Matarrese}, {Mauri}, {McEwen}, {Meinhold}, {Melchiorri}, {Mennella}, {Migliaccio}, {Miville-Desch{\^e}nes}, {Molinari}, {Moneti}, {Montier}, {Morgante}, {Natoli}, {Oppizzi}, {Pagano}, {Paoletti}, {Partridge}, {Peel}, {Pettorino}, {Piacentini}, {Polenta}, {Puget}, {Rachen}, {Reinecke}, {Remazeilles}, {Renzi}, {Rocha}, {Roudier}, {Rubi{\~n}o-Mart{\'\i}n}, {Ruiz-Granados}, {Salvati}, {Sandri}, {Savelainen}, {Scott}, {Seljebotn}, {Sirignano}, {Spencer}, {Suur-Uski}, {Tauber}, {Tavagnacco}, {Tenti}, {Thommesen}, {Toffolatti}, {Tomasi}, {Trombetti}, {Valiviita}, {Van Tent}, {Vielva}, {Villa}, {Vittorio}, {Wandelt}, {Wehus}, {Zacchei}, \& {Zonca}}]{P18A4}
{Planck Collaboration}, {Akrami}, Y., {Ashdown}, M., {et~al.} 2020{\natexlab{b}}, \aap, 641, A4, \dodoi{10.1051/0004-6361/201833881}

\bibitem[{{Planck Collaboration} {et~al.}(2020{\natexlab{c}}){Planck Collaboration}, {Aghanim}, {Akrami}, {Ashdown}, {Aumont}, {Baccigalupi}, {Ballardini}, {Banday}, {Barreiro}, {Bartolo}, {Basak}, {Benabed}, {Bernard}, {Bersanelli}, {Bielewicz}, {Bond}, {Borrill}, {Bouchet}, {Boulanger}, {Bucher}, {Burigana}, {Calabrese}, {Cardoso}, {Carron}, {Challinor}, {Chiang}, {Colombo}, {Combet}, {Couchot}, {Crill}, {Cuttaia}, {de Bernardis}, {de Rosa}, {de Zotti}, {Delabrouille}, {Delouis}, {Di Valentino}, {Diego}, {Dor{\'e}}, {Douspis}, {Ducout}, {Dupac}, {Efstathiou}, {Elsner}, {En{\ss}lin}, {Eriksen}, {Falgarone}, {Fantaye}, {Finelli}, {Frailis}, {Fraisse}, {Franceschi}, {Frolov}, {Galeotta}, {Galli}, {Ganga}, {G{\'e}nova-Santos}, {Gerbino}, {Ghosh}, {Gonz{\'a}lez-Nuevo}, {G{\'o}rski}, {Gratton}, {Gruppuso}, {Gudmundsson}, {Handley}, {Hansen}, {Henrot-Versill{\'e}}, {Herranz}, {Hivon}, {Huang}, {Jaffe}, {Jones}, {Karakci}, {Keih{\"a}nen}, {Keskitalo}, {Kiiveri}, {Kim}, {Kisner}, {Krachmalnicoff}, {Kunz},
  {Kurki-Suonio}, {Lagache}, {Lamarre}, {Lasenby}, {Lattanzi}, {Lawrence}, {Levrier}, {Liguori}, {Lilje}, {Lindholm}, {L{\'o}pez-Caniego}, {Ma}, {Mac{\'\i}as-P{\'e}rez}, {Maggio}, {Maino}, {Mandolesi}, {Mangilli}, {Martin}, {Mart{\'\i}nez-Gonz{\'a}lez}, {Matarrese}, {Mauri}, {McEwen}, {Melchiorri}, {Mennella}, {Migliaccio}, {Miville-Desch{\^e}nes}, {Molinari}, {Moneti}, {Montier}, {Morgante}, {Moss}, {Mottet}, {Natoli}, {Pagano}, {Paoletti}, {Partridge}, {Patanchon}, {Patrizii}, {Perdereau}, {Perrotta}, {Pettorino}, {Piacentini}, {Puget}, {Rachen}, {Reinecke}, {Remazeilles}, {Renzi}, {Rocha}, {Roudier}, {Salvati}, {Sandri}, {Savelainen}, {Scott}, {Sirignano}, {Sirri}, {Spencer}, {Sunyaev}, {Suur-Uski}, {Tauber}, {Tavagnacco}, {Tenti}, {Toffolatti}, {Tomasi}, {Tristram}, {Trombetti}, {Valiviita}, {Vansyngel}, {Van Tent}, {Vibert}, {Vielva}, {Villa}, {Vittorio}, {Wandelt}, {Wehus}, \& {Zonca}}]{P18A3}
{Planck Collaboration}, {Aghanim}, N., {Akrami}, Y., {et~al.} 2020{\natexlab{c}}, \aap, 641, A3, \dodoi{10.1051/0004-6361/201832909}

\bibitem[{{Planck Collaboration} {et~al.}(2020{\natexlab{d}}){Planck Collaboration}, {Aghanim}, {Akrami}, {Ashdown}, {Aumont}, {Baccigalupi}, {Ballardini}, {Banday}, {Barreiro}, {Bartolo}, {Basak}, {Battye}, {Benabed}, {Bernard}, {Bersanelli}, {Bielewicz}, {Bock}, {Bond}, {Borrill}, {Bouchet}, {Boulanger}, {Bucher}, {Burigana}, {Butler}, {Calabrese}, {Cardoso}, {Carron}, {Challinor}, {Chiang}, {Chluba}, {Colombo}, {Combet}, {Contreras}, {Crill}, {Cuttaia}, {de Bernardis}, {de Zotti}, {Delabrouille}, {Delouis}, {Di Valentino}, {Diego}, {Dor{\'e}}, {Douspis}, {Ducout}, {Dupac}, {Dusini}, {Efstathiou}, {Elsner}, {En{\ss}lin}, {Eriksen}, {Fantaye}, {Farhang}, {Fergusson}, {Fernandez-Cobos}, {Finelli}, {Forastieri}, {Frailis}, {Fraisse}, {Franceschi}, {Frolov}, {Galeotta}, {Galli}, {Ganga}, {G{\'e}nova-Santos}, {Gerbino}, {Ghosh}, {Gonz{\'a}lez-Nuevo}, {G{\'o}rski}, {Gratton}, {Gruppuso}, {Gudmundsson}, {Hamann}, {Handley}, {Hansen}, {Herranz}, {Hildebrandt}, {Hivon}, {Huang}, {Jaffe}, {Jones}, {Karakci},
  {Keih{\"a}nen}, {Keskitalo}, {Kiiveri}, {Kim}, {Kisner}, {Knox}, {Krachmalnicoff}, {Kunz}, {Kurki-Suonio}, {Lagache}, {Lamarre}, {Lasenby}, {Lattanzi}, {Lawrence}, {Le Jeune}, {Lemos}, {Lesgourgues}, {Levrier}, {Lewis}, {Liguori}, {Lilje}, {Lilley}, {Lindholm}, {L{\'o}pez-Caniego}, {Lubin}, {Ma}, {Mac{\'\i}as-P{\'e}rez}, {Maggio}, {Maino}, {Mandolesi}, {Mangilli}, {Marcos-Caballero}, {Maris}, {Martin}, {Martinelli}, {Mart{\'\i}nez-Gonz{\'a}lez}, {Matarrese}, {Mauri}, {McEwen}, {Meinhold}, {Melchiorri}, {Mennella}, {Migliaccio}, {Millea}, {Mitra}, {Miville-Desch{\^e}nes}, {Molinari}, {Montier}, {Morgante}, {Moss}, {Natoli}, {N{\o}rgaard-Nielsen}, {Pagano}, {Paoletti}, {Partridge}, {Patanchon}, {Peiris}, {Perrotta}, {Pettorino}, {Piacentini}, {Polastri}, {Polenta}, {Puget}, {Rachen}, {Reinecke}, {Remazeilles}, {Renzi}, {Rocha}, {Rosset}, {Roudier}, {Rubi{\~n}o-Mart{\'\i}n}, {Ruiz-Granados}, {Salvati}, {Sandri}, {Savelainen}, {Scott}, {Shellard}, {Sirignano}, {Sirri}, {Spencer}, {Sunyaev}, {Suur-Uski},
  {Tauber}, {Tavagnacco}, {Tenti}, {Toffolatti}, {Tomasi}, {Trombetti}, {Valenziano}, {Valiviita}, {Van Tent}, {Vibert}, {Vielva}, {Villa}, {Vittorio}, {Wandelt}, {Wehus}, {White}, {White}, {Zacchei}, \& {Zonca}}]{P18A6}
---. 2020{\natexlab{d}}, \aap, 641, A6, \dodoi{10.1051/0004-6361/201833910}

\bibitem[{{Planck Collaboration} {et~al.}(2020{\natexlab{e}}){Planck Collaboration}, {Aghanim}, {Akrami}, {Ashdown}, {Aumont}, {Baccigalupi}, {Ballardini}, {Banday}, {Barreiro}, {Bartolo}, {Basak}, {Benabed}, {Bernard}, {Bersanelli}, {Bielewicz}, {Bock}, {Bond}, {Borrill}, {Bouchet}, {Boulanger}, {Bucher}, {Burigana}, {Butler}, {Calabrese}, {Cardoso}, {Carron}, {Casaponsa}, {Challinor}, {Chiang}, {Colombo}, {Combet}, {Crill}, {Cuttaia}, {de Bernardis}, {de Rosa}, {de Zotti}, {Delabrouille}, {Delouis}, {Di Valentino}, {Diego}, {Dor{\'e}}, {Douspis}, {Ducout}, {Dupac}, {Dusini}, {Efstathiou}, {Elsner}, {En{\ss}lin}, {Eriksen}, {Fantaye}, {Fernandez-Cobos}, {Finelli}, {Frailis}, {Fraisse}, {Franceschi}, {Frolov}, {Galeotta}, {Galli}, {Ganga}, {G{\'e}nova-Santos}, {Gerbino}, {Ghosh}, {Giraud-H{\'e}raud}, {Gonz{\'a}lez-Nuevo}, {G{\'o}rski}, {Gratton}, {Gruppuso}, {Gudmundsson}, {Hamann}, {Handley}, {Hansen}, {Herranz}, {Hivon}, {Huang}, {Jaffe}, {Jones}, {Keih{\"a}nen}, {Keskitalo}, {Kiiveri}, {Kim}, {Kisner},
  {Krachmalnicoff}, {Kunz}, {Kurki-Suonio}, {Lagache}, {Lamarre}, {Lasenby}, {Lattanzi}, {Lawrence}, {Le Jeune}, {Levrier}, {Lewis}, {Liguori}, {Lilje}, {Lilley}, {Lindholm}, {L{\'o}pez-Caniego}, {Lubin}, {Ma}, {Mac{\'\i}as-P{\'e}rez}, {Maggio}, {Maino}, {Mandolesi}, {Mangilli}, {Marcos-Caballero}, {Maris}, {Martin}, {Mart{\'\i}nez-Gonz{\'a}lez}, {Matarrese}, {Mauri}, {McEwen}, {Meinhold}, {Melchiorri}, {Mennella}, {Migliaccio}, {Millea}, {Miville-Desch{\^e}nes}, {Molinari}, {Moneti}, {Montier}, {Morgante}, {Moss}, {Natoli}, {N{\o}rgaard-Nielsen}, {Pagano}, {Paoletti}, {Partridge}, {Patanchon}, {Peiris}, {Perrotta}, {Pettorino}, {Piacentini}, {Polenta}, {Puget}, {Rachen}, {Reinecke}, {Remazeilles}, {Renzi}, {Rocha}, {Rosset}, {Roudier}, {Rubi{\~n}o-Mart{\'\i}n}, {Ruiz-Granados}, {Salvati}, {Sandri}, {Savelainen}, {Scott}, {Shellard}, {Sirignano}, {Sirri}, {Spencer}, {Sunyaev}, {Suur-Uski}, {Tauber}, {Tavagnacco}, {Tenti}, {Toffolatti}, {Tomasi}, {Trombetti}, {Valiviita}, {Van Tent}, {Vielva}, {Villa},
  {Vittorio}, {Wandelt}, {Wehus}, {Zacchei}, \& {Zonca}}]{P18A5}
---. 2020{\natexlab{e}}, \aap, 641, A5, \dodoi{10.1051/0004-6361/201936386}

\bibitem[{{Qu} {et~al.}(2023){Qu}, {Sherwin}, {Madhavacheril}, {Han}, {Crowley}, {Abril-Cabezas}, {Ade}, {Aiola}, {Alford}, {Amiri}, {Amodeo}, {An}, {Atkins}, {Austermann}, {Battaglia}, {Battistelli}, {Beall}, {Bean}, {Beringue}, {Bhandarkar}, {Biermann}, {Bolliet}, {Bond}, {Cai}, {Calabrese}, {Calafut}, {Capalbo}, {Carrero}, {Carron}, {Challinor}, {Chesmore}, {Cho}, {Choi}, {Clark}, {C{\'o}rdova Rosado}, {Cothard}, {Coughlin}, {Coulton}, {Dalal}, {Darwish}, {Devlin}, {Dicker}, {Doze}, {Duell}, {Duff}, {Duivenvoorden}, {Dunkley}, {D{\"u}nner}, {Fanfani}, {Fankhanel}, {Farren}, {Ferraro}, {Freundt}, {Fuzia}, {Gallardo}, {Garrido}, {Gluscevic}, {Golec}, {Guan}, {Halpern}, {Harrison}, {Hasselfield}, {Healy}, {Henderson}, {Hensley}, {Herv{\'\i}as-Caimapo}, {Hill}, {Hilton}, {Hilton}, {Hincks}, {Hlo{\v{z}}ek}, {Ho}, {Huber}, {Hubmayr}, {Huffenberger}, {Hughes}, {Irwin}, {Isopi}, {Jense}, {Keller}, {Kim}, {Knowles}, {Koopman}, {Kosowsky}, {Kramer}, {Kusiak}, {La Posta}, {Lague}, {Lakey}, {Lee}, {Li}, {Li}, {Limon},
  {Lokken}, {Louis}, {Lungu}, {MacCrann}, {MacInnis}, {Maldonado}, {Maldonado}, {Mallaby-Kay}, {Marques}, {McMahon}, {Mehta}, {Menanteau}, {Moodley}, {Morris}, {Mroczkowski}, {Naess}, {Namikawa}, {Nati}, {Newburgh}, {Nicola}, {Niemack}, {Nolta}, {Orlowski-Scherer}, {Page}, {Pandey}, {Partridge}, {Prince}, {Puddu}, {Radiconi}, {Robertson}, {Rojas}, {Sakuma}, {Salatino}, {Schaan}, {Schmitt}, {Sehgal}, {Shaikh}, {Sierra}, {Sievers}, {Sif{\'o}n}, {Simon}, {Sonka}, {Spergel}, {Staggs}, {Storer}, {Switzer}, {Tampier}, {Thornton}, {Trac}, {Treu}, {Tucker}, {Ulluom}, {Vale}, {Van Engelen}, {Van Lanen}, {van Marrewijk}, {Vargas}, {Vavagiakis}, {Wagoner}, {Wang}, {Wenzl}, {Wollack}, {Xu}, {Zago}, \& {Zhang}}]{QSM+23}
{Qu}, F.~J., {Sherwin}, B.~D., {Madhavacheril}, M.~S., {et~al.} 2023, arXiv e-prints, arXiv:2304.05202, \dodoi{10.48550/arXiv.2304.05202}

\bibitem[{{Raveri} \& {Doux}(2021)}]{RD21}
{Raveri}, M., \& {Doux}, C. 2021, \prd, 104, 043504, \dodoi{10.1103/PhysRevD.104.043504}

\bibitem[{{Reeves} {et~al.}(2023){Reeves}, {Nicola}, {Refregier}, {Kacprzak}, \& {Machado Poletti Valle}}]{RNR+23}
{Reeves}, A., {Nicola}, A., {Refregier}, A., {Kacprzak}, T., \& {Machado Poletti Valle}, L.~F. 2023, arXiv e-prints, arXiv:2309.03258, \dodoi{10.48550/arXiv.2309.03258}

\bibitem[{{Robertson} {et~al.}(2021){Robertson}, {Alonso}, {Harnois-D{\'e}raps}, {Darwish}, {Kannawadi}, {Amon}, {Asgari}, {Bilicki}, {Calabrese}, {Choi}, {Devlin}, {Dunkley}, {Dvornik}, {Erben}, {Ferraro}, {Fortuna}, {Giblin}, {Han}, {Heymans}, {Hildebrandt}, {Hill}, {Hilton}, {Ho}, {Hoekstra}, {Hubmayr}, {Hughes}, {Joachimi}, {Joudaki}, {Knowles}, {Kuijken}, {Madhavacheril}, {Moodley}, {Miller}, {Namikawa}, {Nati}, {Niemack}, {Page}, {Partridge}, {Schaan}, {Schillaci}, {Schneider}, {Sehgal}, {Sherwin}, {Sif{\'o}n}, {Staggs}, {Tr{\"o}ster}, {van Engelen}, {Valentijn}, {Wollack}, {Wright}, \& {Xu}}]{RAH+21}
{Robertson}, N.~C., {Alonso}, D., {Harnois-D{\'e}raps}, J., {et~al.} 2021, \aap, 649, A146, \dodoi{10.1051/0004-6361/202039975}

\bibitem[{{Ross} {et~al.}(2015){Ross}, {Samushia}, {Howlett}, {Percival}, {Burden}, \& {Manera}}]{RSH+15}
{Ross}, A.~J., {Samushia}, L., {Howlett}, C., {et~al.} 2015, \mnras, 449, 835, \dodoi{10.1093/mnras/stv154}

\bibitem[{{Rozo} {et~al.}(2016){Rozo}, {Rykoff}, {Abate}, {Bonnett}, {Crocce}, {Davis}, {Hoyle}, {Leistedt}, {Peiris}, {Wechsler}, {Abbott}, {Abdalla}, {Banerji}, {Bauer}, {Benoit-L{\'e}vy}, {Bernstein}, {Bertin}, {Brooks}, {Buckley-Geer}, {Burke}, {Capozzi}, {Rosell}, {Carollo}, {Kind}, {Carretero}, {Castander}, {Childress}, {Cunha}, {D'Andrea}, {Davis}, {DePoy}, {Desai}, {Diehl}, {Dietrich}, {Doel}, {Eifler}, {Evrard}, {Neto}, {Flaugher}, {Fosalba}, {Frieman}, {Gaztanaga}, {Gerdes}, {Glazebrook}, {Gruen}, {Gruendl}, {Honscheid}, {James}, {Jarvis}, {Kim}, {Kuehn}, {Kuropatkin}, {Lahav}, {Lidman}, {Lima}, {Maia}, {March}, {Martini}, {Melchior}, {Miller}, {Miquel}, {Mohr}, {Nichol}, {Nord}, {O'Neill}, {Ogando}, {Plazas}, {Romer}, {Roodman}, {Sako}, {Sanchez}, {Santiago}, {Schubnell}, {Sevilla-Noarbe}, {Smith}, {Soares-Santos}, {Sobreira}, {Suchyta}, {Swanson}, {Thaler}, {Thomas}, {Uddin}, {Vikram}, {Walker}, {Wester}, {Zhang}, \& {da Costa}}]{R+_redmagic}
{Rozo}, E., {Rykoff}, E.~S., {Abate}, A., {et~al.} 2016, \mnras, 461, 1431, \dodoi{10.1093/mnras/stw1281}

\bibitem[{{Rudd} {et~al.}(2008){Rudd}, {Zentner}, \& {Kravtsov}}]{RZK08}
{Rudd}, D.~H., {Zentner}, A.~R., \& {Kravtsov}, A.~V. 2008, \apj, 672, 19, \dodoi{10.1086/523836}

\bibitem[{{Salcido} {et~al.}(2023){Salcido}, {McCarthy}, {Kwan}, {Upadhye}, \& {Font}}]{SMK+23}
{Salcido}, J., {McCarthy}, I.~G., {Kwan}, J., {Upadhye}, A., \& {Font}, A.~S. 2023, \mnras, \dodoi{10.1093/mnras/stad1474}

\bibitem[{{Schaye} {et~al.}(2015){Schaye}, {Crain}, {Bower}, {Furlong}, {Schaller}, {Theuns}, {Dalla Vecchia}, {Frenk}, {McCarthy}, {Helly}, {Jenkins}, {Rosas-Guevara}, {White}, {Baes}, {Booth}, {Camps}, {Navarro}, {Qu}, {Rahmati}, {Sawala}, {Thomas}, \& {Trayford}}]{SCB+15}
{Schaye}, J., {Crain}, R.~A., {Bower}, R.~G., {et~al.} 2015, \mnras, 446, 521, \dodoi{10.1093/mnras/stu2058}

\bibitem[{{Schneider} {et~al.}(2022){Schneider}, {Giri}, {Amodeo}, \& {Refregier}}]{SGA+22}
{Schneider}, A., {Giri}, S.~K., {Amodeo}, S., \& {Refregier}, A. 2022, \mnras, 514, 3802, \dodoi{10.1093/mnras/stac1493}

\bibitem[{{Schneider} \& {Teyssier}(2015)}]{ST15}
{Schneider}, A., \& {Teyssier}, R. 2015, \jcap, 2015, 049, \dodoi{10.1088/1475-7516/2015/12/049}

\bibitem[{{Schneider} {et~al.}(2019){Schneider}, {Teyssier}, {Stadel}, {Chisari}, {Le Brun}, {Amara}, \& {Refregier}}]{STS+19}
{Schneider}, A., {Teyssier}, R., {Stadel}, J., {et~al.} 2019, \jcap, 2019, 020, \dodoi{10.1088/1475-7516/2019/03/020}

\bibitem[{{Schneider} {et~al.}(2002){Schneider}, {van Waerbeke}, {Kilbinger}, \& {Mellier}}]{SWK+02}
{Schneider}, P., {van Waerbeke}, L., {Kilbinger}, M., \& {Mellier}, Y. 2002, \aap, 396, 1, \dodoi{10.1051/0004-6361:20021341}

\bibitem[{{Scolnic} {et~al.}(2018){Scolnic}, {Jones}, {Rest}, {Pan}, {Chornock}, {Foley}, {Huber}, {Kessler}, {Narayan}, {Riess}, {Rodney}, {Berger}, {Brout}, {Challis}, {Drout}, {Finkbeiner}, {Lunnan}, {Kirshner}, {Sanders}, {Schlafly}, {Smartt}, {Stubbs}, {Tonry}, {Wood-Vasey}, {Foley}, {Hand}, {Johnson}, {Burgett}, {Chambers}, {Draper}, {Hodapp}, {Kaiser}, {Kudritzki}, {Magnier}, {Metcalfe}, {Bresolin}, {Gall}, {Kotak}, {McCrum}, \& {Smith}}]{SJR+18}
{Scolnic}, D.~M., {Jones}, D.~O., {Rest}, A., {et~al.} 2018, \apj, 859, 101, \dodoi{10.3847/1538-4357/aab9bb}

\bibitem[{{Secco} {et~al.}(2022){Secco}, {Samuroff}, {Krause}, {Jain}, {Blazek}, {Raveri}, {Campos}, {Amon}, {Chen}, {Doux}, {Choi}, {Gruen}, {Bernstein}, {Chang}, {DeRose}, {Myles}, {Fert{\'e}}, {Lemos}, {Huterer}, {Prat}, {Troxel}, {MacCrann}, {Liddle}, {Kacprzak}, {Fang}, {S{\'a}nchez}, {Pandey}, {Dodelson}, {Chintalapati}, {Hoffmann}, {Alarcon}, {Alves}, {Andrade-Oliveira}, {Baxter}, {Bechtol}, {Becker}, {Brandao-Souza}, {Camacho}, {Carnero Rosell}, {Carrasco Kind}, {Cawthon}, {Cordero}, {Crocce}, {Davis}, {Di Valentino}, {Drlica-Wagner}, {Eckert}, {Eifler}, {Elidaiana}, {Elsner}, {Elvin-Poole}, {Everett}, {Fosalba}, {Friedrich}, {Gatti}, {Giannini}, {Gruendl}, {Harrison}, {Hartley}, {Herner}, {Huang}, {Huff}, {Jarvis}, {Jeffrey}, {Kuropatkin}, {Leget}, {Muir}, {Mccullough}, {Navarro Alsina}, {Omori}, {Park}, {Porredon}, {Rollins}, {Roodman}, {Rosenfeld}, {Ross}, {Rykoff}, {Sanchez}, {Sevilla-Noarbe}, {Sheldon}, {Shin}, {Troja}, {Tutusaus}, {Varga}, {Weaverdyck}, {Wechsler}, {Yanny}, {Yin}, {Zhang},
  {Zuntz}, {Abbott}, {Aguena}, {Allam}, {Annis}, {Bacon}, {Bertin}, {Bhargava}, {Bridle}, {Brooks}, {Buckley-Geer}, {Burke}, {Carretero}, {Costanzi}, {da Costa}, {De Vicente}, {Diehl}, {Dietrich}, {Doel}, {Ferrero}, {Flaugher}, {Frieman}, {Garc{\'\i}a-Bellido}, {Gaztanaga}, {Gerdes}, {Giannantonio}, {Gschwend}, {Gutierrez}, {Hinton}, {Hollowood}, {Honscheid}, {Hoyle}, {James}, {Jeltema}, {Kuehn}, {Lahav}, {Lima}, {Lin}, {Maia}, {Marshall}, {Martini}, {Melchior}, {Menanteau}, {Miquel}, {Mohr}, {Morgan}, {Ogando}, {Palmese}, {Paz-Chinch{\'o}n}, {Petravick}, {Pieres}, {Plazas Malag{\'o}n}, {Rodriguez-Monroy}, {Romer}, {Sanchez}, {Scarpine}, {Schubnell}, {Scolnic}, {Serrano}, {Smith}, {Soares-Santos}, {Suchyta}, {Swanson}, {Tarle}, {Thomas}, {To}, \& {DES Collaboration}}]{DES_Y3_WL2}
{Secco}, L.~F., {Samuroff}, S., {Krause}, E., {et~al.} 2022, \prd, 105, 023515, \dodoi{10.1103/PhysRevD.105.023515}

\bibitem[{{Sellentin} {et~al.}(2018){Sellentin}, {Heymans}, \& {Harnois-D{\'e}raps}}]{SHH18}
{Sellentin}, E., {Heymans}, C., \& {Harnois-D{\'e}raps}, J. 2018, \mnras, 477, 4879, \dodoi{10.1093/mnras/sty988}

\bibitem[{{Semboloni} {et~al.}(2013){Semboloni}, {Hoekstra}, \& {Schaye}}]{SHS13}
{Semboloni}, E., {Hoekstra}, H., \& {Schaye}, J. 2013, \mnras, 434, 148, \dodoi{10.1093/mnras/stt1013}

\bibitem[{{Semboloni} {et~al.}(2011){Semboloni}, {Hoekstra}, {Schaye}, {van Daalen}, \& {McCarthy}}]{SHS+11}
{Semboloni}, E., {Hoekstra}, H., {Schaye}, J., {van Daalen}, M.~P., \& {McCarthy}, I.~G. 2011, \mnras, 417, 2020, \dodoi{10.1111/j.1365-2966.2011.19385.x}

\bibitem[{{Sheldon} \& {Huff}(2017)}]{SH17}
{Sheldon}, E.~S., \& {Huff}, E.~M. 2017, \apj, 841, 24, \dodoi{10.3847/1538-4357/aa704b}

\bibitem[{{Springel} {et~al.}(2018){Springel}, {Pakmor}, {Pillepich}, {Weinberger}, {Nelson}, {Hernquist}, {Vogelsberger}, {Genel}, {Torrey}, {Marinacci}, \& {Naiman}}]{SPP+18}
{Springel}, V., {Pakmor}, R., {Pillepich}, A., {et~al.} 2018, \mnras, 475, 676, \dodoi{10.1093/mnras/stx3304}

\bibitem[{{Sugiyama} {et~al.}(2023){Sugiyama}, {Miyatake}, {More}, {Li}, {Shirasaki}, {Takada}, {Kobayashi}, {Takahashi}, {Nishimichi}, {Nishizawa}, {Rau}, {Zhang}, {Dalal}, {Mandelbaum}, {Strauss}, {Hamana}, {Oguri}, {Osato}, {Kannawadi}, {Armstrong}, {Komiyama}, {Lupton}, {Lust}, {Miyazaki}, {Murayama}, {Okura}, {Price}, {Tait}, {Tanaka}, \& {Wang}}]{HSC_Y3_SMM+23}
{Sugiyama}, S., {Miyatake}, H., {More}, S., {et~al.} 2023, arXiv e-prints, arXiv:2304.00705, \dodoi{10.48550/arXiv.2304.00705}

\bibitem[{{Takada} \& {Hu}(2013)}]{TH13}
{Takada}, M., \& {Hu}, W. 2013, \prd, 87, 123504, \dodoi{10.1103/PhysRevD.87.123504}

\bibitem[{{Takahashi} {et~al.}(2012){Takahashi}, {Sato}, {Nishimichi}, {Taruya}, \& {Oguri}}]{TSN+12}
{Takahashi}, R., {Sato}, M., {Nishimichi}, T., {Taruya}, A., \& {Oguri}, M. 2012, \apj, 761, 152, \dodoi{10.1088/0004-637X/761/2/152}

\bibitem[{{Tenneti} {et~al.}(2015){Tenneti}, {Mandelbaum}, {Di Matteo}, {Kiessling}, \& {Khandai}}]{TMM+15}
{Tenneti}, A., {Mandelbaum}, R., {Di Matteo}, T., {Kiessling}, A., \& {Khandai}, N. 2015, \mnras, 453, 469, \dodoi{10.1093/mnras/stv1625}

\bibitem[{{To} {et~al.}(2021){To}, {Krause}, {Rozo}, {Wu}, {Gruen}, {Wechsler}, {Eifler}, {Rykoff}, {Costanzi}, {Becker}, {Bernstein}, {Blazek}, {Bocquet}, {Bridle}, {Cawthon}, {Choi}, {Crocce}, {Davis}, {DeRose}, {Drlica-Wagner}, {Elvin-Poole}, {Fang}, {Farahi}, {Friedrich}, {Gatti}, {Gaztanaga}, {Giannantonio}, {Hartley}, {Hoyle}, {Jarvis}, {MacCrann}, {McClintock}, {Miranda}, {Pereira}, {Park}, {Porredon}, {Prat}, {Rau}, {Ross}, {Samuroff}, {S{\'a}nchez}, {Sevilla-Noarbe}, {Sheldon}, {Troxel}, {Varga}, {Vielzeuf}, {Zhang}, {Zuntz}, {Abbott}, {Aguena}, {Amon}, {Annis}, {Avila}, {Bertin}, {Bhargava}, {Brooks}, {Burke}, {Carnero Rosell}, {Carrasco Kind}, {Carretero}, {Chang}, {Conselice}, {da Costa}, {Davis}, {Desai}, {Diehl}, {Dietrich}, {Everett}, {Evrard}, {Ferrero}, {Flaugher}, {Fosalba}, {Frieman}, {Garc{\'\i}a-Bellido}, {Gruendl}, {Gutierrez}, {Hinton}, {Hollowood}, {Honscheid}, {Huterer}, {James}, {Jeltema}, {Kron}, {Kuehn}, {Kuropatkin}, {Lima}, {Maia}, {Marshall}, {Menanteau}, {Miquel}, {Morgan},
  {Muir}, {Myles}, {Palmese}, {Paz-Chinch{\'o}n}, {Plazas}, {Romer}, {Roodman}, {Sanchez}, {Santiago}, {Scarpine}, {Serrano}, {Smith}, {Suchyta}, {Swanson}, {Tarle}, {Thomas}, {Tucker}, {Weller}, {Wester}, {Wilkinson}, \& {DES Collaboration}}]{TKR+21}
{To}, C., {Krause}, E., {Rozo}, E., {et~al.} 2021, \prl, 126, 141301, \dodoi{10.1103/PhysRevLett.126.141301}

\bibitem[{{Torrado} \& {Lewis}(2019)}]{TL19}
{Torrado}, J., \& {Lewis}, A. 2019, {Cobaya: Bayesian analysis in cosmology}, Astrophysics Source Code Library, record ascl:1910.019.
\newblock \doeprint{1910.019}

\bibitem[{Torrado \& Lewis(2021)}]{TL21}
Torrado, J., \& Lewis, A. 2021, JCAP, 05, 057, \dodoi{10.1088/1475-7516/2021/05/057}

\bibitem[{{Tr{\"o}ster} {et~al.}(2022){Tr{\"o}ster}, {Mead}, {Heymans}, {Yan}, {Alonso}, {Asgari}, {Bilicki}, {Dvornik}, {Hildebrandt}, {Joachimi}, {Kannawadi}, {Kuijken}, {Schneider}, {Shan}, {van Waerbeke}, \& {Wright}}]{TMH+22}
{Tr{\"o}ster}, T., {Mead}, A.~J., {Heymans}, C., {et~al.} 2022, \aap, 660, A27, \dodoi{10.1051/0004-6361/202142197}

\bibitem[{{Troxel} {et~al.}(2018){Troxel}, {Krause}, {Chang}, {Eifler}, {Friedrich}, {Gruen}, {MacCrann}, {Chen}, {Davis}, {DeRose}, {Dodelson}, {Gatti}, {Hoyle}, {Huterer}, {Jarvis}, {Lacasa}, {Lemos}, {Peiris}, {Prat}, {Samuroff}, {S{\'a}nchez}, {Sheldon}, {Vielzeuf}, {Wang}, {Zuntz}, {Lahav}, {Abdalla}, {Allam}, {Annis}, {Avila}, {Bertin}, {Brooks}, {Burke}, {Carnero Rosell}, {Carrasco Kind}, {Carretero}, {Crocce}, {Cunha}, {D'Andrea}, {da Costa}, {De Vicente}, {Diehl}, {Doel}, {Evrard}, {Flaugher}, {Fosalba}, {Frieman}, {Garc{\'\i}a-Bellido}, {Gaztanaga}, {Gerdes}, {Gruendl}, {Gschwend}, {Gutierrez}, {Hartley}, {Hollowood}, {Honscheid}, {James}, {Kirk}, {Kuehn}, {Kuropatkin}, {Li}, {Lima}, {March}, {Menanteau}, {Miquel}, {Mohr}, {Ogando}, {Plazas}, {Roodman}, {Sanchez}, {Scarpine}, {Schindler}, {Sevilla-Noarbe}, {Smith}, {Soares-Santos}, {Sobreira}, {Suchyta}, {Swanson}, {Thomas}, {Walker}, \& {Wechsler}}]{TKC+18}
{Troxel}, M.~A., {Krause}, E., {Chang}, C., {et~al.} 2018, \mnras, 479, 4998, \dodoi{10.1093/mnras/sty1889}

\bibitem[{{van Daalen} {et~al.}(2020){van Daalen}, {McCarthy}, \& {Schaye}}]{DMS20}
{van Daalen}, M.~P., {McCarthy}, I.~G., \& {Schaye}, J. 2020, \mnras, 491, 2424, \dodoi{10.1093/mnras/stz3199}

\bibitem[{{van Uitert} {et~al.}(2018){van Uitert}, {Joachimi}, {Joudaki}, {Amon}, {Heymans}, {K{\"o}hlinger}, {Asgari}, {Blake}, {Choi}, {Erben}, {Farrow}, {Harnois-D{\'e}raps}, {Hildebrandt}, {Hoekstra}, {Kitching}, {Klaes}, {Kuijken}, {Merten}, {Miller}, {Nakajima}, {Schneider}, {Valentijn}, \& {Viola}}]{vUJJ+18}
{van Uitert}, E., {Joachimi}, B., {Joudaki}, S., {et~al.} 2018, \mnras, 476, 4662, \dodoi{10.1093/mnras/sty551}

\bibitem[{{Villaescusa-Navarro} {et~al.}(2021){Villaescusa-Navarro}, {Angl{\'e}s-Alc{\'a}zar}, {Genel}, {Spergel}, {Somerville}, {Dave}, {Pillepich}, {Hernquist}, {Nelson}, {Torrey}, {Narayanan}, {Li}, {Philcox}, {La Torre}, {Maria Delgado}, {Ho}, {Hassan}, {Burkhart}, {Wadekar}, {Battaglia}, {Contardo}, \& {Bryan}}]{VAG+21}
{Villaescusa-Navarro}, F., {Angl{\'e}s-Alc{\'a}zar}, D., {Genel}, S., {et~al.} 2021, \apj, 915, 71, \dodoi{10.3847/1538-4357/abf7ba}

\bibitem[{{Vogelsberger} {et~al.}(2014){Vogelsberger}, {Genel}, {Springel}, {Torrey}, {Sijacki}, {Xu}, {Snyder}, {Nelson}, \& {Hernquist}}]{VGS+14}
{Vogelsberger}, M., {Genel}, S., {Springel}, V., {et~al.} 2014, \mnras, 444, 1518, \dodoi{10.1093/mnras/stu1536}

\bibitem[{{White}(2004)}]{M04}
{White}, M. 2004, Astroparticle Physics, 22, 211, \dodoi{10.1016/j.astropartphys.2004.06.001}

\bibitem[{{Yan} {et~al.}(2021){Yan}, {van Waerbeke}, {Tr{\"o}ster}, {Wright}, {Alonso}, {Asgari}, {Bilicki}, {Erben}, {Gu}, {Heymans}, {Hildebrandt}, {Hinshaw}, {Koukoufilippas}, {Kannawadi}, {Kuijken}, {Mead}, \& {Shan}}]{YWT+21}
{Yan}, Z., {van Waerbeke}, L., {Tr{\"o}ster}, T., {et~al.} 2021, \aap, 651, A76, \dodoi{10.1051/0004-6361/202140568}

\bibitem[{{Zentner} {et~al.}(2008){Zentner}, {Rudd}, \& {Hu}}]{ZRH08}
{Zentner}, A.~R., {Rudd}, D.~H., \& {Hu}, W. 2008, \prd, 77, 043507, \dodoi{10.1103/PhysRevD.77.043507}

\bibitem[{{Zentner} {et~al.}(2013){Zentner}, {Semboloni}, {Dodelson}, {Eifler}, {Krause}, \& {Hearin}}]{ZSD+13}
{Zentner}, A.~R., {Semboloni}, E., {Dodelson}, S., {et~al.} 2013, \prd, 87, 043509, \dodoi{10.1103/PhysRevD.87.043509}

\bibitem[{{Zhong} {et~al.}(2023){Zhong}, {Saraivanov}, {Miranda}, {Xu}, {Eifler}, \& {Krause}}]{ZSM+23}
{Zhong}, K., {Saraivanov}, E., {Miranda}, V., {et~al.} 2023, \prd, 107, 123529, \dodoi{10.1103/PhysRevD.107.123529}

\bibitem[{Zonca {et~al.}(2019)Zonca, Singer, Lenz, Reinecke, Rosset, Hivon, \& Gorski}]{Zonca2019}
Zonca, A., Singer, L., Lenz, D., {et~al.} 2019, Journal of Open Source Software, 4, 1298, \dodoi{10.21105/joss.01298}

\bibitem[{{Zuntz} {et~al.}(2018){Zuntz}, {Sheldon}, {Samuroff}, {Troxel}, {Jarvis}, {MacCrann}, {Gruen}, {Prat}, {S{\'a}nchez}, {Choi}, {Bridle}, {Bernstein}, {Dodelson}, {Drlica-Wagner}, {Fang}, {Gruendl}, {Hoyle}, {Huff}, {Jain}, {Kirk}, {Kacprzak}, {Krawiec}, {Plazas}, {Rollins}, {Rykoff}, {Sevilla-Noarbe}, {Soergel}, {Varga}, {Abbott}, {Abdalla}, {Allam}, {Annis}, {Bechtol}, {Benoit-L{\'e}vy}, {Bertin}, {Buckley-Geer}, {Burke}, {Carnero Rosell}, {Carrasco Kind}, {Carretero}, {Castander}, {Crocce}, {Cunha}, {D'Andrea}, {da Costa}, {Davis}, {Desai}, {Diehl}, {Dietrich}, {Doel}, {Eifler}, {Estrada}, {Evrard}, {Fausti Neto}, {Fernandez}, {Flaugher}, {Fosalba}, {Frieman}, {Garc{\'\i}a-Bellido}, {Gaztanaga}, {Gerdes}, {Giannantonio}, {Gschwend}, {Gutierrez}, {Hartley}, {Honscheid}, {James}, {Jeltema}, {Johnson}, {Johnson}, {Kuehn}, {Kuhlmann}, {Kuropatkin}, {Lahav}, {Li}, {Lima}, {Maia}, {March}, {Martini}, {Melchior}, {Menanteau}, {Miller}, {Miquel}, {Mohr}, {Neilsen}, {Nichol}, {Ogando}, {Roe}, {Romer},
  {Roodman}, {Sanchez}, {Scarpine}, {Schindler}, {Schubnell}, {Smith}, {Smith}, {Soares-Santos}, {Sobreira}, {Suchyta}, {Swanson}, {Tarle}, {Thomas}, {Tucker}, {Vikram}, {Walker}, {Wechsler}, {Zhang}, \& {DES Collaboration}}]{ZSS+18}
{Zuntz}, J., {Sheldon}, E., {Samuroff}, S., {et~al.} 2018, \mnras, 481, 1149, \dodoi{10.1093/mnras/sty2219}

\end{thebibliography}
\bibliographystyle{aasjournal}



\end{document}